\documentclass[12pt,preprint]{aastex}

 \usepackage{epstopdf}
\newcommand\WISE{\emph{WISE}}
\newcommand\NEOWISE{\emph{NEOWISE}}

\newcommand{\ie}{{\it i.e.}}
\newcommand{\eg}{{\it e.g.}}

\begin{document}

 \DeclareGraphicsExtensions{.pdf,.gif,.jpg}

\title{Binary Candidates in the Jovian Trojan and Hilda \\Populations from \NEOWISE\, Lightcurves}
\author{S. Sonnett\altaffilmark{1}, A. Mainzer\altaffilmark{1}, T. Grav\altaffilmark{2}, J. Masiero\altaffilmark{1}, J. Bauer\altaffilmark{1}}

\altaffiltext{1}{Jet Propulsion Laboratory, California Institute of Technology, Pasadena, CA 91109 USA}
\altaffiltext{2}{Planetary Science Institute, Tucson, AZ USA}

 \email{Sarah.Sonnett@jpl.nasa.gov}

 \begin{abstract}
 
Determining the binary fraction for a population of asteroids, particularly as a function of separation between the two components, helps describe the dynamical environment at the time the binaries formed, which in turn offers constraints on the dynamical evolution of the solar system.  We searched the \NEOWISE\, archival dataset for close and contact binary Trojans and Hildas via their diagnostically large lightcurve amplitudes.  We present 48 out of 554 Hilda and 34 out of 953 Trojan binary candidates in need of follow-up to confirm their large lightcurve amplitudes and subsequently constrain the binary orbit and component sizes.  From these candidates, we calculate a preliminary estimate of the binary fraction without confirmation or debiasing of $14-23$\% for Trojans larger than $\sim 12$ km and $30-51$\% for Hildas larger than $\sim 4$ km.  Once the binary candidates have been confirmed, it should be possible to infer the underlying, debiased binary fraction through estimation of survey biases.

 \end{abstract}

 \section{Introduction}

Trojan asteroids lie in stable orbits at the L4 (leading) and L5 (trailing) Lagrange points of a planet.  There are currently $\sim 5,500$ Jovian Trojan asteroids known, making them the most numerous known Trojan population and thus one of the most useful for constraining the dynamical processes that shaped their orbits and physical states (size, structure, etc.).  Just inward of the Jovian Trojans are the Hildas in 3:2 orbital resonance with Jupiter.  In early solar system formation models of minimal planetary migration, Jovian Trojans (hereafter, Trojans) and Hildas were captured relatively gently in situ \citep{1989aste.conf..487S,1998A&A...339..278M}.  

The Nice model instead proposes that when Jupiter and Saturn reached 2:1 orbital resonance, a violent scattering episode was ignited, with Neptune moving into the Trans-Neptunian region and chaotically scattering planetesimals \citep[\eg,][]{2005Natur.435..462M,2005Natur.435..459T,2005Natur.435..466G,2011AJ....142..152L}.  The Nice model thus predicts that Trojans were captured from the trans-Neptunian region and experienced a turbulent dynamical environment relative to previous formation models.  A later version of the Nice model suggests that if one of the ice giants traversed one of the Trojan clouds during migration, the clouds would undergo asymmetric depletion, producing the difference in population ratio observed today \citep[$N_{L4} : N_{L5} = 1.4 \pm 0.2$;][]{2011ApJ...742...40G,2013ApJ...768...45N}.  Combined with the Grand Tack model of inner solar system mixing through Jupiter's migration, the Nice model also predicts Hildas to have similar origins as Trojans \citep{2009Natur.460..364L,2011Natur.475..206W}.

In order to help discern the Trojans' formation location, their present dynamical state should be well-characterized and compared with those of other small body populations like the Hildas.  Determining the fraction of Trojans and Hildas in binary or multiple systems is one of the fundamental modes of constraining dynamical and collisional history.  For example, a turbulent environment like the one described in the Nice model might imply more interaction between small bodies and consequently a higher probability of either disrupting more weakly bound wide binaries or causing wide binaries to spiral inward, in which case we should see a low wide binary fraction but perhaps a high tight binary fraction \citep[separations less than five times the Hill radius of the primary][]{2011ApJ...727L...3P}.  Several binary formation models exist, each of which make a set of predictions about the binary's synodic orbit and sometimes its physical properties (mass ratio, similarity between component surfaces, etc.).  For example, dynamical friction tends to produce tight binaries while exchange reactions and three-body interactions increases mutual separation, favoring production of wide binaries \citep{2002Natur.420..643G,2002Icar..160..212W,2004Natur.427..518F,2005MNRAS.360..401A}.  

Some tight binaries can be identified by their lightcurves.  If the binary components are near-fluid rubble piles and not monoliths, they become tidally elongated toward each other, distorting into Jacobi ellipsoid shapes stretched along the semi major axis of the system \citep{1969efe..book.....C}.  The lightcurve of a binary made of two elongated components can have an amplitude so high that it cannot be explained by a singular equilibrium rubble pile.  Very large amplitude lightcurves therefore offer a means of identifying candidate rubble pile binaries \citep[$\Delta m > 0.9$ magnitudes compared to an average lightcurve amplitude of 0.3 magnitudes for Trojans;][]{2004AJ....127.3023S,2009Icar..202..134W}.  This technique of identifying binary candidates is limited to systems oriented such that the variation can be observed and with mass ratios high enough ($\geq 0.6$ mags) to cause sufficiently diagnostic elongation of the components \citep{1984A&A...140..265L}.  Still, \cite{2007AJ....134.1133M} successfully used it to identify two L5 Trojan binaries (17365 and 29314 Eurydamas).

Apart from 17365 (1978 VF$_{11}$) and 29314 Eurydamas, two other Trojans binaries are known: 617 Patroclus-Menoetius and 624 Hektor \citep{2001IAUC.7741....2M,2006DPS....38.6507M}.  624 Hektor is in fact a triple system possibly formed through a low-velocity collision, with a bilobate primary and a moderately separated satellite \citep{2014ApJ...783L..37M}.  617 Patroclus-Menoetius is a moderately separated system consisting of two spheroids nearly equal in size and with a low bulk density \citep{2006Natur.439..565M}. 

In addition to the \cite{2007AJ....134.1133M} survey for tight binaries, three other dedicated observational surveys for wide Trojan binaries have been conducted.  \cite{2006DPS....38.6507M} used high-resolution direct imaging to search for L4 binaries, detecting one (624 Hektor) out of 55 objects observed.  \cite{2007DPS....39.6009M} also used direct imaging on a sample of 35 Trojans, finding no binaries.  Lastly, \cite{2014LPI....45.1703N} directly imaged 8 Trojans, finding no binaries.  Surveys that could inadvertently detect tight binaries by being aimed at determining rotation periods and amplitudes of Trojans have mostly sampled only large objects ($\gtrsim 30$ km) and found no bound pairs \citep[\eg,][]{1992Icar...95..222B,2011AJ....141..170M}.  The Hildas have never been explicitly searched for binaries, though several have well-constrained lightcurves, some of which exhibit large amplitudes typical of contact binaries \citep{1988Icar...73..487H,1998Icar..133..247D,1999Icar..138..259D}.  

In this work, we seek to more fully explore the tight binary fraction in an effort to understand how Trojans are dynamically linked to other small body populations.  To that end, we harvested Trojan and Hilda lightcurves identified by the solar system data processing portion of the Wide-field Infrared Survey Explorer mission \citep[\WISE;][]{2010AJ....140.1868W}, known as NEOWISE \citep{2011ApJ...731...53M}.  Here, we present the candidates identified by our binary search algorithm for objects within our sensitivity range in the 12 $\mu$m band (roughly corresponding to diameters $\gtrsim 12$ km).  Follow-up is needed on each of these 29 candidates previously not known to be binary in order to: (i) reduce the uncertainty in their photometric ranges, which in some cases is needed to confirm their high amplitudes; and (ii) enable characterization of the system through detailed Roche binary modeling of the component sizes and orientations.  In an upcoming publication, we will report the binary fraction that can be extrapolated from these candidates as a function of dynamical class (Trojans vs. Hildas), Trojan cloud designation, taxonomic type, and separation between components. 

\section{Observations}

Lightcurve data were taken by the \WISE\, spacecraft, which conducted a space-based all-sky survey that operated in four bandpasses simultaneously: 3.4, 4.6, 12, and 22 $\mu$m \citep[denoted W1, W2, W3, and W4;][]{2010AJ....140.1868W}.  The \WISE\, observing cadence typically provided 12 observations per object per bandpass spanning $\sim 36$ hours.  Several Trojans and Hildas were also observed at multiple epochs \citep{2012ApJ...759...49G}.  Profile-fitting photometry was done using the \WISE\, science data processing pipeline described in \cite{2012wise.rept....1C}.  

\begin{figure}
\figurenum{1}
\includegraphics[width=6.5in]{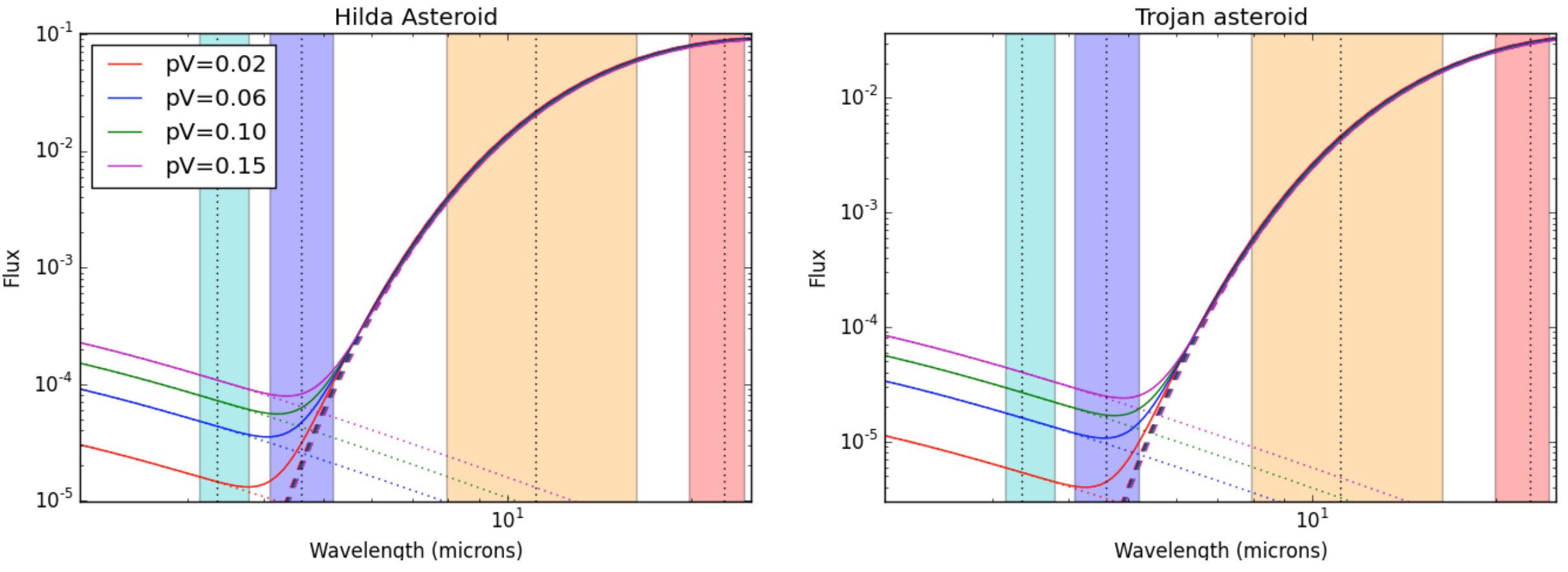}
\caption{\label{Fig:THSED} Theoretical spectral energy distributions for Trojans and Hildas with four different albedos. Fluxes are in Jy. Dotted curves show the reflected light component, dashed curves show the emitted light, and solid curves are the combined flux. The vertical black dotted lines show the band centers for the four WISE filters, and the vertical colored bands show the effective wavelength coverage of the W1, W2, W3, and W4 bands \citep[cyan, purple, orange, and red, respectively;][]{2010AJ....140.1868W}.  We used the W3 filter at 12 $\mu$m to identify diagnostically large lightcurve amplitudes in this work.  For the nominal range of Trojan and Hilda albedos (0.02, 0.06, 0.10, 0.15 for red, blue, green, magenta, respectively) the W3 filter is dominated by thermal emission.}
\end{figure}

The \WISE\, cadence with 3 hour spacing covering a 1.5 day span cannot be repeated from ground-based observatories unless telescopes at multiple longitudes are coordinated, so NEOWISE lightcurves offer a nearly unique advantage in sampling periodicities on the order of $\sim1-2$ days \citep[\eg, Main Belt large-amplitude eclipsing binaries 854 Frostia, 1313 Berna, and 4492 Debussy with synodic periods 37.728, 25.464, and 26.606 hours, respectively;][]{2006A&A...446.1177B}.  
Also, the W3 and W4 bandpasses contain almost purely thermal emission from Trojans and Hildas, somewhat isolating shape as the cause of brightness variations (Fig. \ref{Fig:THSED}).  Moreover, the peak of the Trojan and Hilda black bodies lie between W3 and W4, making them relatively bright at those wavelengths.  Of these two bandpasses, W3 is more sensitive, making it an ideal filter choice for calculating the lightcurve amplitude.  

\section{Sample and Analysis\label{Sec:Sample}}

\begin{figure}
\figurenum{2}
\includegraphics[width=6in]{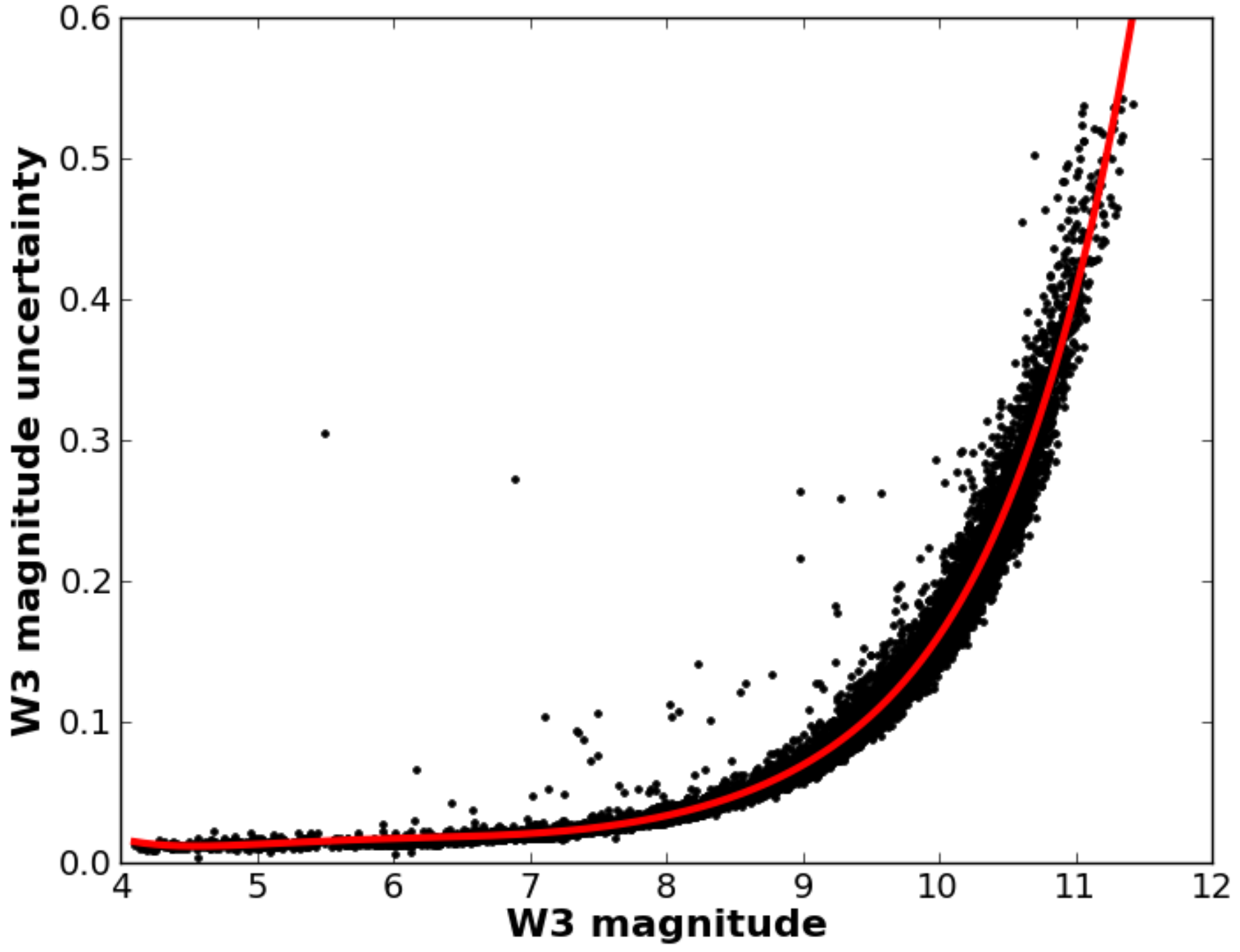}
\caption{\label{Fig:W3magVsW3sigma} Six-term polynomial fit (solid red line) to the Trojan and Hilda W3 magnitudes versus their photometric uncertainties (black dots).  This relationship was used in helping determine the sensitivity limits of our binary search technique. }
\end{figure}

In total, \WISE\, observed $\sim 1800$ Trojans and $\sim 1100$ Hildas, with diameters between $4-150$ km and $1-220$ km, respectively \citep{2011ApJ...742...40G,2012ApJ...744..197G}.  We chose to limit our sample to objects that could have binary lightcurve minima with a signal-to-noise (S/N) greater than five (\ie, magnitude uncertainty $\leq 0.2$ magnitudes).  We explored the relationship between W3 magnitude of our sample and its uncertainty by fitting polynomials with various numbers of terms (Fig. \ref{Fig:W3magVsW3sigma}).  We found that a six-term polynomial produced the best quality fit (lowest $\chi_{\nu}^{2}$).  The typical W3 magnitude corresponding to a W3 uncertainty of 0.2 was 10.3 mag, which roughly corresponds to $\sim12$ km for the Trojans and $\sim 4$ km for the Hildas.  Therefore, we limited our sample to only objects with mean W3 magnitudes such that their binary lightcurve minima would be brighter than 10.3 magnitudes in W3, meaning that at no point will a typical large-amplitude binary have S/N $< 5$.  We were left with 953 Trojans and 554 Hildas in our sample that met these criteria.

Choosing a sample with this physical size range meant including objects with a range of possible structures, which could affect the applicability of our search technique.  This method of identifying binary candidates through large lightcurve amplitudes relies on the components being near-fluid rubble piles that tidally distort as opposed to rigid monoliths.  Shape and structure of an asteroid is thought to correlate with its size, with larger objects being massive enough to remain gravitationally bound after successive impacts and smaller objects likely being monolithic collisional remnants \citep[\eg,][]{1982Icar...52..409F}.  The diameter below which a small body population starts to become strength-dominated monoliths is not well known.  Observational studies found that sub-kilometer sized near-Earth asteroids contain a large fraction of objects spinning faster than the critical period (below which rubble piles spin apart), suggesting that asteroids start to become strength-dominated at $\lesssim 1$km \citep[\eg,][]{2000Icar..148...12P,2013Icar..225..141S}.  In between rubble piles and monoliths, an object can be fractured but with a lower porosity than pure rubble piles, allowing them to take on more elongated shapes than rubble piles, which have a limiting axis ratio of $a:b \sim 2.3$ \citep[\eg,][]{1984A&A...140..265L}.  The non-binary near-Earth asteroid 433 Eros is an example of such a fractured body elongated beyond the rubble pile limit \citep[$a\sim 31$ km, $b \sim 14$ km;][]{2000Sci...289.2088V,2001LPI....32.1721W}.  Eros' large size violates the suggestion that objects larger than the sub-kilometer range cannot have $a:b > 2.3$.  These results' relevance to the Trojan region, which may have formed in the more ice-rich outer solar system, has not been developed.  Tidal disruption caused by close flybys with planets might also change the shape of an asteroid \citep{1999AJ....117.1921B}.  This distortion mechanism has been studied for near-Earth asteroids through simulations by \cite{2014Icar..239..118B}, but it has not been explored in the context of Trojans or Hildas.  It is therefore possible that some of the high-amplitude candidates presented here may in fact be monolithic shards or elongated low-porosity fractured bodies.

For each object, we excluded measurements with field sources within a radius of $10\arcsec$ of their centroid (the beam size being $\sim 6\arcsec$ in W1-W3 and $\sim 12\arcsec$ in W4) by comparing the single-frame extracted source lists with the AllWISE Catalog \citep{2013yCat.2328....0C}, which coadds together all exposures available at each point on the sky.  We estimate that most sources outside this radius would not affect profile-fitting photometry.  However, a very extended source could still significantly affect the background determination, effectively raising the background threshold and underestimating the target flux.  Source or background flux can also be affected by observations taken inside the South Atlantic Anomaly, which would mean a significant increase in cosmic ray hits, or by observations taken close to the Moon, which would introduce a significant (usually non-linear) gradient to the background.  To guard against these contaminants and also against bad pixels affecting the lightcurve, the images for every candidate were visually inspected.  
We constrained the W3 lightcurve amplitude by calculating the photometric range for each epoch and each object observed from the maximum and minimum usable data points.  Corresponding range uncertainties are the sum in quadrature of the maximum's and minimum's uncertainties.  The values reported are lower limits to the amplitudes since the observations may have missed the intrinsic lightcurve extrema.  

\section{Results and Discussion}

The Trojan and Hilda binary candidates identified after performing the analysis described in \S\ref{Sec:Sample} are reported in Tables \ref{Table:TrojanCandidates} and \ref{Table:HildaCandidates}, respectively.  We found that 38 of the 953 Trojans in our sample had photometric ranges larger than 0.9 mag, 35 of which were not known binaries, making them new candidate binary objects (Fig. \ref{Fig:TrojanCandidates}).  The three known binaries flagged by our binary search technique were 624 Hektor, 17365 (1978 VF$_{11}$), and 29314 Eurydamas.  As described in the introduction, observations of 624 Hektor are consistent with a triple system, where a small satellite is moderately separated from the large, bilobate primary \citep{2014ApJ...783L..37M}.  Our technique is only sensitive to detecting the bilobate primary (itself being a binary), since the satellite is too small to produce significant effect on the lightcurve.  WISE obtained two epochs of data on 624 Hektor.  Using the ephemeris generator from the Institut de M\'{e}canique C\'{e}leste et de Calcul des \'{E}ph\'{e}m\'{e}rides (IMCCE), we determined that 624 Hektor was only oriented such that the primary would produce a large lightcurve amplitude during one of our epochs of coverage, which was indeed flagged as a binary in our search.  Binary Trojans 17365 and 29314 Eurydamas are consistent with contact binaries, though their binary orbital elements are not as well known, preventing us from checking their orientation at the time of our coverage epochs \citep{2007AJ....134.1133M}.  

The only other known Trojan binary, 617 Patroclus-Menoetius, is moderately separated with roughly spherical components, consequently giving it a lower thermal lightcurve amplitude than would be detected by our algorithm.  We used the IMCCE ephemeris generator again to determine the configuration of the Patroclus-Menoetius system and found that is was neither fully nor partially eclipsing during either of our epochs of coverage, giving it a very low WISE thermal lightcurve amplitude of $0.18 \pm 0.02$ magnitudes.  
Of the 554 Hildas explored here, 48 had photometric ranges larger than the binary candidate limit (Fig. \ref{Fig:HildaCandidates}).  There are 503 L4 Trojans in the sample, 21 of which are binary candidates, compared to 16 of the 446 L5 Trojans being binary candidates.  We note that the sum of the L4 and L5 sample does not equal the complete Trojan sample explored because 4 Trojans do not yet have cloud designations due to their unusual orbits (\ie, spending much of their time opposite Jupiter or traversing both clouds equally).


We can estimate the observed binary fraction before debiasing (assuming all candidates are true binaries) by dividing the number of candidates by the total number of objects in that sample, then dividing by the probability that the system will be oriented such that a large amplitude is projected back to the observer \citep[$17-29$\% depending on the angularity, or ``boxiness'' of the components' shapes; \eg,][]{2007AJ....134.1133M}.  This approach gives us first-order observed binary fractions of $13-23$\% for all Trojans, $14-25$\% for L4 Trojans, $12-21$\% for L5 Trojans, and $30-51$\% for Hildas.  However, proper debiasing for incompleteness of the survey, Poisson statistics, and consideration for the probability of detecting a large amplitude given the observing cadence, uncertainties, and theoretical distribution of rotation periods and amplitudes is needed to determine the true binary fraction.  This debiasing is the subject of an upcoming publication by the authors.

Compared to other large-sample lightcurve surveys of Trojans and Hildas, we found a greater number of photometric ranges indicative of possible binarity.  Of the 47 Hildas whose lightcurves were constrained by \cite{1998Icar..133..247D}, one showed a very large lightcurve amplitude -- 3923 Radzievskij -- an object also flagged as having a large amplitude in our results.  \cite{2007AJ....134.1133M} found that 2 of their 114 Trojans had large amplitudes, giving a binary fraction estimate of $6-10$\% when computed as described in the preceding paragraph.  The differences between their observing cadence and ours could explain the discrepancy in the fraction of Trojans observed to have large amplitudes since this factor was not accounted for in the binary fraction estimate.  They sampled each object's lightcurve five times whereas we observed each object an average of 12 times per epoch, affording us more opportunities to catch the lightcurve extrema.  

Another possible contributor to the discrepancy in estimated binary fractions is the physical size range of the survey sample.  \cite{2007AJ....134.1133M} were able to extend their search down to $\sim 20$ km objects, whereas our Trojan sample reached $\sim 12$ km, introducing a possible bias in our results toward detecting smaller highly elongated bodies with a fractured structure instead of a rubble pile (see \S1 for discussion of asteroid structure versus size).  Lastly, if only half of our binary candidates are true binaries and not highly elongated fractured bodies like 433 Eros, then these fractions will decrease by a factor of two, making our results consistent with the \cite{2007AJ....134.1133M} survey.  

\cite{2011AJ....141..170M} densely sampled the lightcurves of 80 Trojans with diameters ranging $\sim 60-150$ km, finding none with large amplitudes.  Their cadence and sample diameter range may have similarly affected their null detection, especially after noting that only one of our 34 Trojans with high lightcurve amplitudes (the bilobate primary of the 624 Hektor system) has a diameter in the range they explored \citep{2014ApJ...783L..37M}.  However, another explanation for our high observed binary fraction could be that the binary fraction amongst smaller Trojans is higher than larger Trojans.  Debiasing and follow-up of our candidates is needed to confirm that possibility.  We therefore encourage the observing community to obtain densely sampled lightcurves of the binary candidates presented here in order to confirm their nature, allowing tight constraints to be set on the binary fraction as a function of orbital elements, size, taxonomy, dynamical classification, and Trojan cloud designation.  

\section{Acknowledgments}

\acknowledgments{This publication makes use of data products from NEOWISE, which is a project of the Jet Propulsion Laboratory/California Institute of Technology, funded by the Planetary Science Division of the National Aeronautics and Space Administration.  This research has made use of the NASA/IPAC Infrared Science Archive, which is operated by the California Institute of Technology, under contract with the National Aeronautics and Space Administration.  S. S. gratefully acknowledges the support of the NASA Postdoctoral Program.}

\clearpage


 \clearpage
 
\begin{deluxetable}{lccrrcrcl}
\tabletypesize{\scriptsize}
\tablecolumns{9}
\tablecaption{Trojans with high-amplitude lightcurves from the fully cryogenic portion of the \WISE\, mission, containing binary candidates and known binaries.  Orbital elements are taken from the Minor Planet Center.\label{Table:TrojanCandidates}}
\tablehead{\colhead{Designation} & \colhead{Diam. (km)\tablenotemark{a}} &\colhead{$a$ (AU)} & \colhead{$e$} & \colhead{$i$ ($^{\circ}$)}  & \colhead{L4/L5} & \colhead{$N$\tablenotemark{b}} & \colhead{$\Delta m$\tablenotemark{c}} & \colhead{Comments} }
\startdata
624		&	150	$\pm$	2	&	5.249	&	0.024	&	18.2	&	L4	&	11	&	1.33	$\pm$	0.02	&	Known binary Hektor	 \\
9431		&	38	$\pm$	3	&	5.126	&	0.084	&	21.3	&	L4	&	11	&	0.91	$\pm$	0.31	&		\\
11429	&	38	$\pm$	1	&	5.272	&	0.029	&	17.1	&	L4	&	13	&	0.96	$\pm$	0.10	&		\\
13323	&	23.2	$\pm$	0.6	&	5.112	&	0.089	&	0.9	&	L4	&	12	&	0.92	$\pm$	0.08	&		\\
15398	&	36	$\pm$	1	&	5.128	&	0.027	&	28.5	&	L4	&	11	&	0.98	$\pm$	0.19	&		\\
16152	&	16.2	$\pm$	0.6	&	5.125	&	0.096	&	3.5	&	L4	&	9	&	0.97	$\pm$	0.15	&		\\
17365	&	44.9	$\pm$	0.5	&	5.268	&	0.079	&	11.6	&	L5	&	9	&	1.17	$\pm$	0.04	&	Known binary 1978 VF11	\\
17414	&	21.6	$\pm$	0.3	&	5.130	&	0.032	&	16.6	&	L5	&	9	&	1.02	$\pm$	0.12	&		\\
20428	&	27	$\pm$	3	&	5.219	&	0.145	&	21.0	&	L4	&	11	&	0.94	$\pm$	0.26	&		\\
25911	&	18	$\pm$	1	&	5.226	&	0.044	&	21.4	&	L4	&	11	&	1.08	$\pm$	0.19	&		\\
29314	&	21.4	$\pm$	0.8	&	5.280	&	0.073	&	15.2	&	L5	&	17	&	0.99	$\pm$	0.15	&	Known binary Eurydamas	\\
51357	&	19.3	$\pm$	0.8	&	5.201	&	0.070	&	9.0	&	L5	&	8	&	1.23	$\pm$	0.19	&		\\
55474	&	21	$\pm$	1	&	5.204	&	0.095	&	18.0	&	L5	&	10	&	0.93	$\pm$	0.19	&		\\
63241	&	22.6	$\pm$	0.9	&	5.250	&	0.049	&	25.8	&	L4	&	8	&	1.21	$\pm$	0.18	&		\\
64270	&	16.5	$\pm$	0.7	&	5.159	&	0.095	&	12.9	&	L5	&	10	&	1.25	$\pm$	0.15	&		\\
65225	&	16.7	$\pm$	0.2	&	5.287	&	0.081	&	7.0	&	L4	&	11	&	1.07	$\pm$	0.15	&		\\
76820	&	17.5	$\pm$	0.6	&	5.164	&	0.097	&	18.4	&	L5	&	10	&	0.91	$\pm$	0.17	&		\\
76836	&	18.3	$\pm$	0.6	&	5.242	&	0.099	&	23.8	&	L5	&	10	&	0.97	$\pm$	0.12	&		\\
114141	&	20.9	$\pm$	0.6	&	5.131	&	0.069	&	19.7	&	L5	&	10	&	1.08	$\pm$	0.13	&		\\
129135	&	20.0	$\pm$	0.7	&	5.302	&	0.038	&	33.1	&	L5	&	11	&	1.41	$\pm$	0.26	&		\\
130190	&	17.2	$\pm$	0.7	&	5.238	&	0.044	&	14.7	&	L4	&	13	&	1.13	$\pm$	0.18	&		\\
155337	&	17.2	$\pm$	0.8	&	5.230	&	0.089	&	17.0	&	L5	&	10	&	1.15	$\pm$	0.21	&		\\
160140	&	19.3	$\pm$	0.6	&	5.200	&	0.057	&	24.5	&	L4	&	12	&	1.13	$\pm$	0.13	&		\\
161018	&	19.2	$\pm$	0.7	&	5.098	&	0.054	&	12.0	&	L4	&	12	&	1.05	$\pm$	0.15	&		\\
182178	&	15.1	$\pm$	0.7	&	5.200	&	0.115	&	25.5	&	L5	&	7	&	1.25	$\pm$	0.18	&		\\
182445	&	14.3	$\pm$	0.7	&	5.150	&	0.059	&	17.3	&	L5	&	7	&	1.66	$\pm$	0.51	&		\\
192221	&	21.4	$\pm$	0.7	&	5.188	&	0.045	&	27.3	&	L4	&	16	&	1.21	$\pm$	0.22	&		\\
192389	&	16.1	$\pm$	0.8	&	5.253	&	0.013	&	22.8	&	L4	&	12	&	0.94	$\pm$	0.22	&		\\
222861	&	13.0	$\pm$	0.9	&	5.167	&	0.100	&	6.7	&	L4	&	5	&	0.93	$\pm$	0.21	&		\\
228114	&	17.0	$\pm$	0.9	&	5.132	&	0.018	&	14.0	&	L4	&	12	&	1.03	$\pm$	0.21	&		\\
231631	&	13.5	$\pm$	0.9	&	5.102	&	0.059	&	9.5	&	L4	&	7	&	1.03	$\pm$	0.18	&		\\
246550	&	15.2	$\pm$	0.5	&	5.146	&	0.223	&	6.7	&	L4	&	11	&	1.04	$\pm$	0.15	&		\\
247969	&	15	$\pm$	1	&	5.279	&	0.098	&	17.0	&	L5	&	8	&	1.03	$\pm$	0.26	&		\\
321611	&	16.0 $\pm$	0.8	&	5.188	&	0.058	&	26.9	&	L4	&	11	&	1.00	$\pm$	0.20	&		\\
341880	&	15.6	$\pm$	0.6	&	5.175	&	0.137	&	35.8	&	L5	&	15	&	1.13	$\pm$	0.18	&		\\
343993	&	13.4	$\pm$	0.7	&	5.236	&	0.202	&	19.5	&	L5	&	11	&	0.95	$\pm$	0.13	&		\\
356261	&	17.8	$\pm$	0.8	&	5.338	&	0.055	&	22.7	&	L4	&	10	&	1.48	$\pm$	0.26	&		\\
\enddata
\tablenotetext{a}{Diameters calculated from thermal fits using the NEOWISE observations as reported in \cite{2011ApJ...742...40G}.}
\tablenotetext{b}{Number of usable data points in the NEOWISE W3 bandpass.}
\tablenotetext{c}{The photometric ranges ($\Delta m$) and their uncertainties come from the W3 photometry and their corresponding uncertainties, not from fits to the W3 lightcurves.}
\end{deluxetable}

\begin{deluxetable}{lccrrrc}
\tabletypesize{\footnotesize}
\tablecolumns{8}
\tablecaption{High-amplitude Hildas from the fully cryogenic portion of the \WISE\, mission.  Orbital elements are taken from the Minor Planet Center.\label{Table:HildaCandidates}}
\tablehead{\colhead{Designation} & \colhead{Diam. (km)\tablenotemark{a}} &\colhead{$a$ (AU)} & \colhead{$e$} & \colhead{$i$ ($^{\circ}$)} & \colhead{$N$\tablenotemark{b}} & \colhead{$\Delta m$\tablenotemark{c}} }
\startdata
2483\tablenotemark{d}	&	35.7	$\pm$	0.2	&	3.972	&	0.278	&	4.5	&	13	&	1.51	$\pm$	0.02	\\
3923\tablenotemark{e}	&	29.9	$\pm$	0.2	&	3.963	&	0.223	&	3.5	&	15	&	0.98	$\pm$	0.02	\\
4230		&	28.5	$\pm$	0.8	&	3.946	&	0.133	&	3.1	&	9	&	1.77	$\pm$	0.04	\\
15626	&	18.6	$\pm$	0.4	&	3.950	&	0.112	&	1.8	&	10	&	1.04	$\pm$	0.07	\\
16927	&	22.6	$\pm$	0.1	&	3.980	&	0.139	&	12.9	&	11	&	0.92	$\pm$	0.04	\\
21047	&	16.0	$\pm$	0.3	&	3.982	&	0.176	&	4.5	&	12	&	1.1	$\pm$	0.2	\\
22070	&	18.8	$\pm$	0.5	&	3.977	&	0.276	&	13.8	&	11	&	1.0	$\pm$	0.2	\\
23405	&	14.5	$\pm$	0.5	&	3.952	&	0.130	&	5.2	&	23	&	1.9	$\pm$	0.1	\\
31097	&	15.3	$\pm$	0.7	&	3.960	&	0.109	&	2.7	&	11	&	1.33	$\pm$	0.09	\\
39405	&	14.6	$\pm$	0.7	&	3.960	&	0.222	&	1.8	&	11	&	1.1	$\pm$	0.3	\\
39415	&	9.4	$\pm$	0.4	&	3.925	&	0.209	&	2.4	&	21	&	1.0	$\pm$	0.2	\\
45862	&	8.6	$\pm$	0.1	&	3.972	&	0.162	&	3.2	&	11	&	1.0	$\pm$	0.2	\\
46629	&	15.7	$\pm$	0.1	&	3.953	&	0.243	&	1.7	&	14	&	1.35	$\pm$	0.03	\\
54630	&	16.07$\pm$     0.09	&	3.981	&	0.139	&	9.0	&	12	&	0.91	$\pm$	0.04	\\
60398	&	11.5	$\pm$	0.2	&	3.931	&	0.146	&	1.8	&	8	&	1.03	$\pm$	0.09	\\
64390	&	6.60	$\pm$	0.06	&	3.936	&	0.250	&	2.5	&	16	&	1.5	$\pm$	0.2	\\
65389	&	9.78	$\pm$	0.02	&	3.948	&	0.260	&	2.3	&	14	&	1.02	$\pm$	0.04	\\
83900	&	7.5	$\pm$	0.5	&	3.935	&	0.101	&	3.4	&	13	&	1.1	$\pm$	0.2	\\
88230	&	18.5	$\pm$	0.4	&	3.974	&	0.150	&	7.4	&	26	&	1.19	$\pm$	0.04	\\
94266	&	10.9	$\pm$	0.5	&	3.933	&	0.100	&	8.6	&	23	&	1.5	$\pm$	0.1	\\
112822	&	10.2	$\pm$	0.3	&	3.935	&	0.185	&	10.5	&	13	&	1.04	$\pm$	0.08	\\
121005	&	10.0	$\pm$	0.2	&	3.932	&	0.171	&	7.9	&	17	&	1.2	$\pm$	0.2	\\
132868	&	8.9	$\pm$	0.3	&	3.959	&	0.242	&	2.0	&	14	&	1.12	$\pm$	0.06	\\
141557	&	10.5	$\pm$	0.4	&	3.971	&	0.117	&	4.1	&	22	&	1.6	$\pm$	0.1	\\
186649	&	8.9	$\pm$	0.2	&	3.973	&	0.289	&	5.4	&	11	&	1.5	$\pm$	0.2	\\
193291	&	7.7	$\pm$	0.5	&	3.960	&	0.307	&	9.3	&	10	&	1.2	$\pm$	0.2	\\
197558	&	8.5	$\pm$	0.5	&	4.000	&	0.084	&	7.7	&	13	&	1.0	$\pm$	0.1	\\
209512	&	8.1	$\pm$	0.6	&	3.969	&	0.076	&	8.0	&	21	&	1.1	$\pm$	0.3	\\
222490	&	6.2	$\pm$	0.3	&	3.975	&	0.269	&	3.5	&	11	&	1.5	$\pm$	0.2	\\
233939	&	5.1	$\pm$	0.1	&	3.960	&	0.185	&	6.8	&	12	&	1.1	$\pm$	0.2	\\
241528	&	7.3	$\pm$	0.7	&	3.953	&	0.138	&	3.6	&	18	&	1.3	$\pm$	0.2	\\
241994	&	7.77	$\pm$	0.01	&	3.937	&	0.289	&	5.5	&	12	&	1.1	$\pm$	0.1	\\
247405	&	6.4	$\pm$	0.3	&	3.977	&	0.236	&	10.0	&	16	&	1.2	$\pm$	0.2	\\
249416	&	6.2	$\pm$	0.3	&	3.938	&	0.115	&	4.2	&	22	&	0.9	$\pm$	0.2	\\
250139	&	8.3	$\pm$	0.1	&	3.933	&	0.178	&	8.0	&	14	&	1.5	$\pm$	0.1	\\
251338	&	7.2	$\pm$	0.4	&	3.921	&	0.114	&	12.3	&	12	&	1.1	$\pm$	0.2	\\
263793	&	6.34	$\pm$	0.04	&	3.939	&	0.202	&	2.6	&	11	&	1.4	$\pm$	0.2	\\
288443	&	6.7	$\pm$	0.2	&	3.955	&	0.133	&	8.6	&	11	&	0.9	$\pm$	0.2	\\
307321	&	5.9	$\pm$	0.4	&	3.923	&	0.167	&	4.2	&	12	&	1.0	$\pm$	0.2	\\
310756	&	6.2	$\pm$	0.1	&	3.949	&	0.266	&	8.3	&	16	&	1.05	$\pm$	0.09	\\
317150	&	7.0	$\pm$	0.4	&	3.962	&	0.216	&	2.8	&	8	&	1.7	$\pm$	0.3	\\
368099	&	5.5	$\pm$	0.3	&	3.933	&	0.205	&	9.7	&	14	&	0.9	$\pm$	0.1	\\
2002 RM	&	5.3	$\pm$	0.1	&	3.953	&	0.279	&	3.4	&	16	&	1.5	$\pm$	0.2	\\
2008 SE$_{268}$	&	4.7	$\pm$	0.2	&	3.955	&	0.223	&	3.2	&	13	&	1.3	$\pm$	0.2	\\
2010 MJ$_{93}$	&	4	$\pm$	1	&	3.967	&	0.240	&	7.5	&	12	&	1.1	$\pm$	0.2	\\
2010 NO$_{52}$	&	4.4	$\pm$	0.5	&	3.964	&	0.252	&	4.0	&	16	&	1.4	$\pm$	0.3	\\
2010 NB$_{115}$	&	5.5	$\pm$	0.2	&	3.979	&	0.274	&	4.1	&	16	&	1.3	$\pm$	0.2	\\
2010 OS$_{14}$	&	6.1	$\pm$	0.2	&	3.913	&	0.253	&	4.0	&	18	&	1.5	$\pm$	0.1	\\
\enddata
\tablenotetext{a}{Diameters reported are thermal fits to the NEOWISE observations as reported in \cite{2012ApJ...744..197G}.}
\tablenotetext{b}{Number of usable W3 observations.}
\tablenotetext{c}{W3 photometric ranges ($\Delta m$) and uncertainties come from the photometry, not a model fit to the lightcurve.}
\tablenotetext{d}{Previous optical studies found amplitude of $\sim 1.3$ \citep{1988Icar...73..487H,1998Icar..133..247D}.}
\tablenotetext{e}{Previous optical studies found amplitude of $> 0.61$ \citep{1998Icar..133..247D}.}
\end{deluxetable}

 \clearpage
 
\begin{figure}
\figurenum{3}
\includegraphics[width=3.5in,height=2.6in]{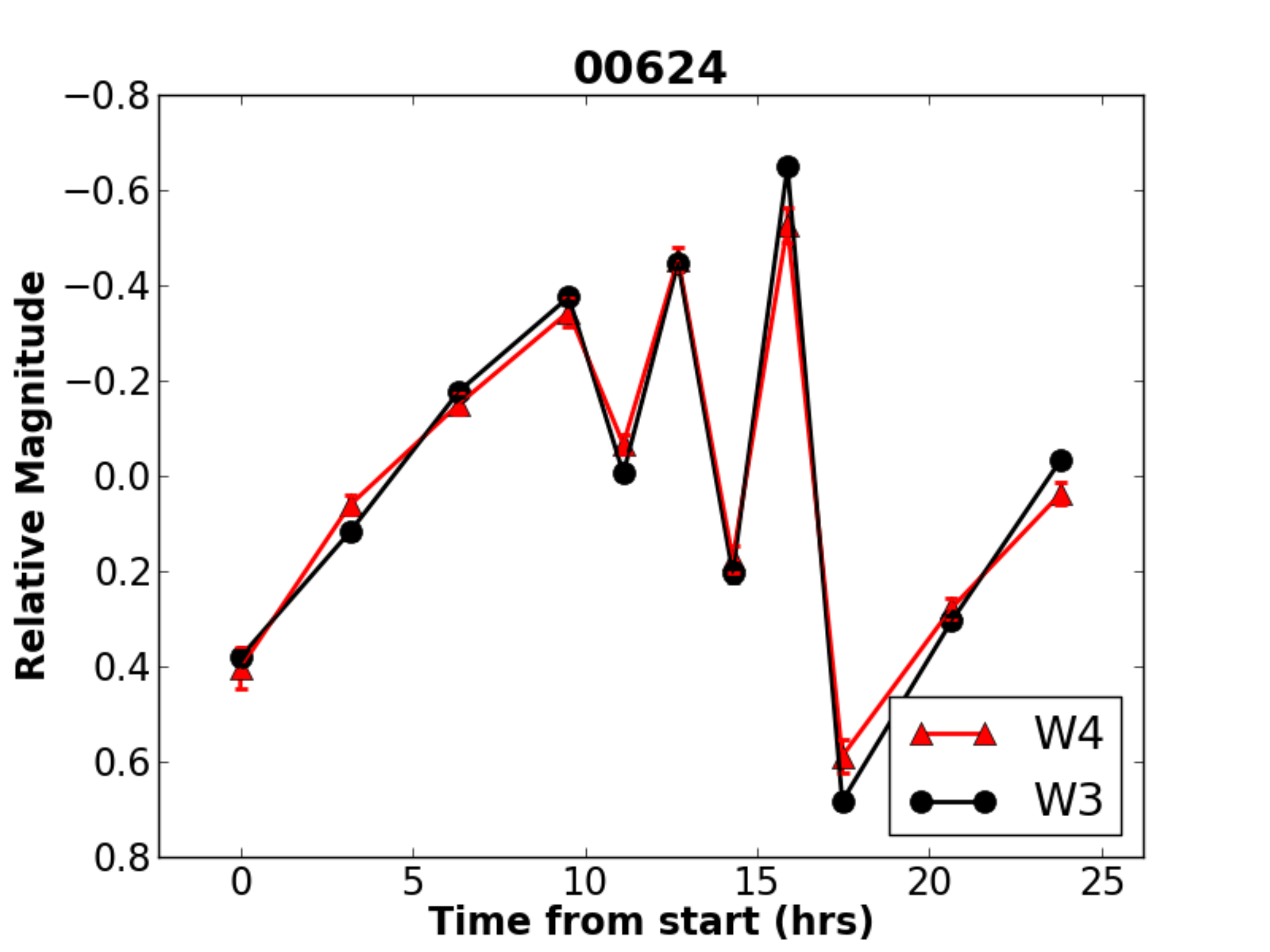}
\includegraphics[width=3.5in,height=2.6in]{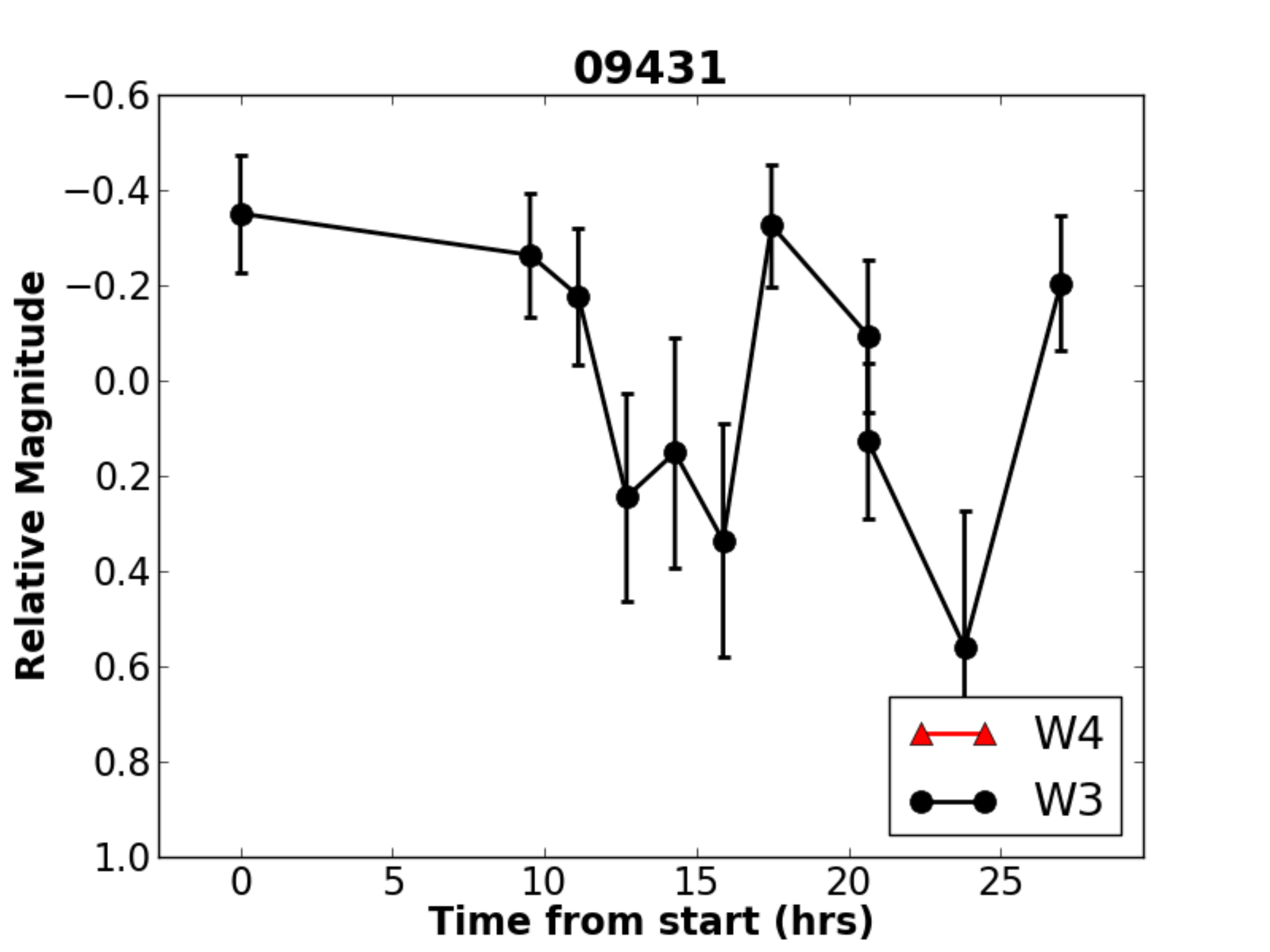}
\includegraphics[width=3.5in,height=2.6in]{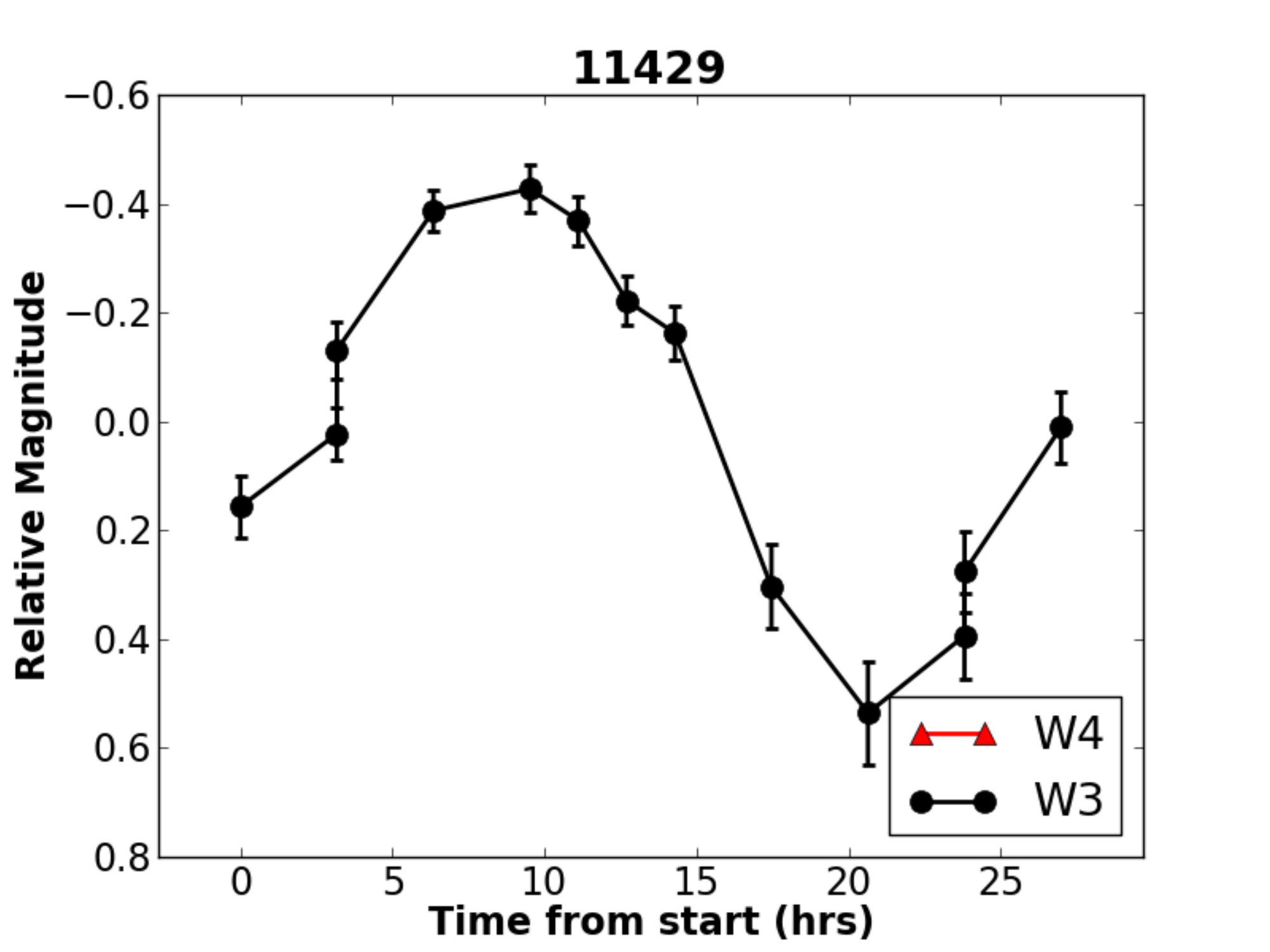}
\includegraphics[width=3.5in,height=2.6in]{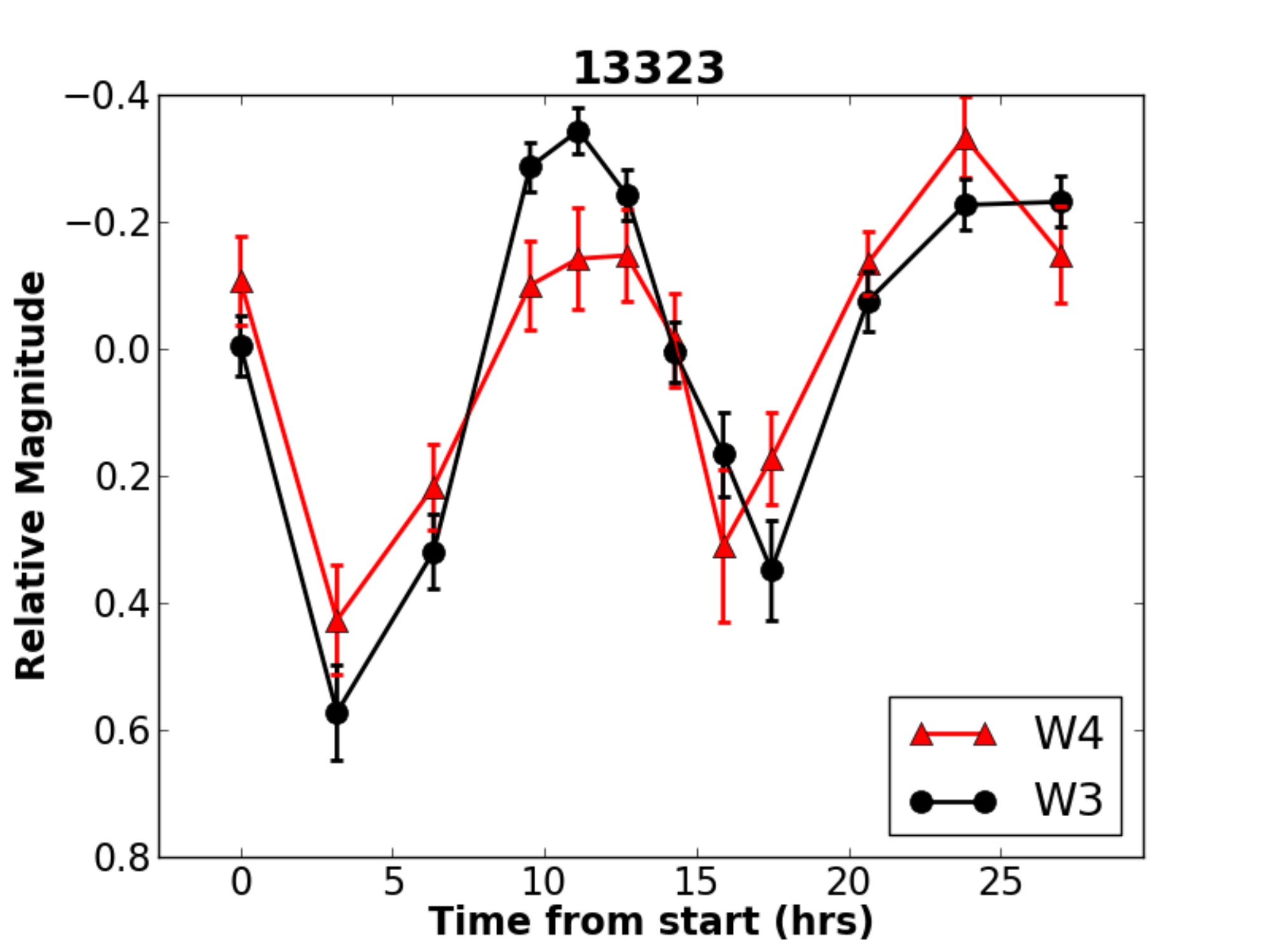}
\includegraphics[width=3.5in,height=2.6in]{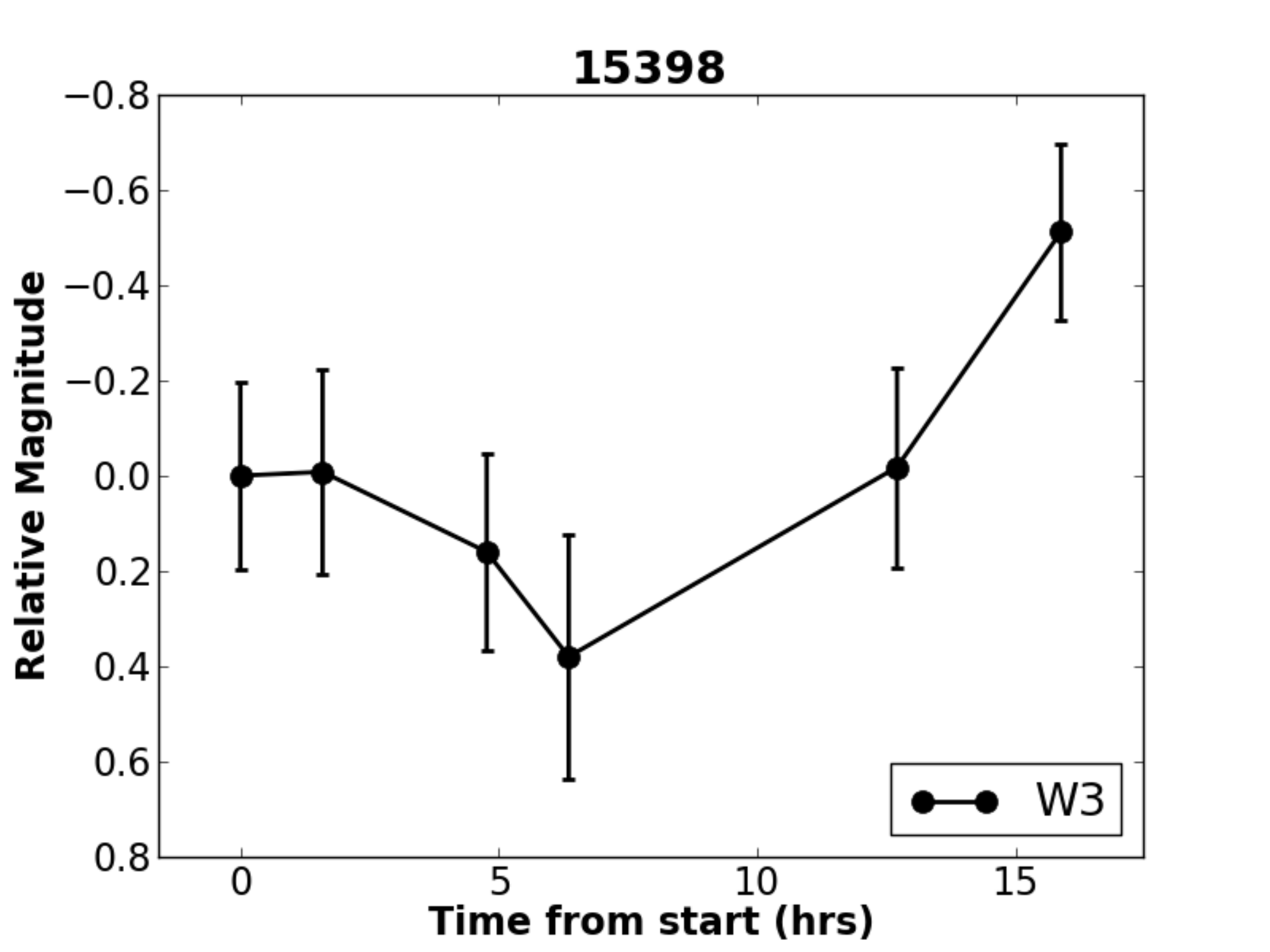}
\includegraphics[width=3.5in,height=2.6in]{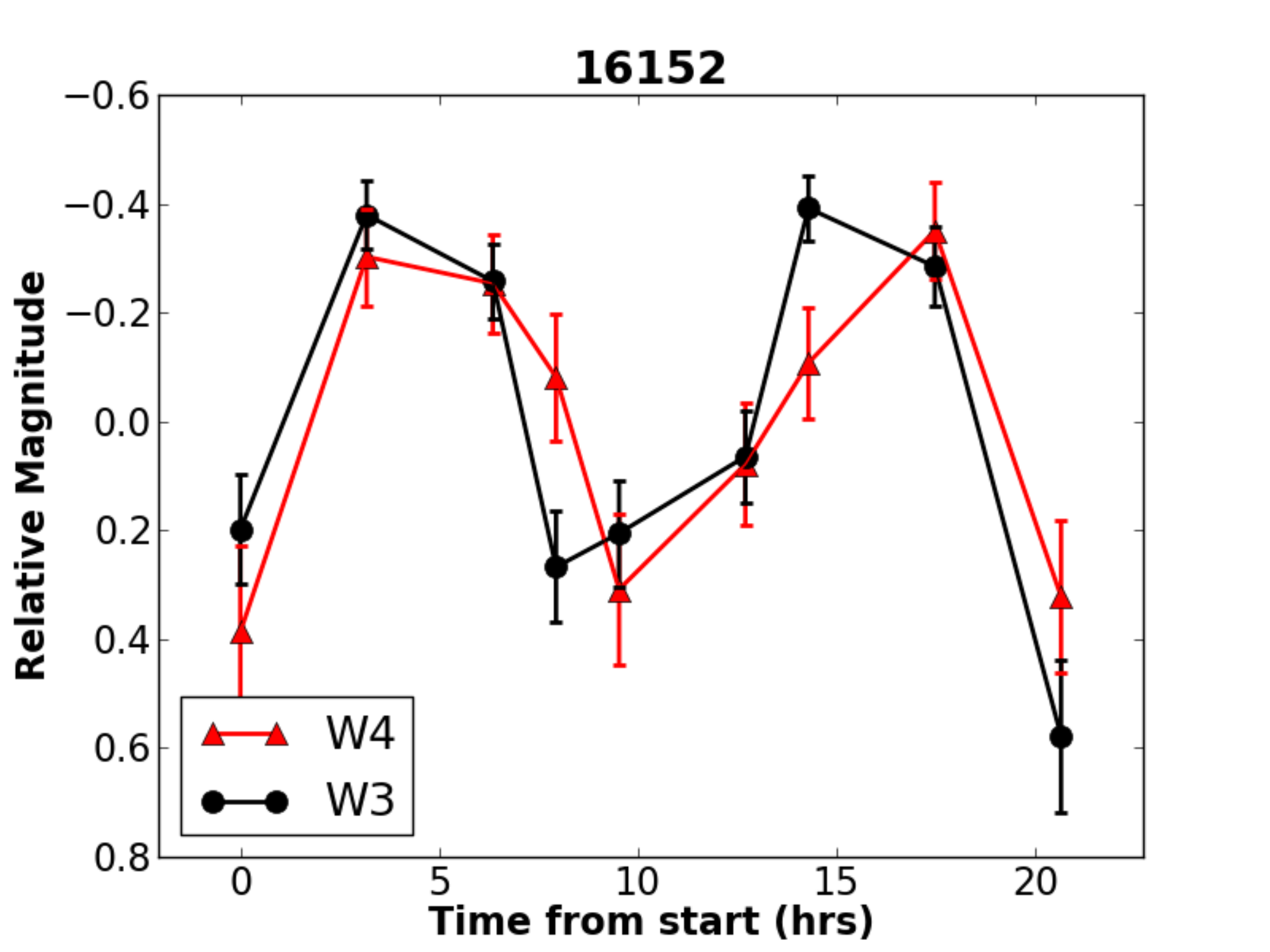}
\caption{\label{Fig:TrojanCandidates} Candidate binary Trojans from our survey identified by their anomalously high lightcurve photometric ranges, including known binaries 624 Hektor, 17365 (1978 VF$_{11}$), and 29314 Eurydamas.}
\end{figure} 
 
\begin{figure}
\figurenum{3}
\includegraphics[width=3.5in,height=2.6in]{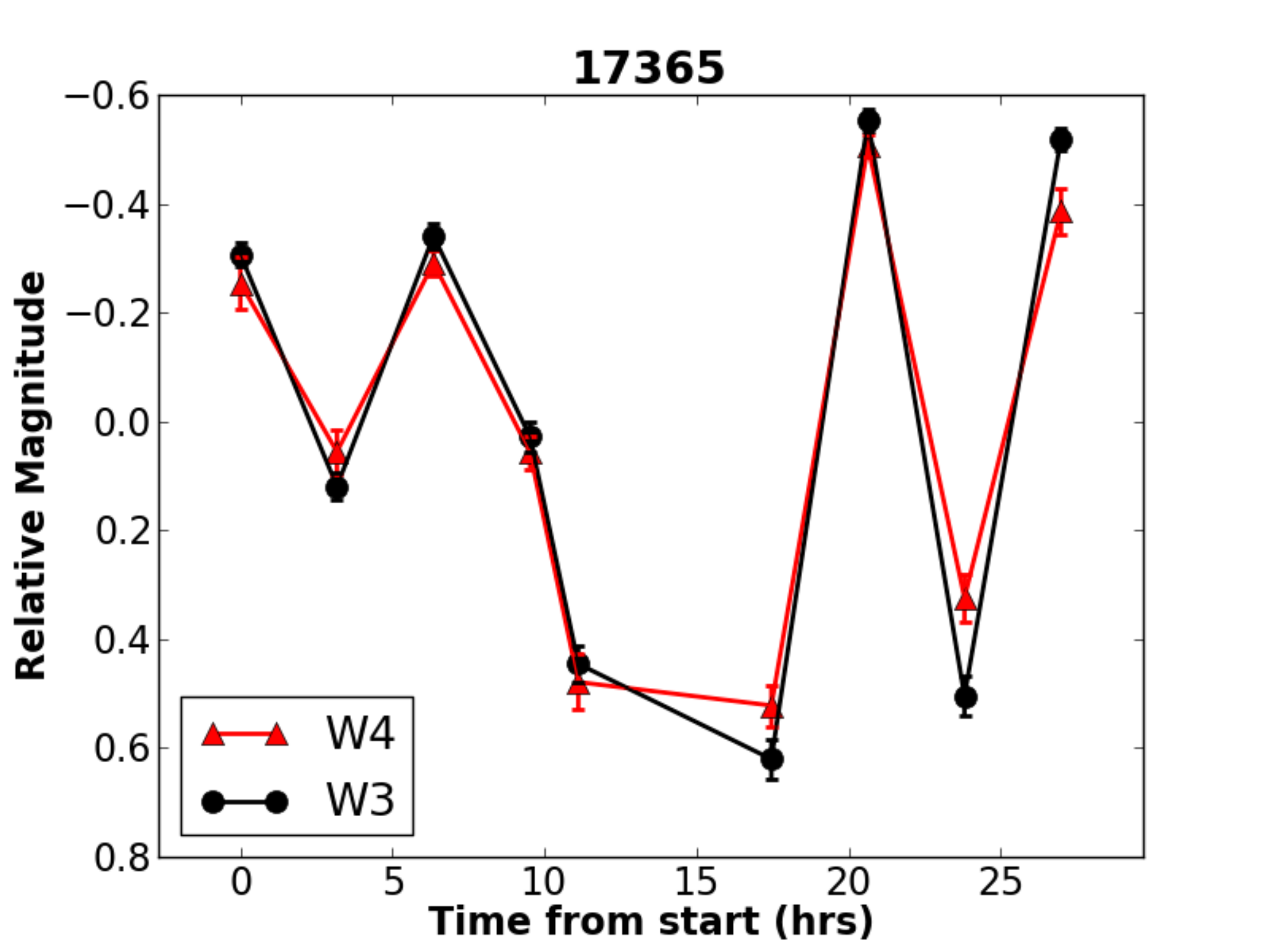}
\includegraphics[width=3.5in,height=2.6in]{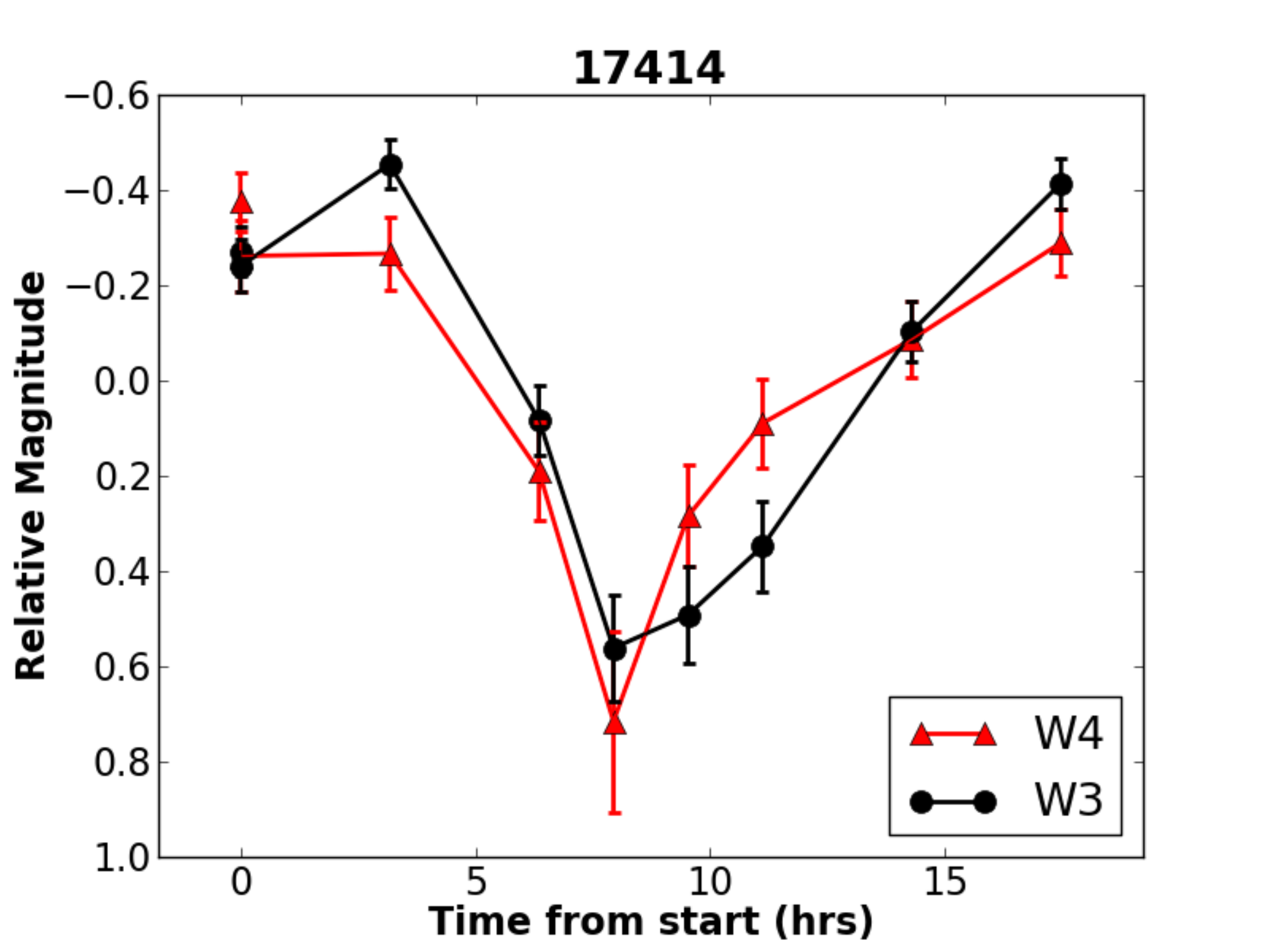}
\includegraphics[width=3.5in,height=2.6in]{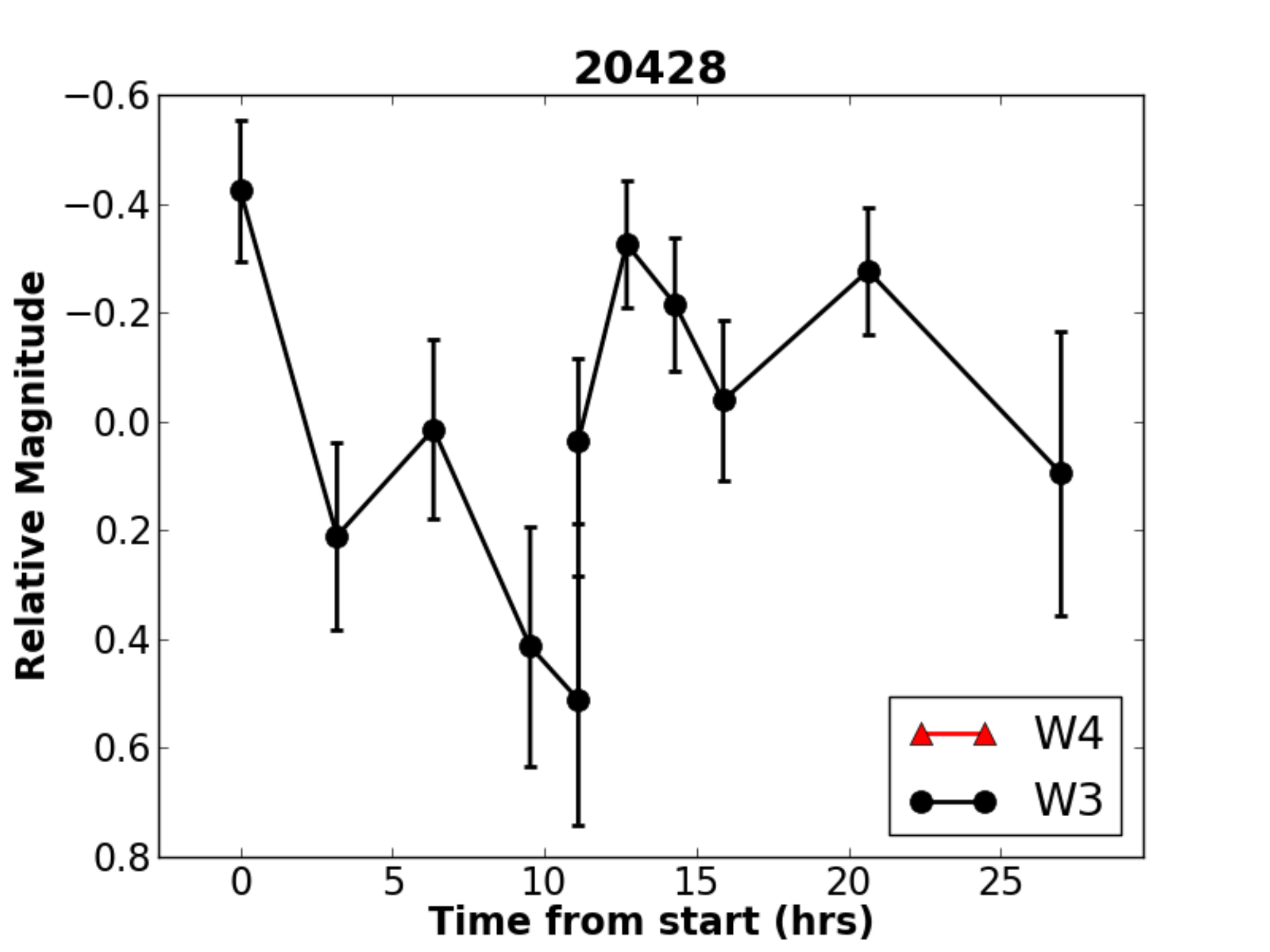}
\includegraphics[width=3.5in,height=2.6in]{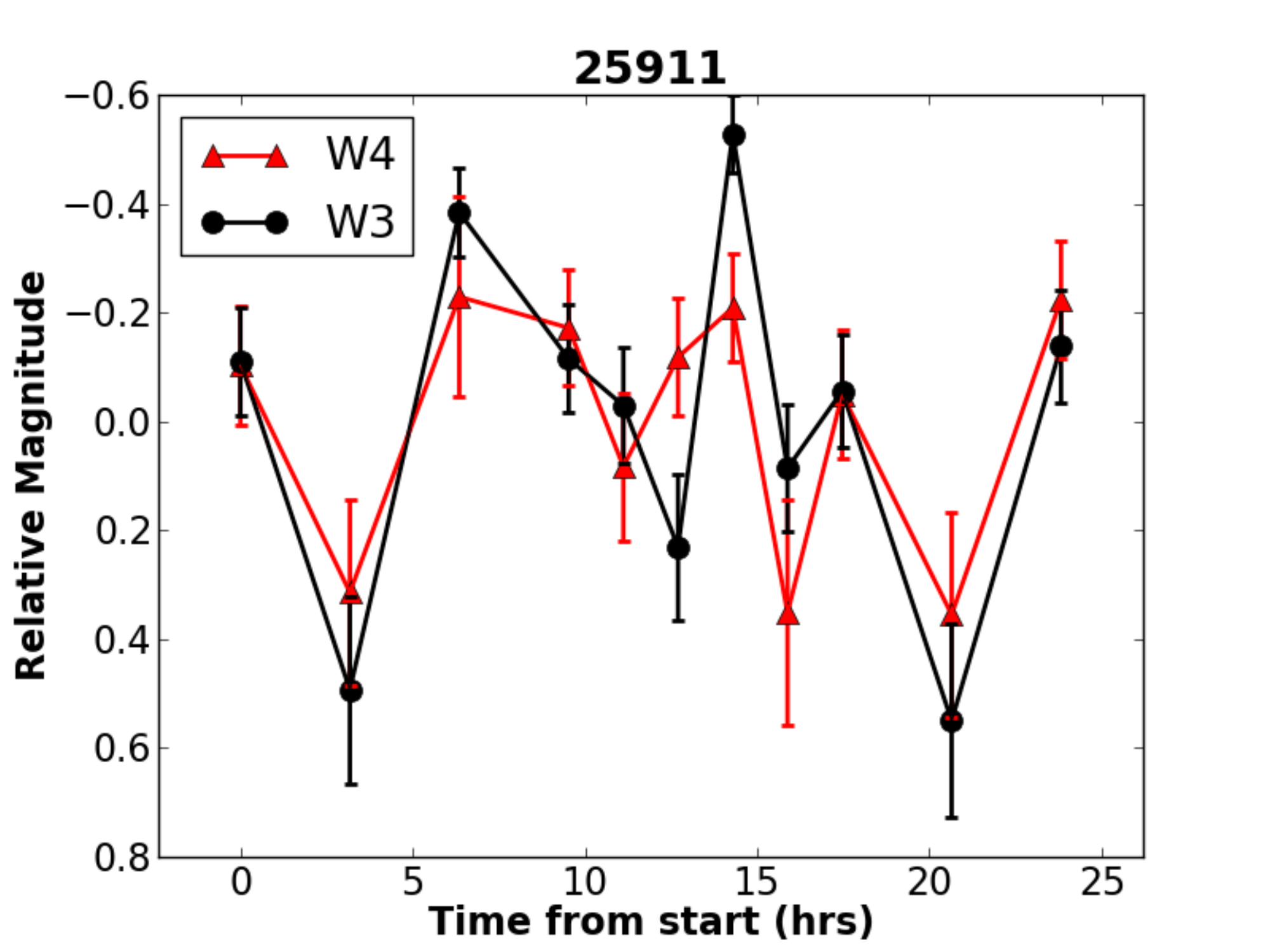}
\includegraphics[width=3.5in,height=2.6in]{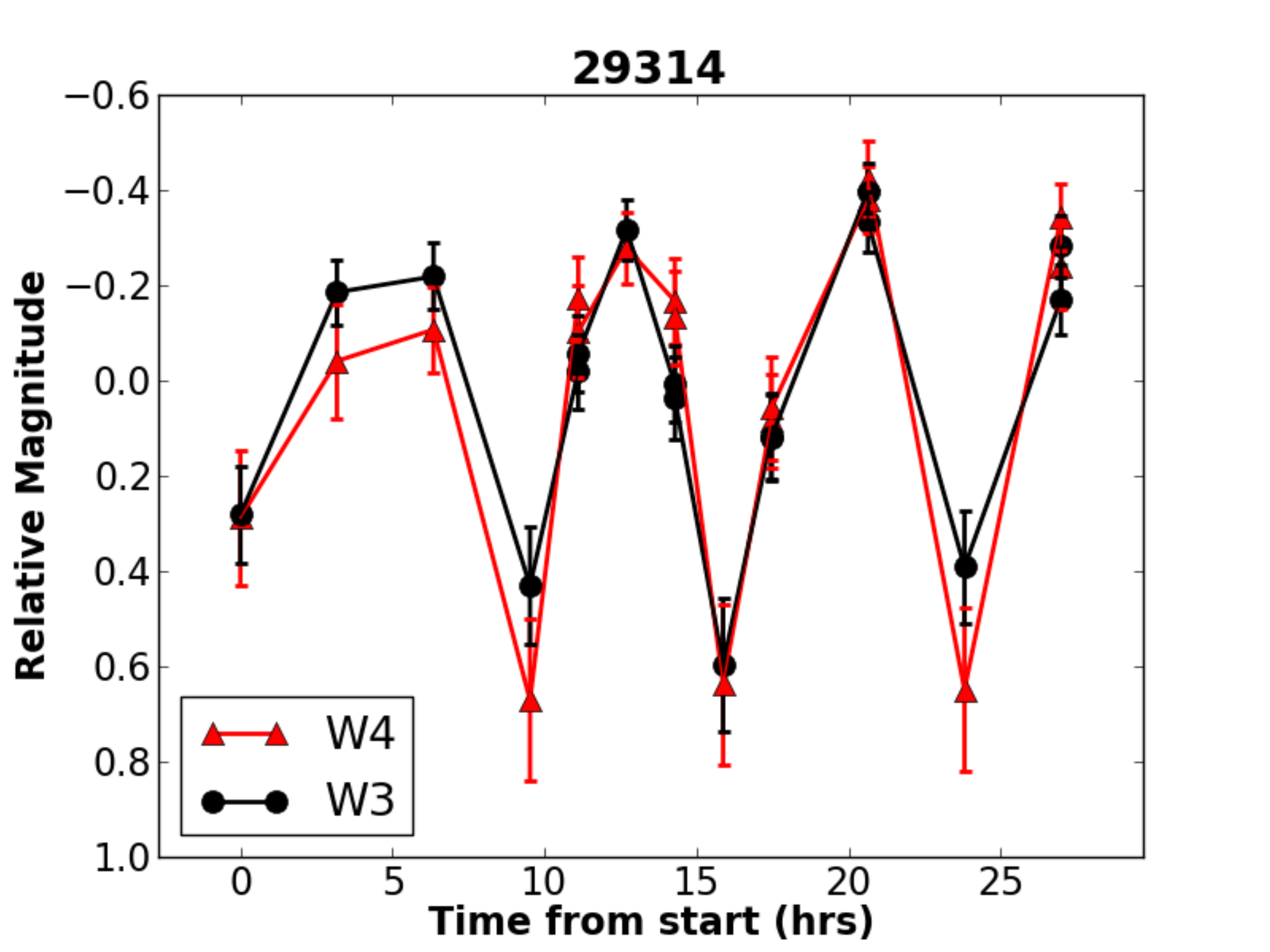}
\includegraphics[width=3.5in,height=2.6in]{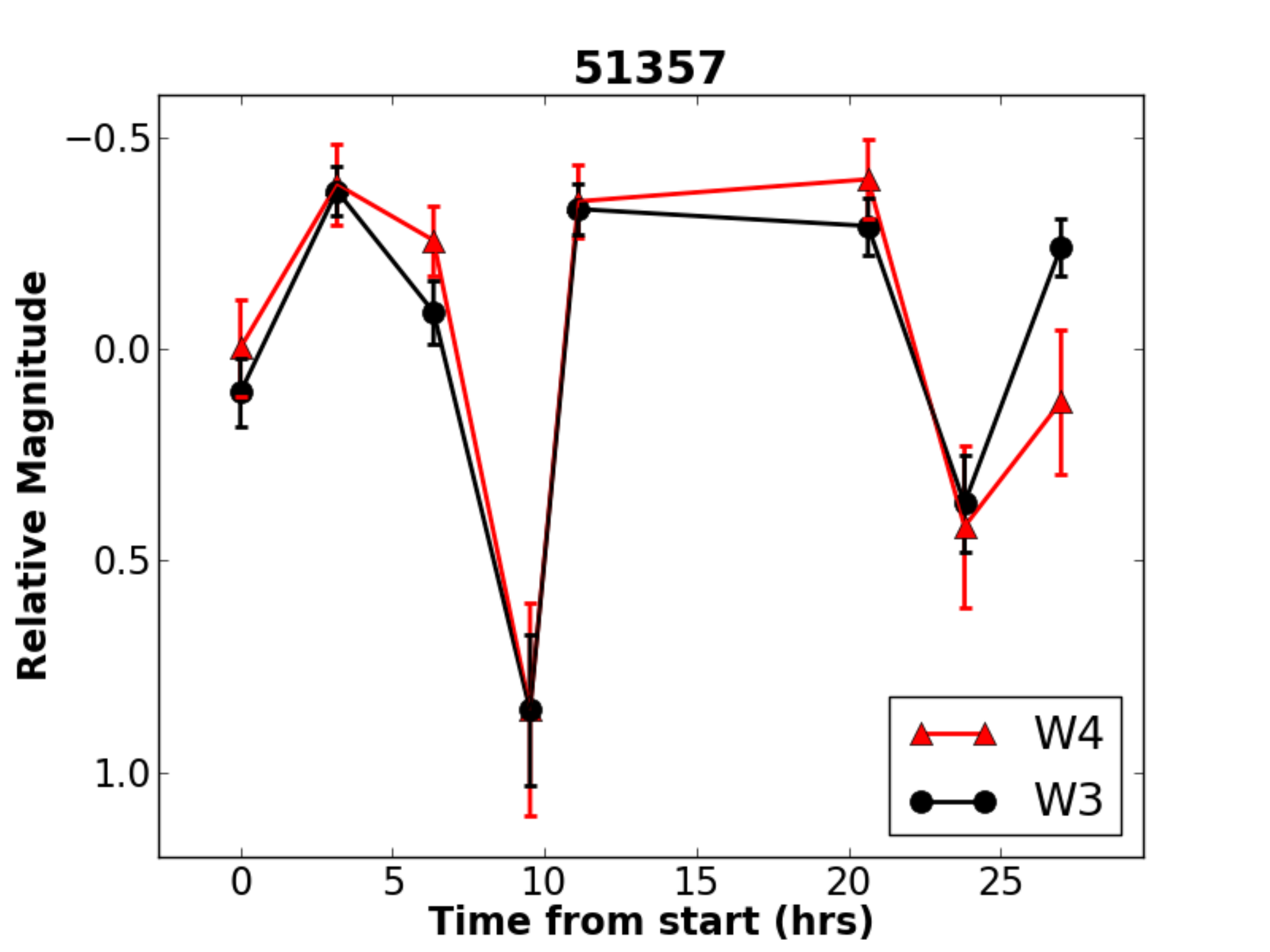}
\caption{\label{Fig:TrojanCandidates} Candidate binary Trojans from our survey identified by their anomalously high lightcurve photometric ranges, including known binaries 624 Hektor, 17365 (1978 VF$_{11}$), and 29314 Eurydamas.}
\end{figure} 

\begin{figure}
\figurenum{3}
\includegraphics[width=3.5in,height=2.6in]{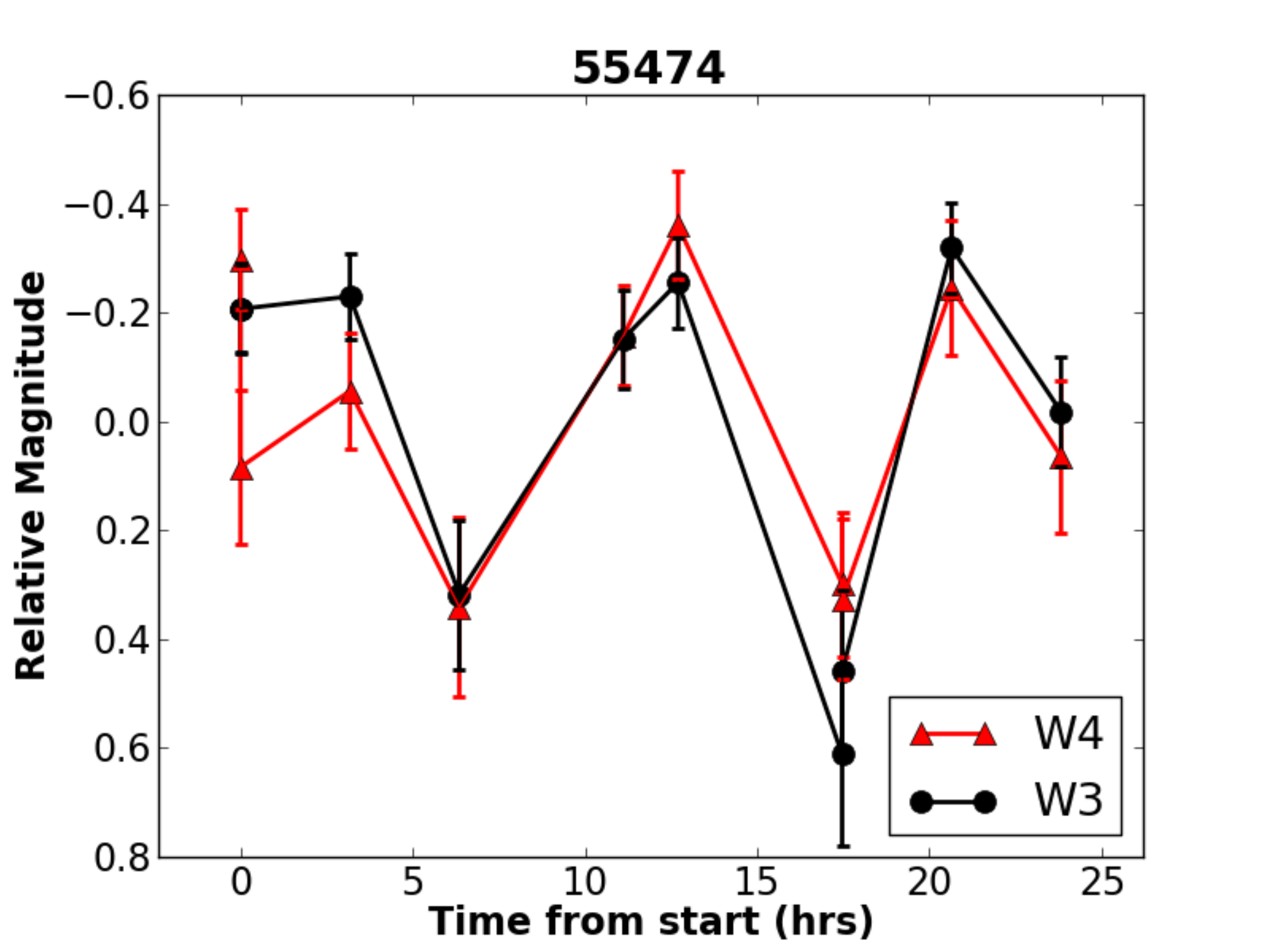}
\includegraphics[width=3.5in,height=2.6in]{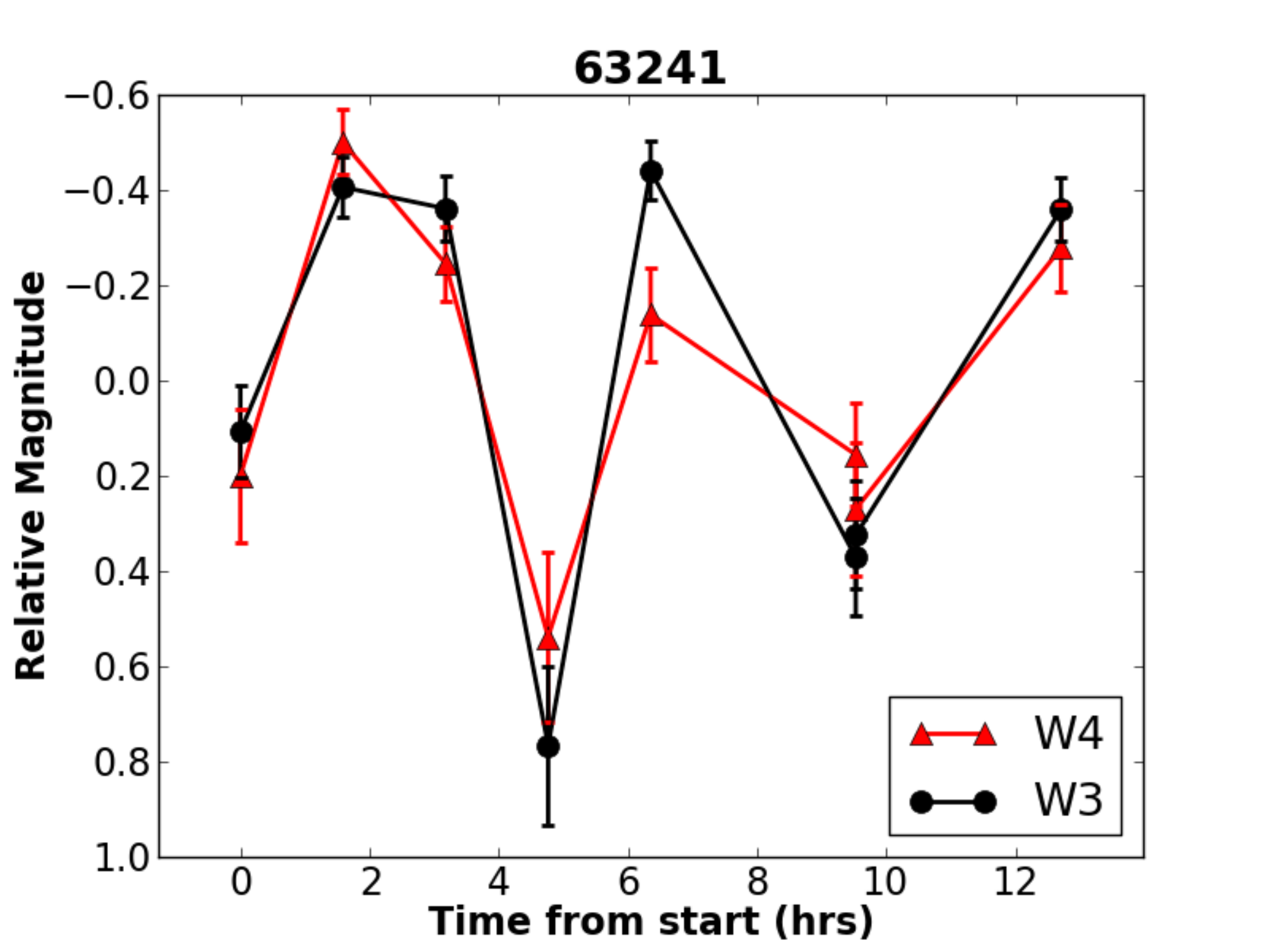}
\includegraphics[width=3.5in,height=2.6in]{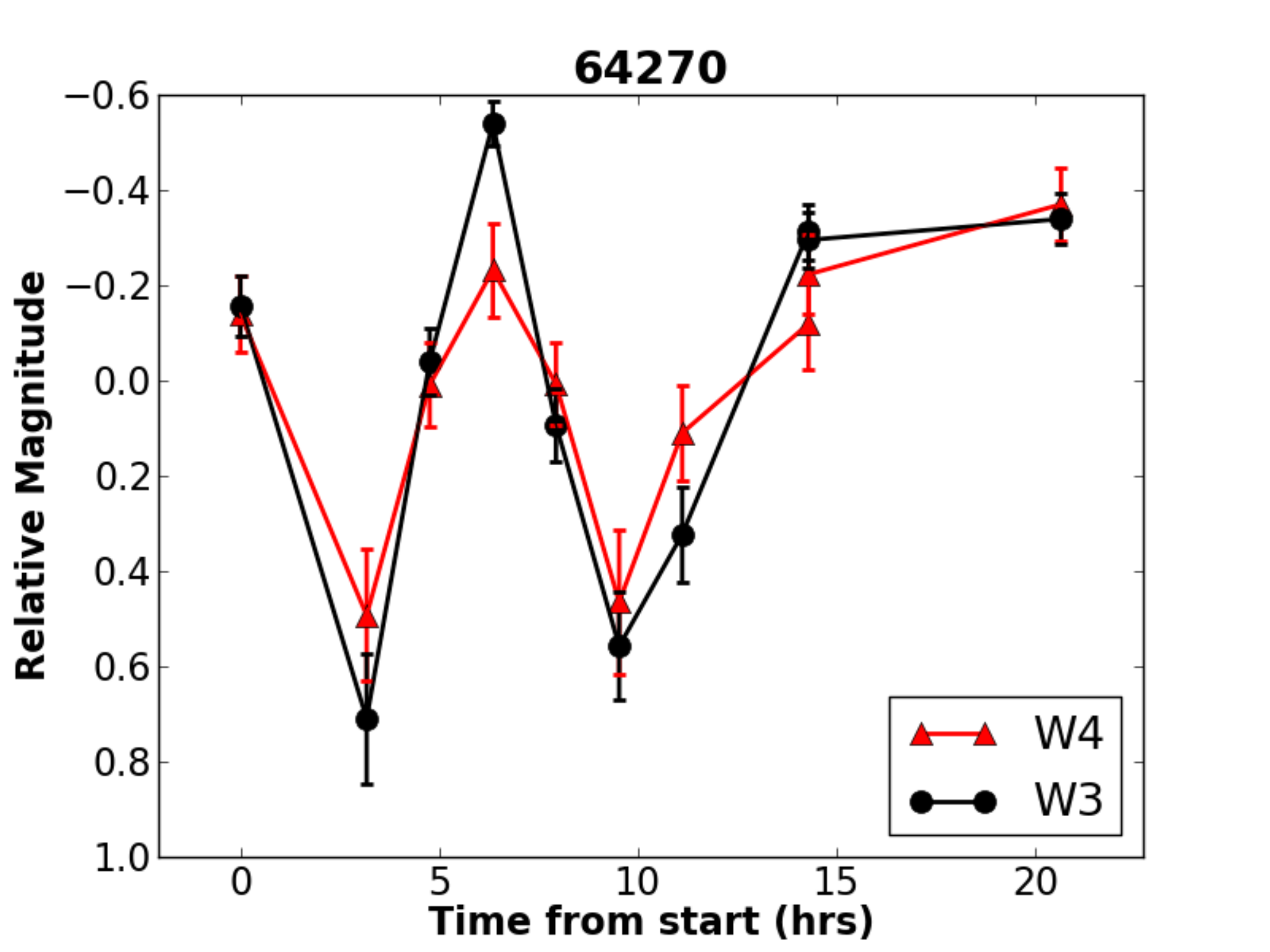}
\includegraphics[width=3.5in,height=2.6in]{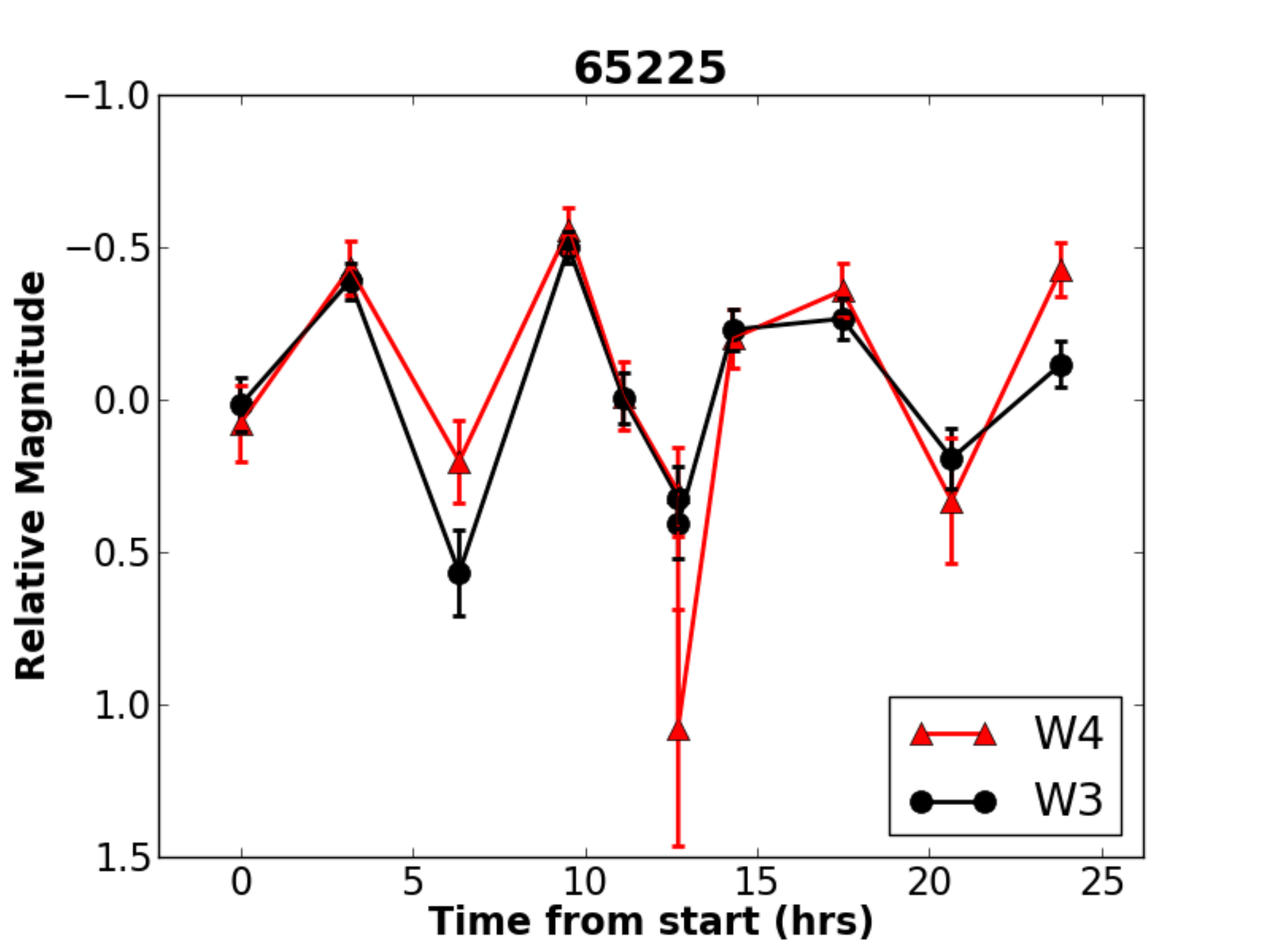}
\includegraphics[width=3.5in,height=2.6in]{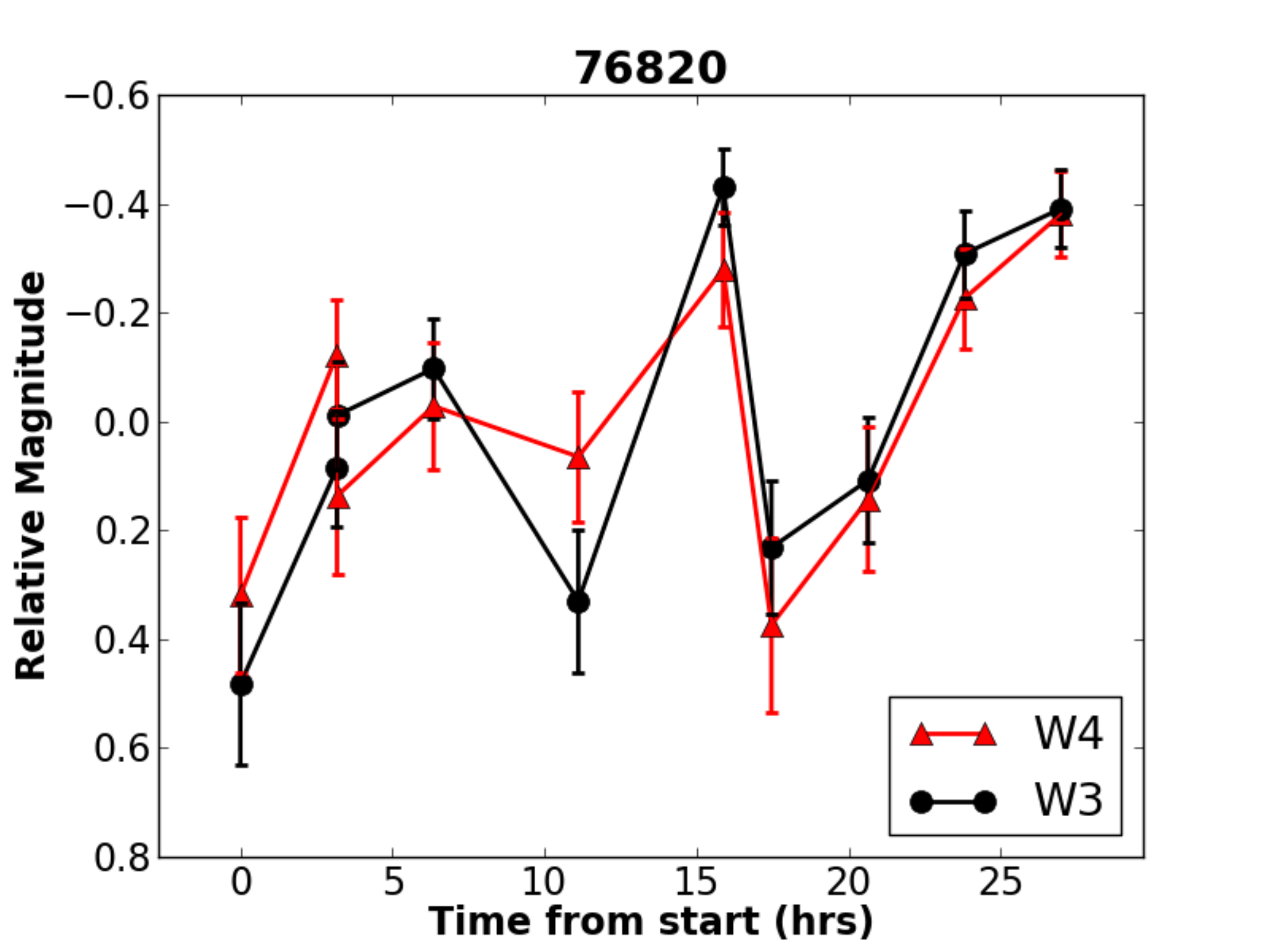}
\includegraphics[width=3.5in,height=2.6in]{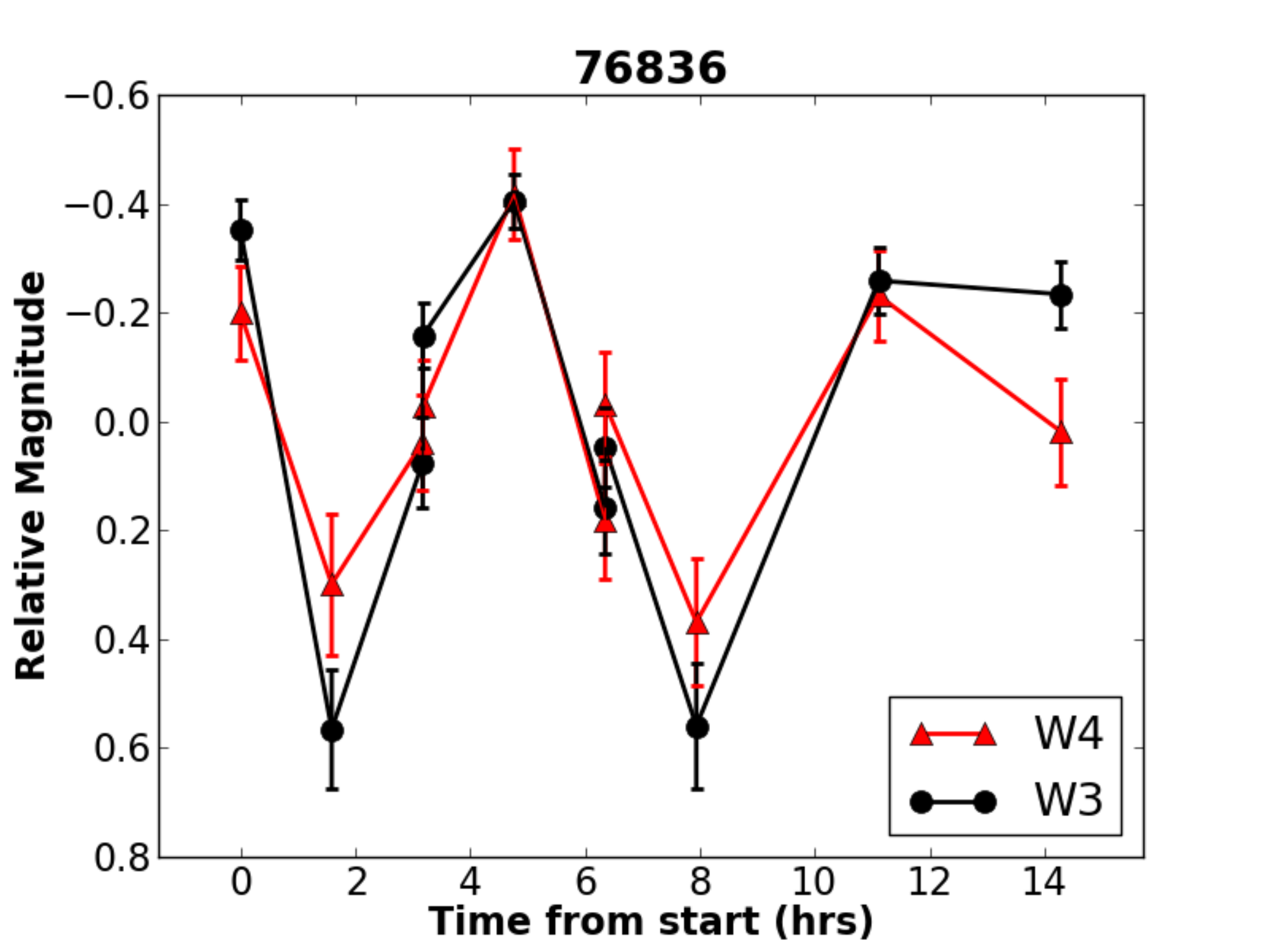}
\caption{\label{Fig:TrojanCandidates} Candidate binary Trojans from our survey identified by their anomalously high lightcurve photometric ranges, including known binaries 624 Hektor, 17365 (1978 VF$_{11}$), and 29314 Eurydamas.}
\end{figure} 

\begin{figure}
\figurenum{3}
\includegraphics[width=3.5in,height=2.6in]{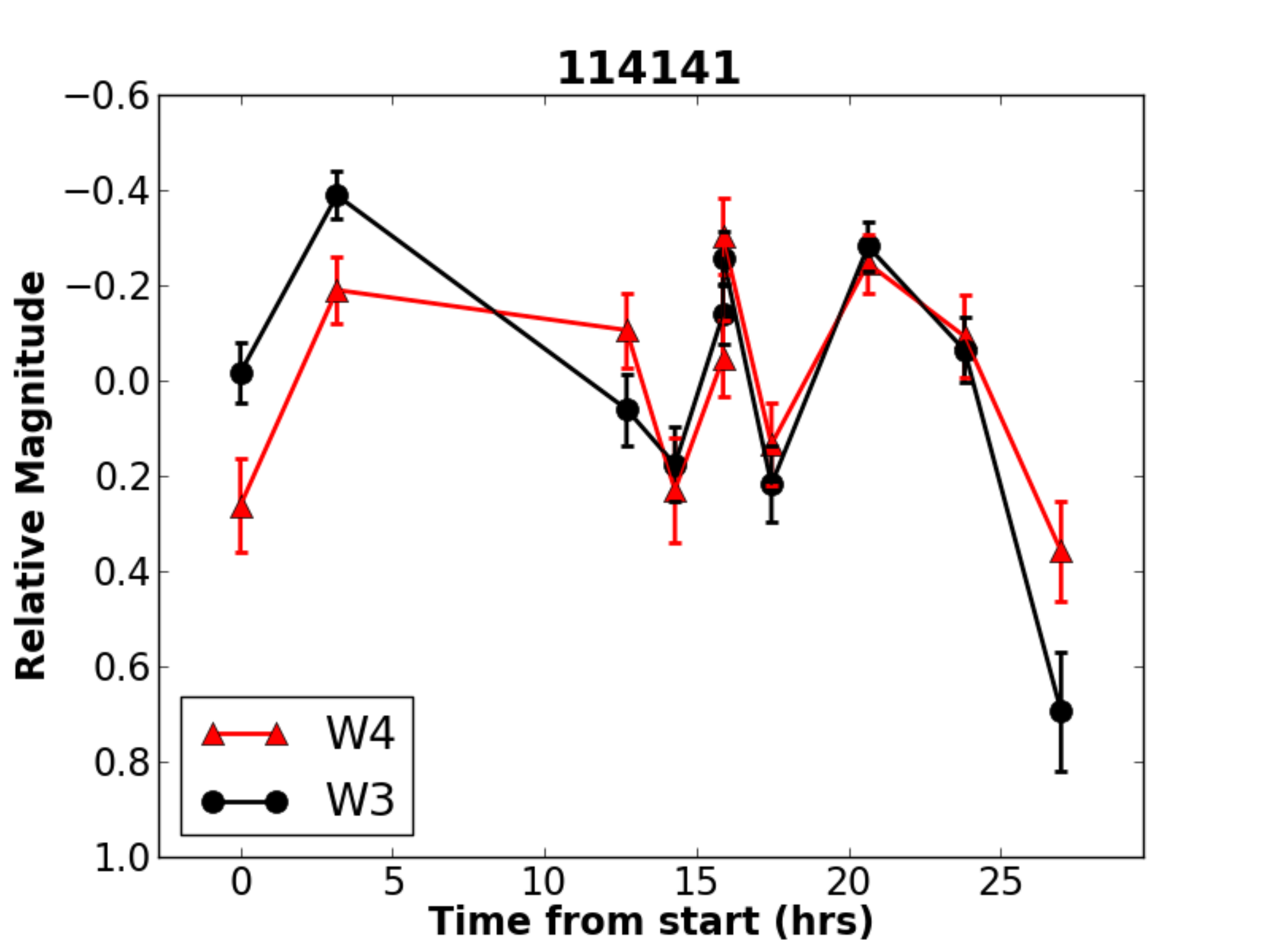}
\includegraphics[width=3.5in,height=2.6in]{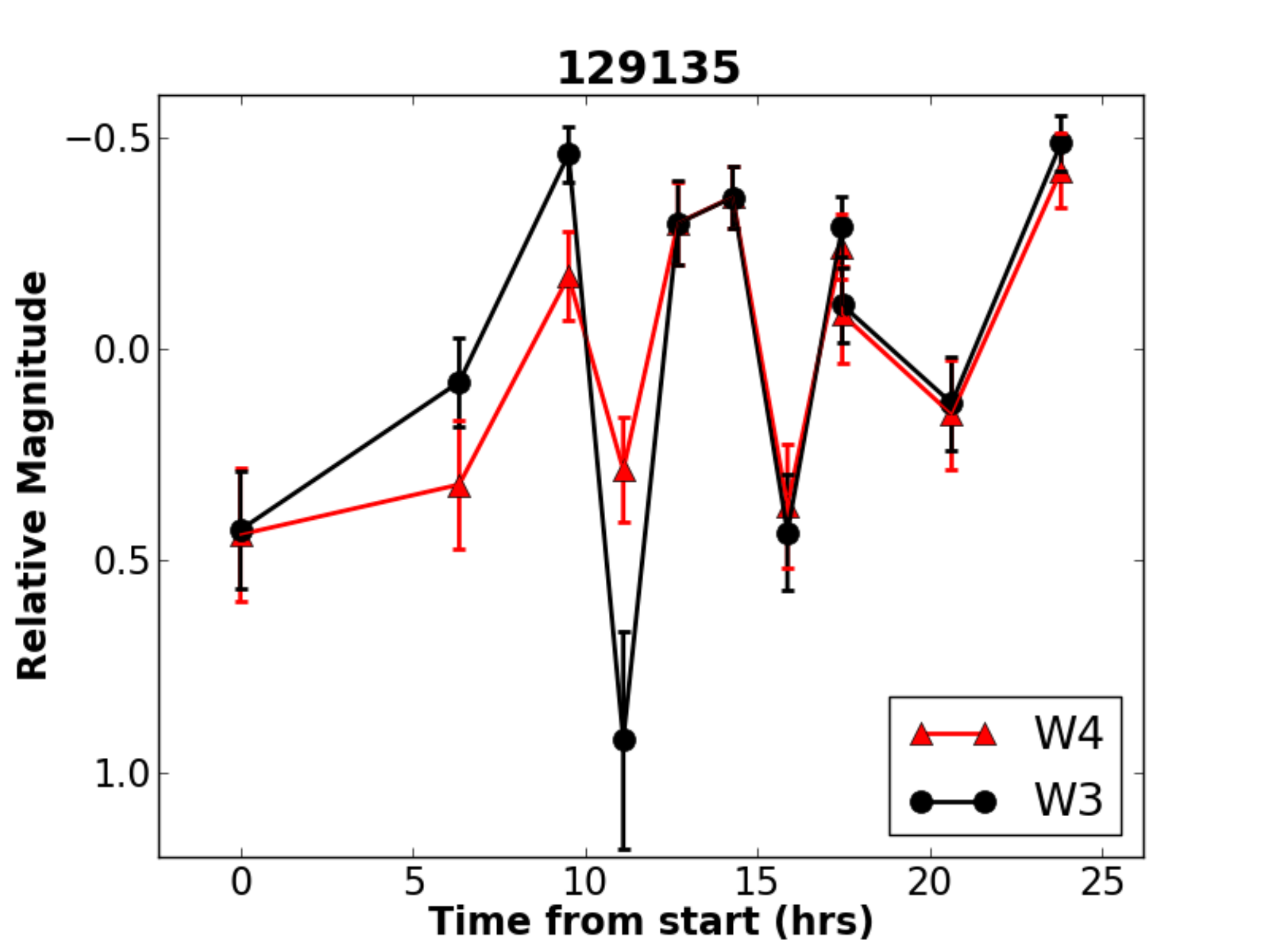}
\includegraphics[width=3.5in,height=2.6in]{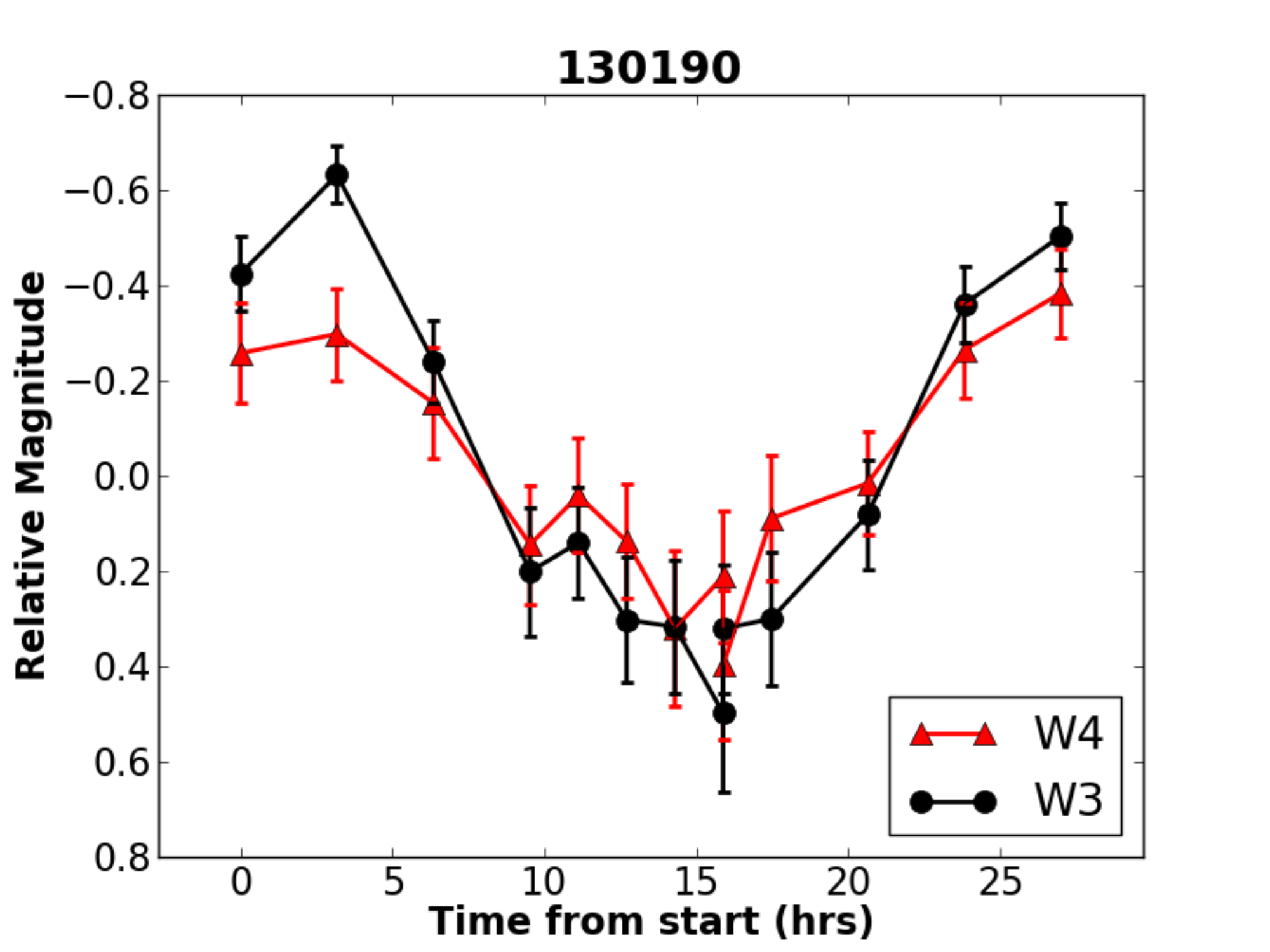}
\includegraphics[width=3.5in,height=2.6in]{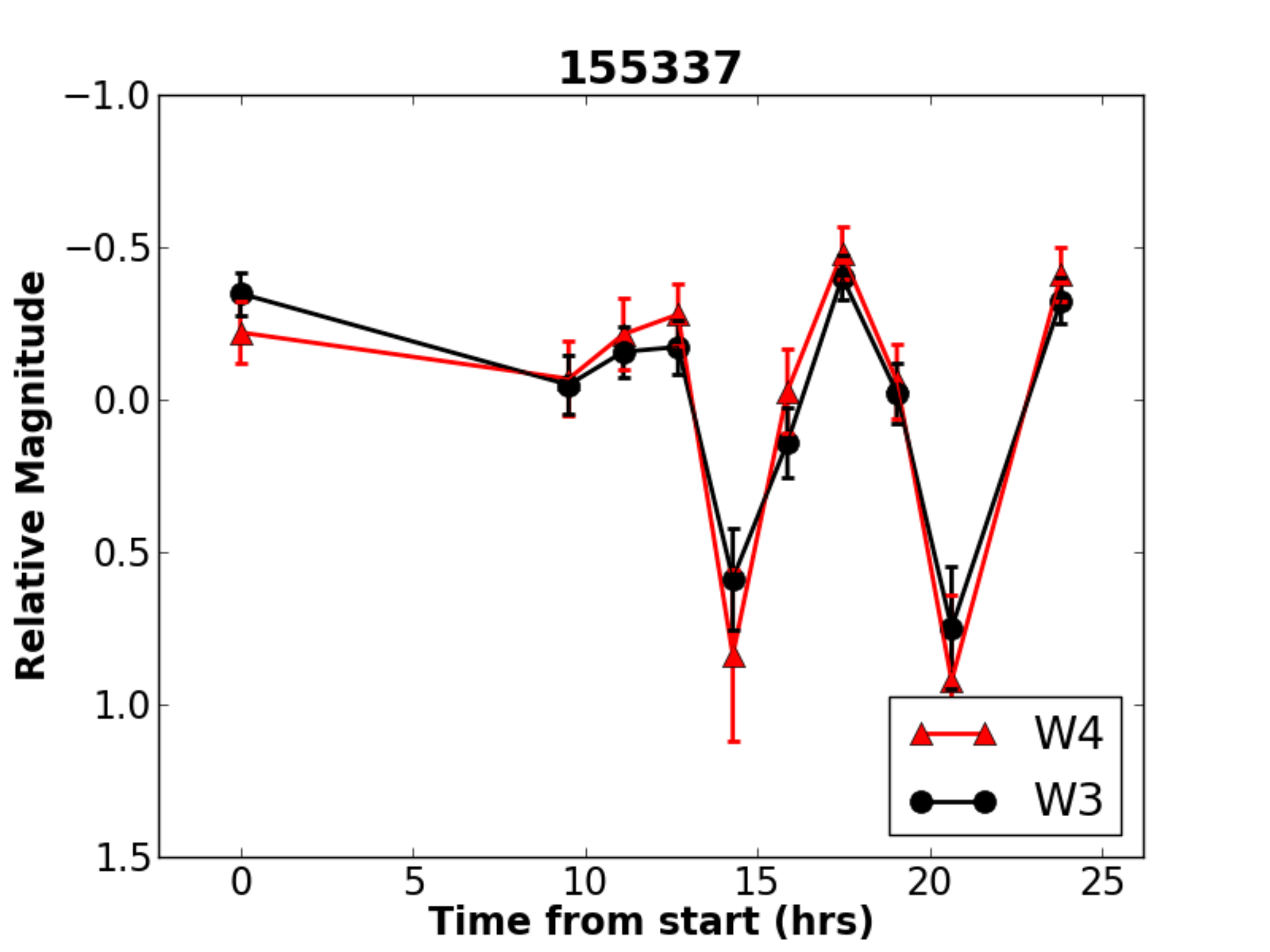}
\includegraphics[width=3.5in,height=2.6in]{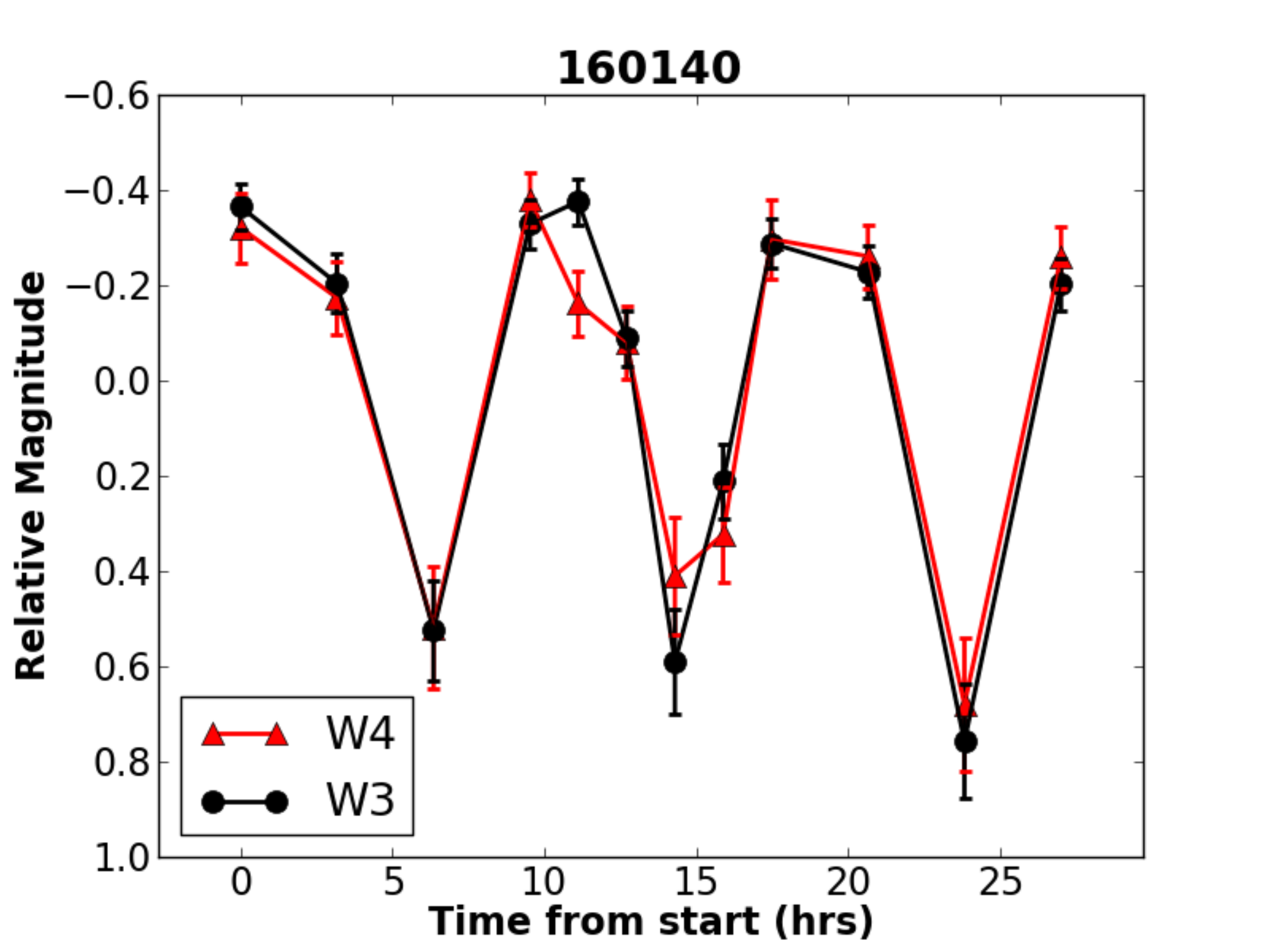}
\includegraphics[width=3.5in,height=2.6in]{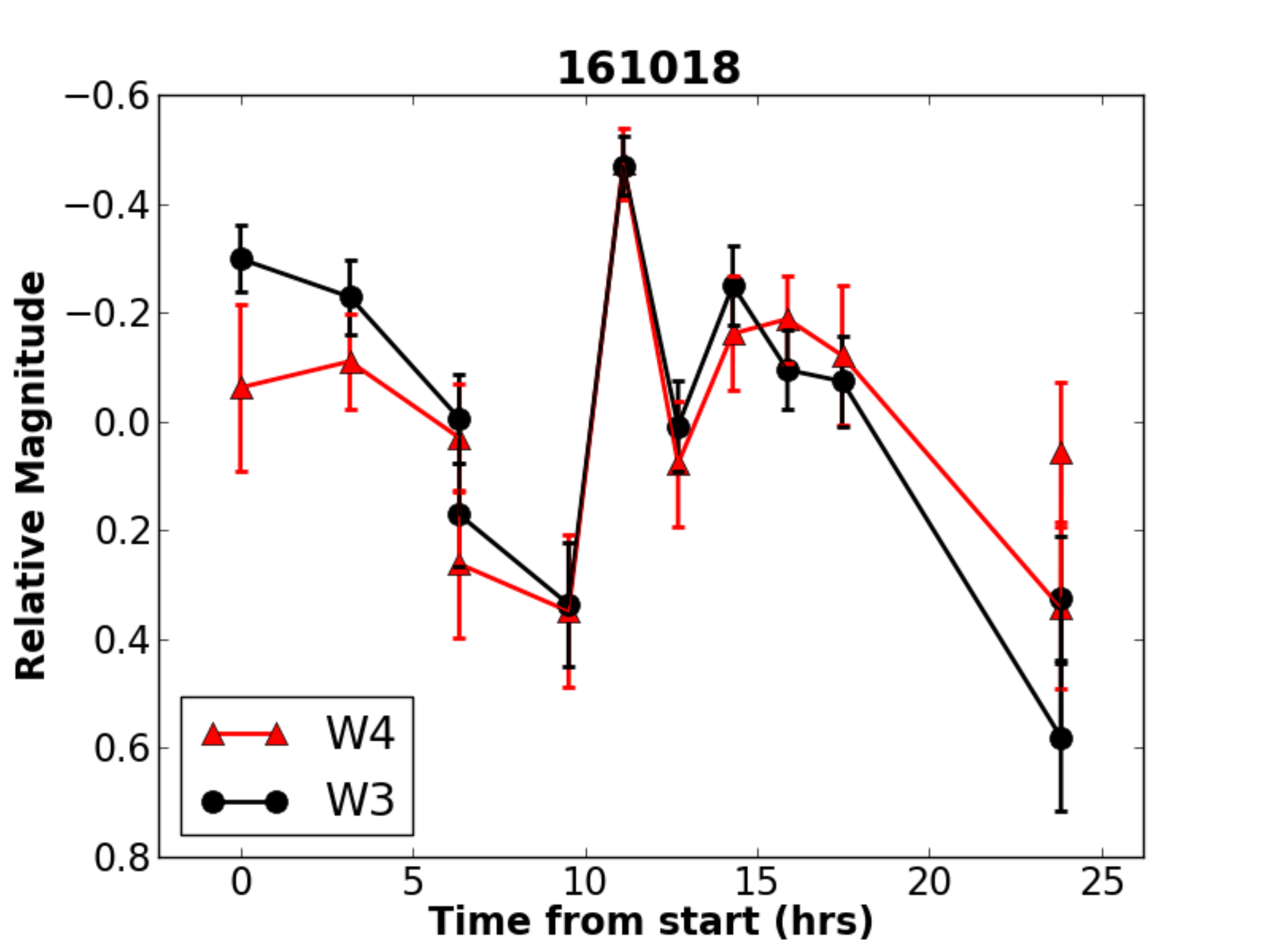}
\caption{\label{Fig:TrojanCandidates} Candidate binary Trojans from our survey identified by their anomalously high lightcurve photometric ranges, including known binaries 624 Hektor, 17365 (1978 VF$_{11}$), and 29314 Eurydamas.}
\end{figure} 

\begin{figure}
\figurenum{3}
\includegraphics[width=3.5in,height=2.6in]{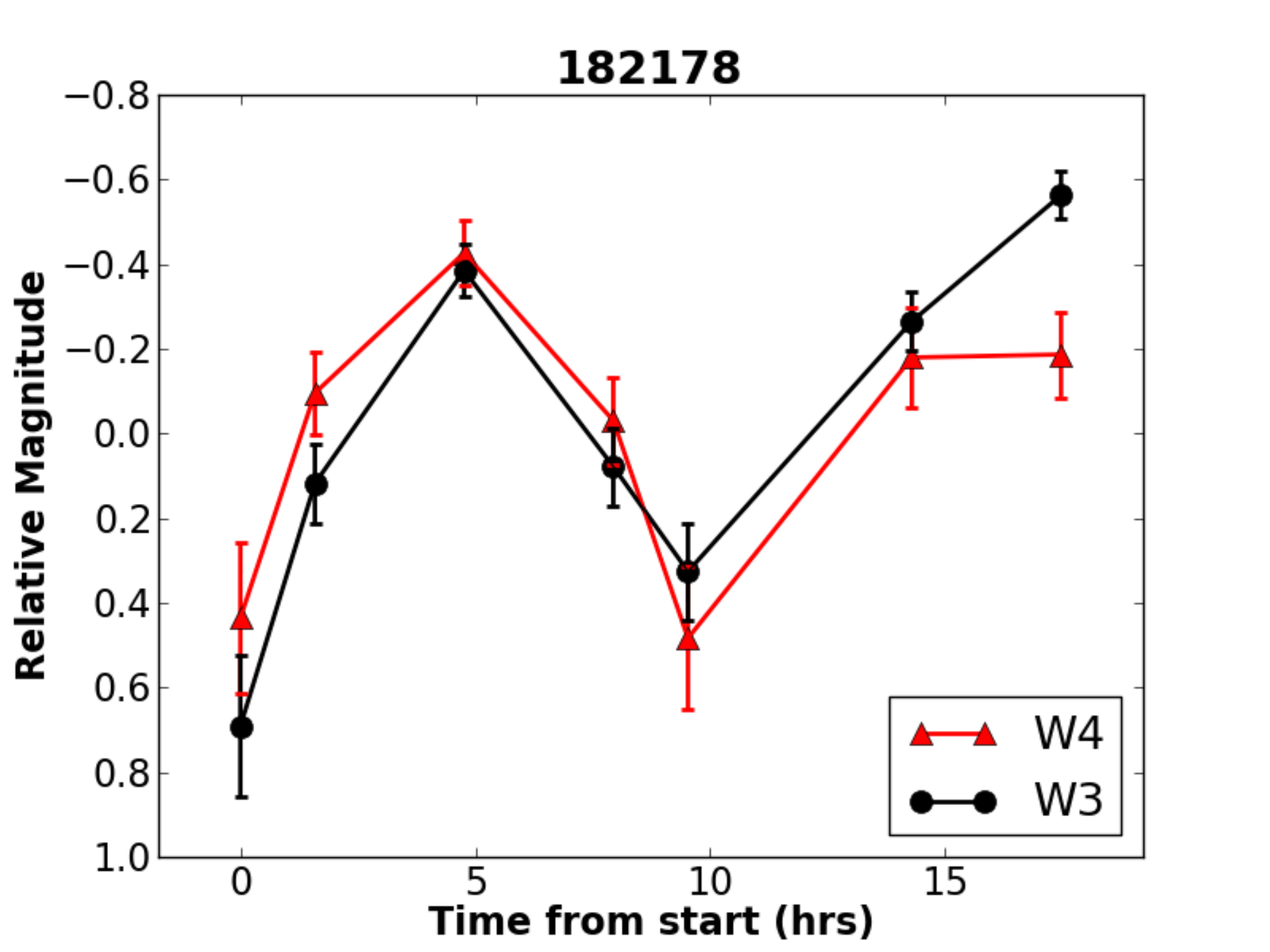}
\includegraphics[width=3.5in,height=2.6in]{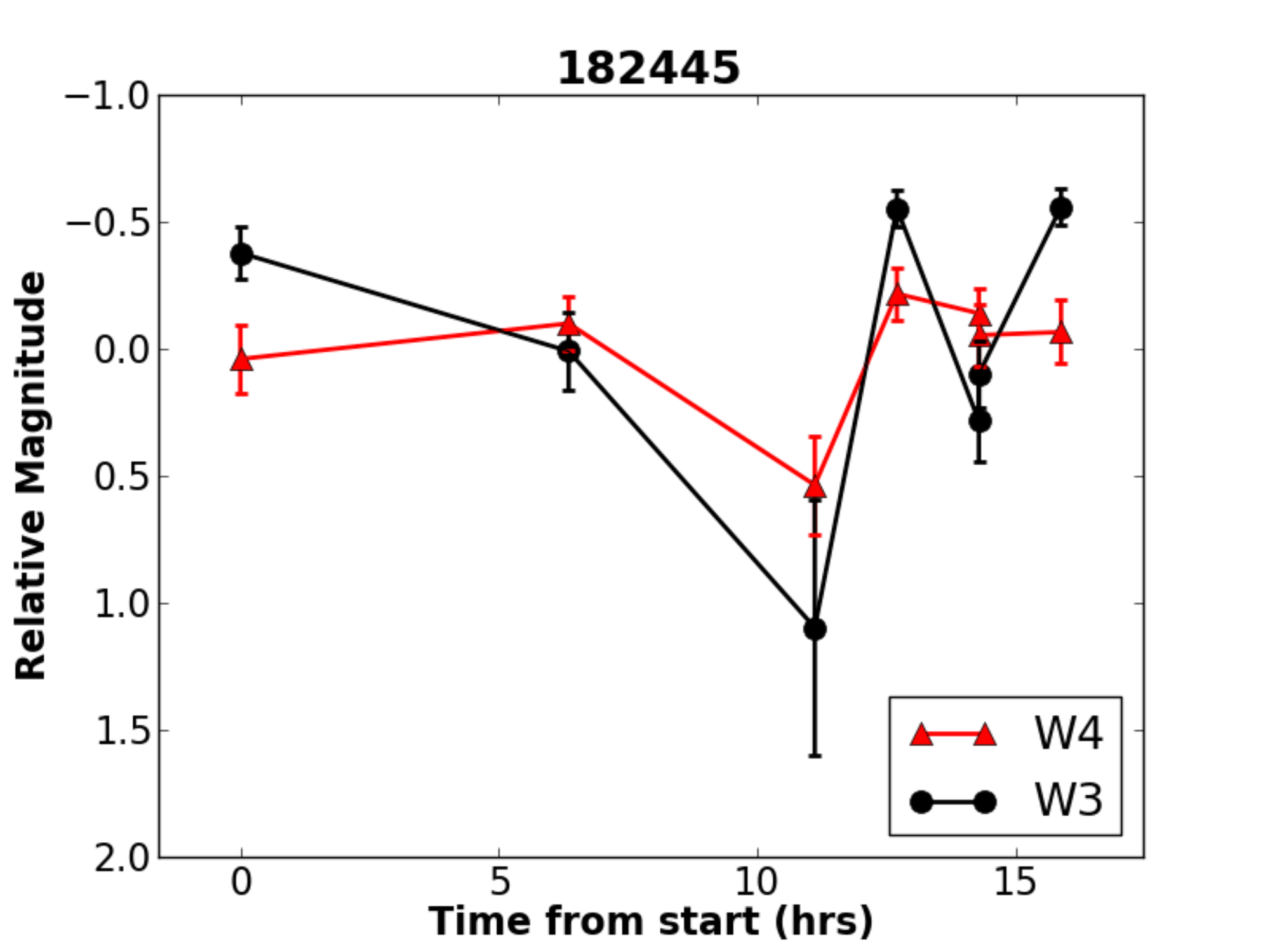}
\includegraphics[width=3.5in,height=2.6in]{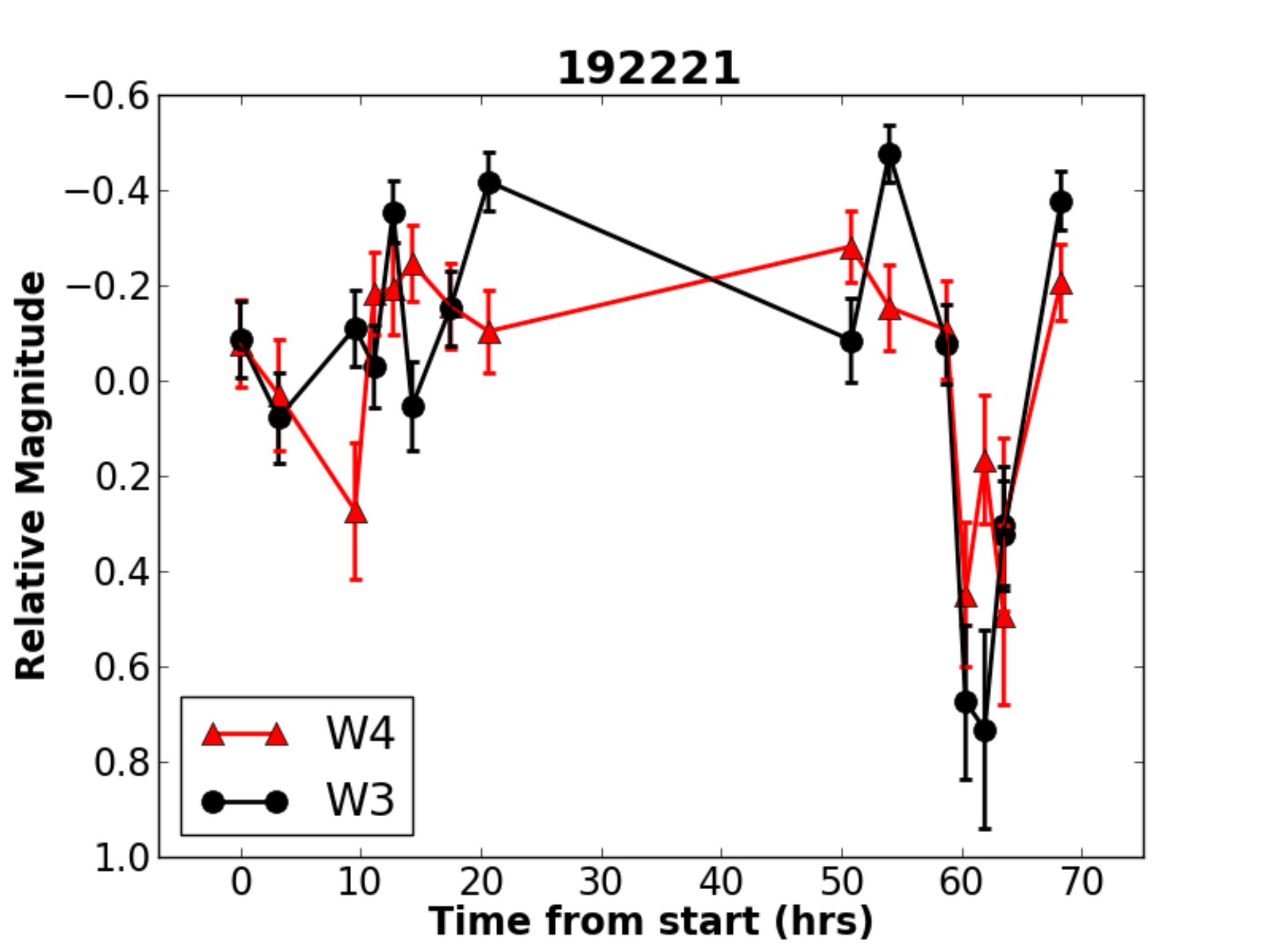}
\includegraphics[width=3.5in,height=2.6in]{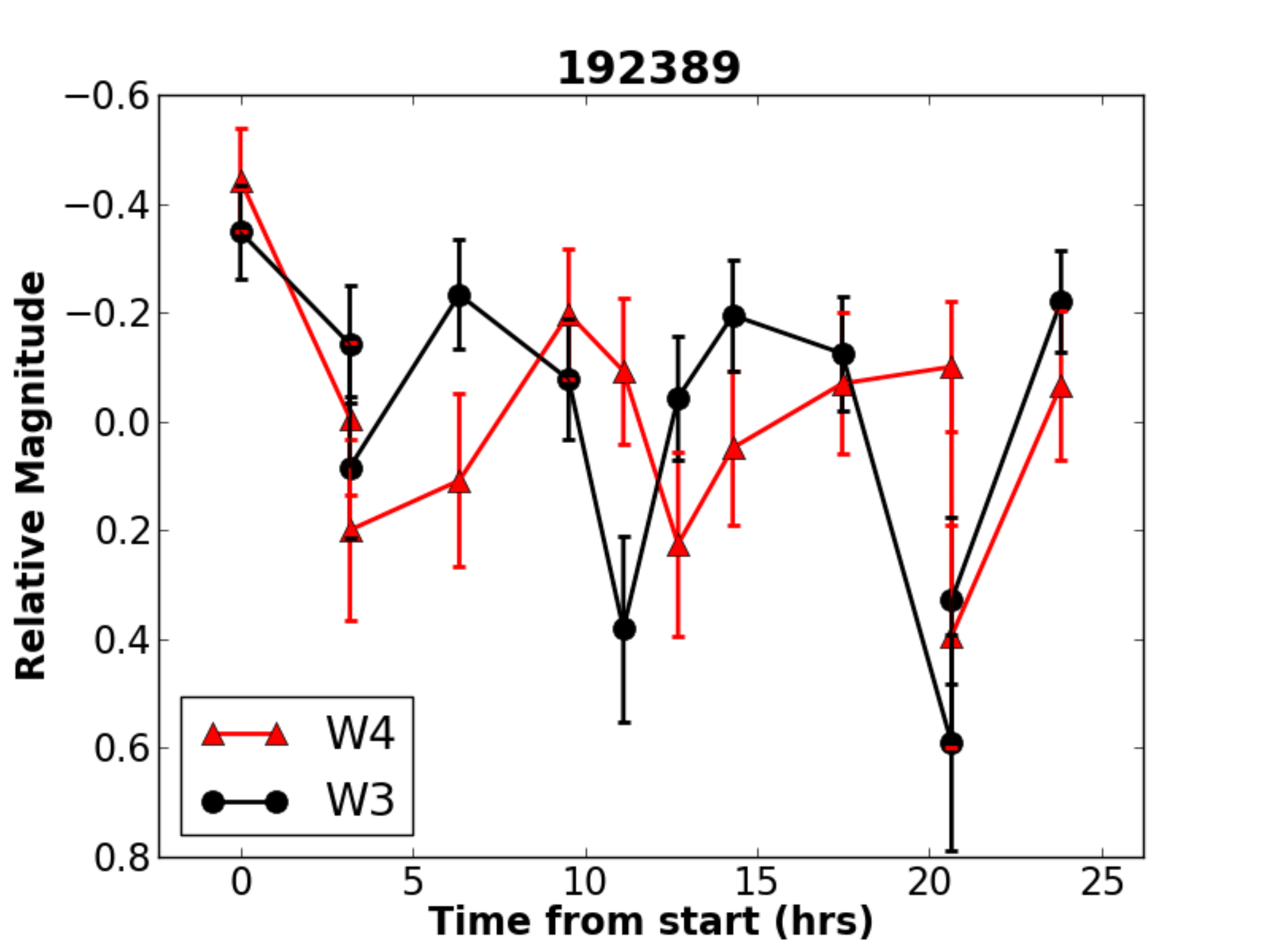}
\includegraphics[width=3.5in,height=2.6in]{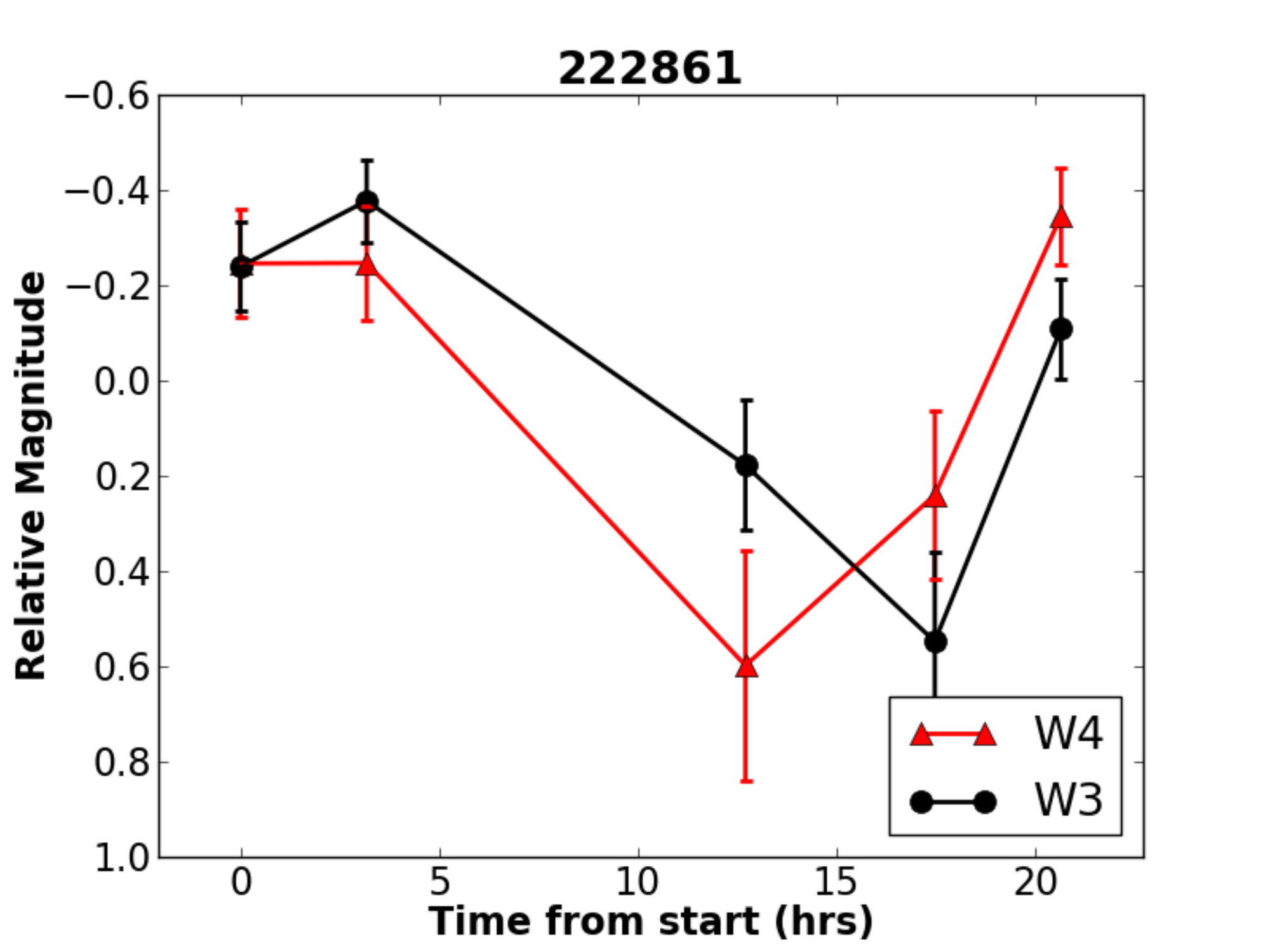}
\includegraphics[width=3.5in,height=2.6in]{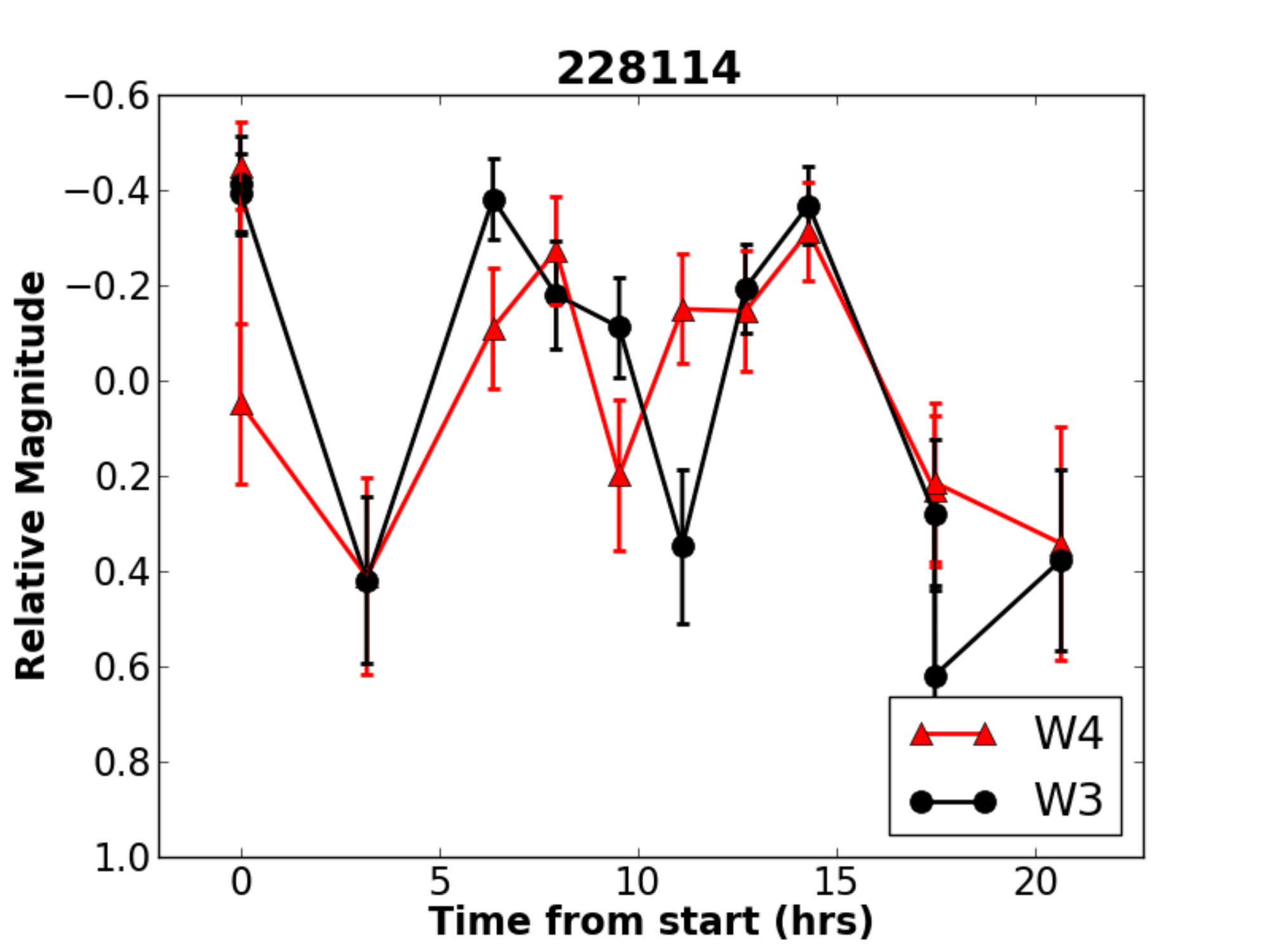}
\caption{\label{Fig:TrojanCandidates} Candidate binary Trojans from our survey identified by their anomalously high lightcurve photometric ranges, including known binaries 624 Hektor, 17365 (1978 VF$_{11}$), and 29314 Eurydamas.}
\end{figure} 

\begin{figure}
\figurenum{3}
\includegraphics[width=3.5in,height=2.6in]{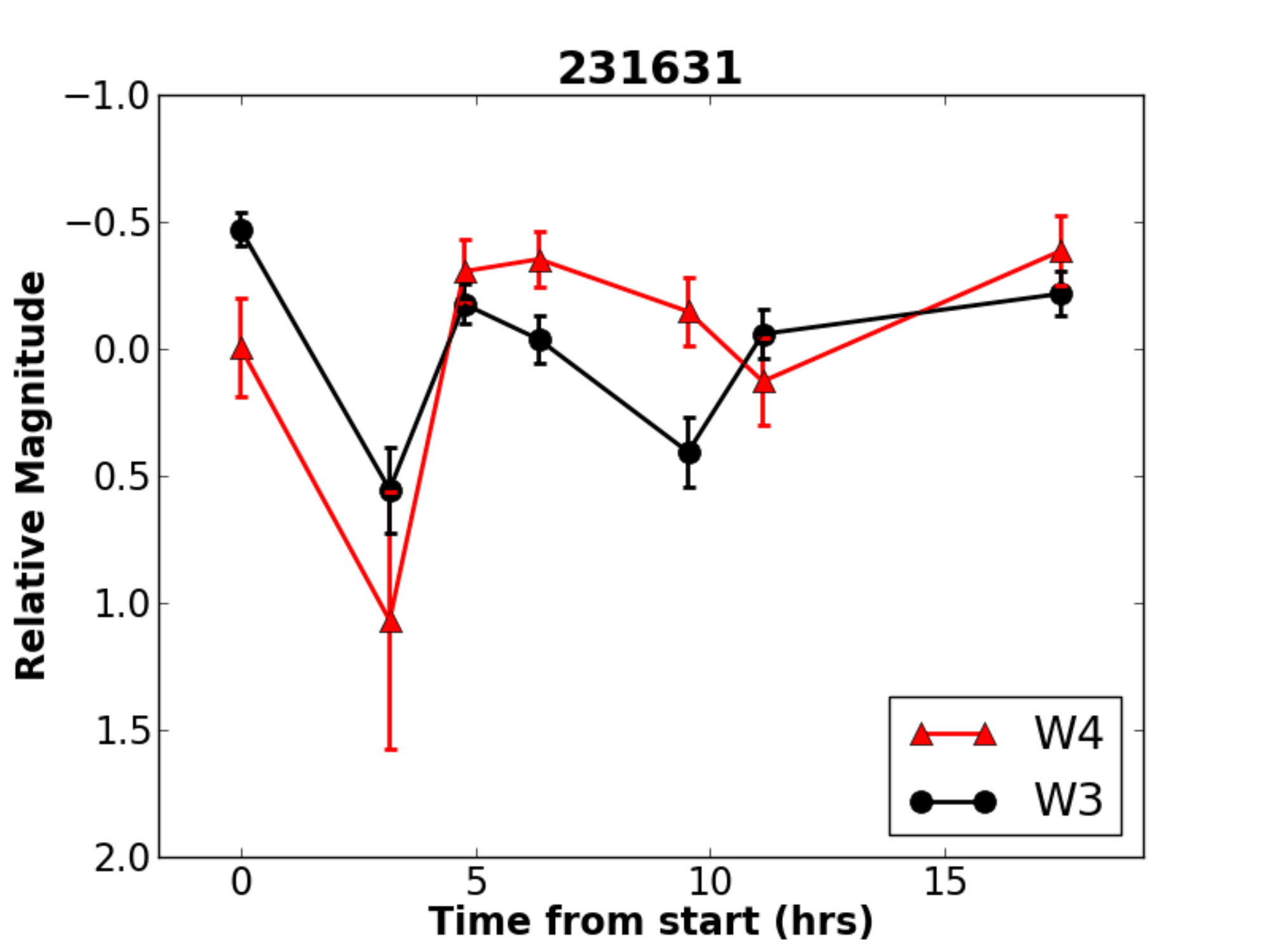}
\includegraphics[width=3.5in,height=2.6in]{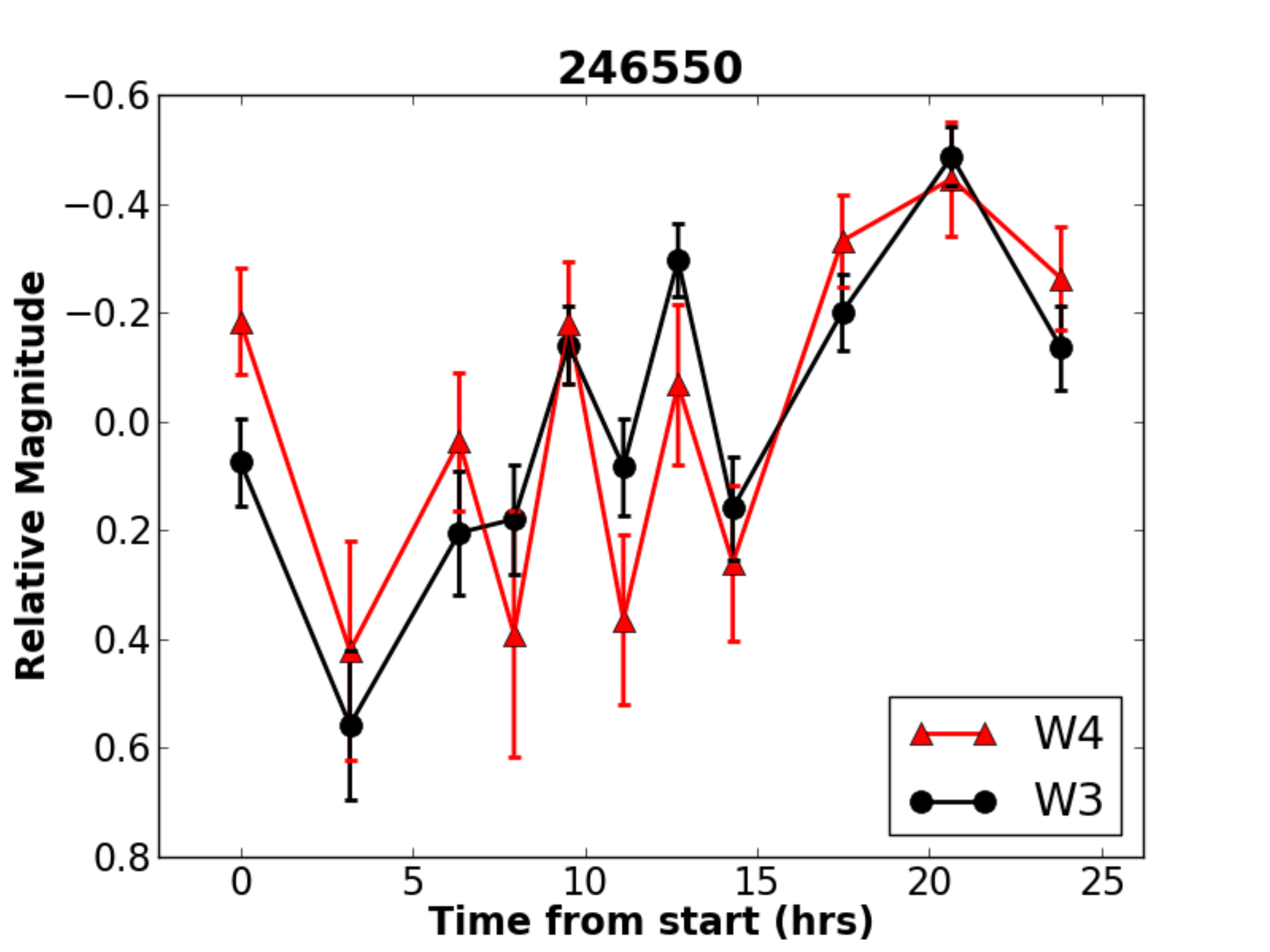}
\includegraphics[width=3.5in,height=2.6in]{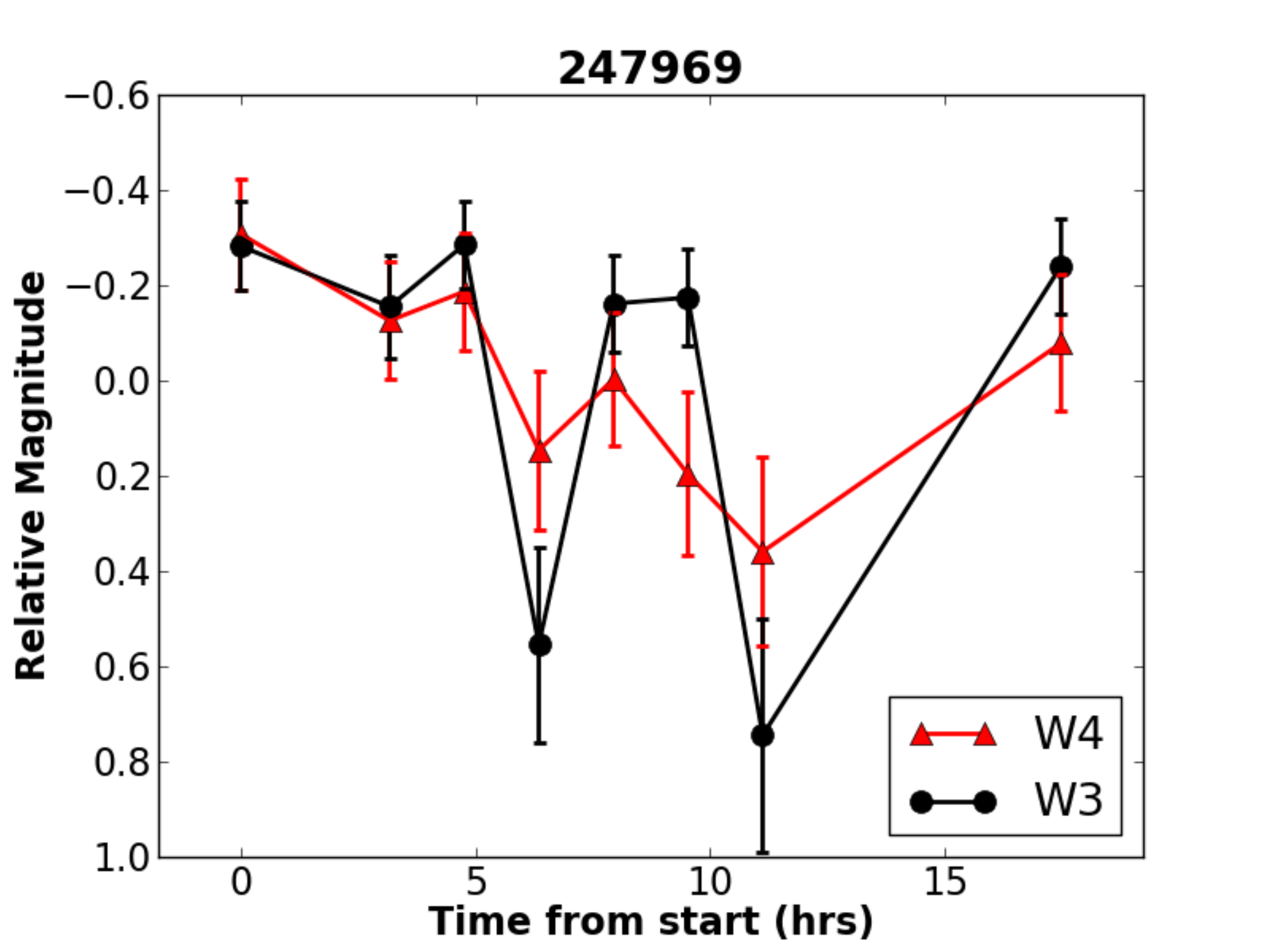}
\includegraphics[width=3.5in,height=2.6in]{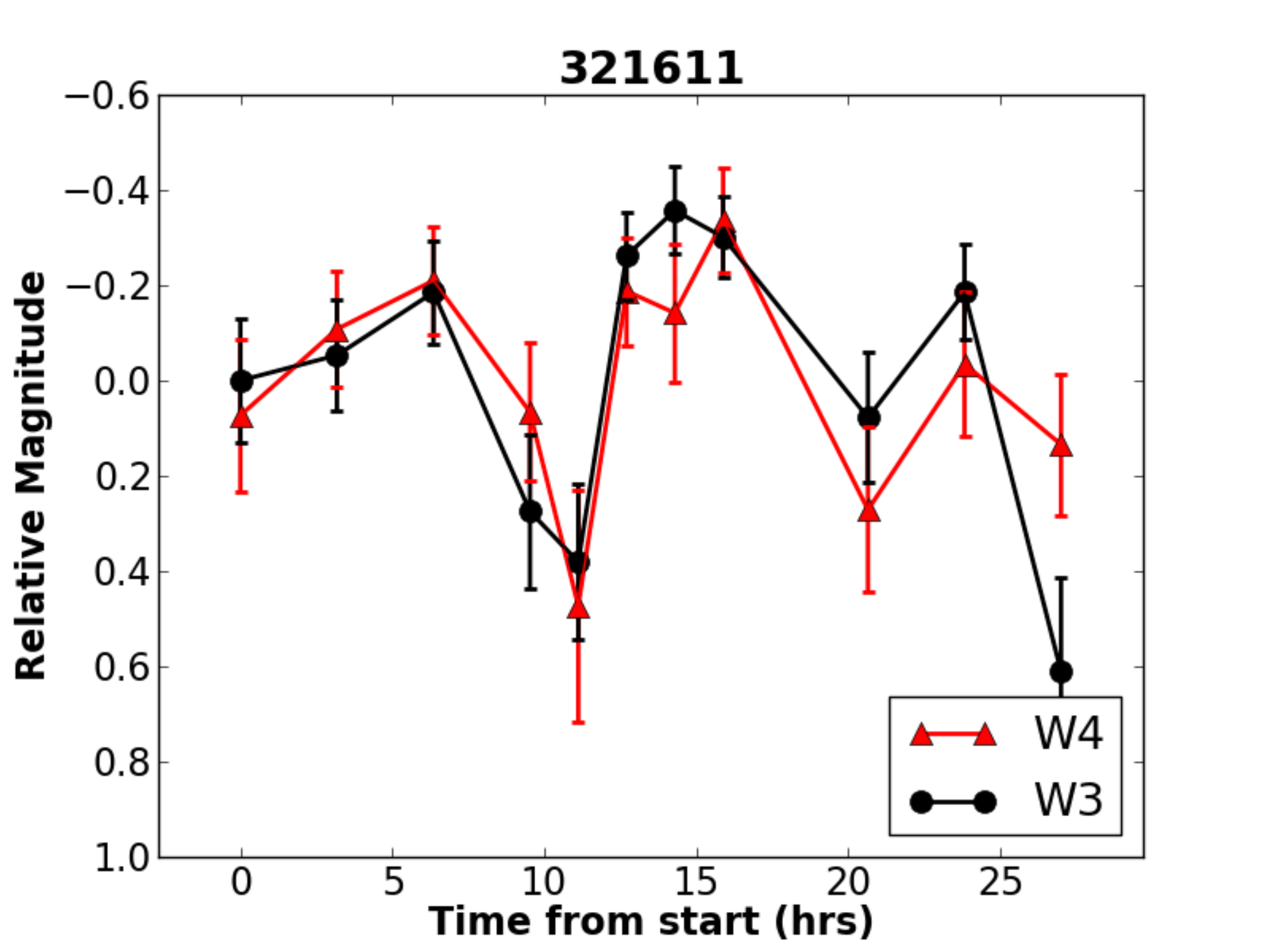}
\includegraphics[width=3.5in,height=2.6in]{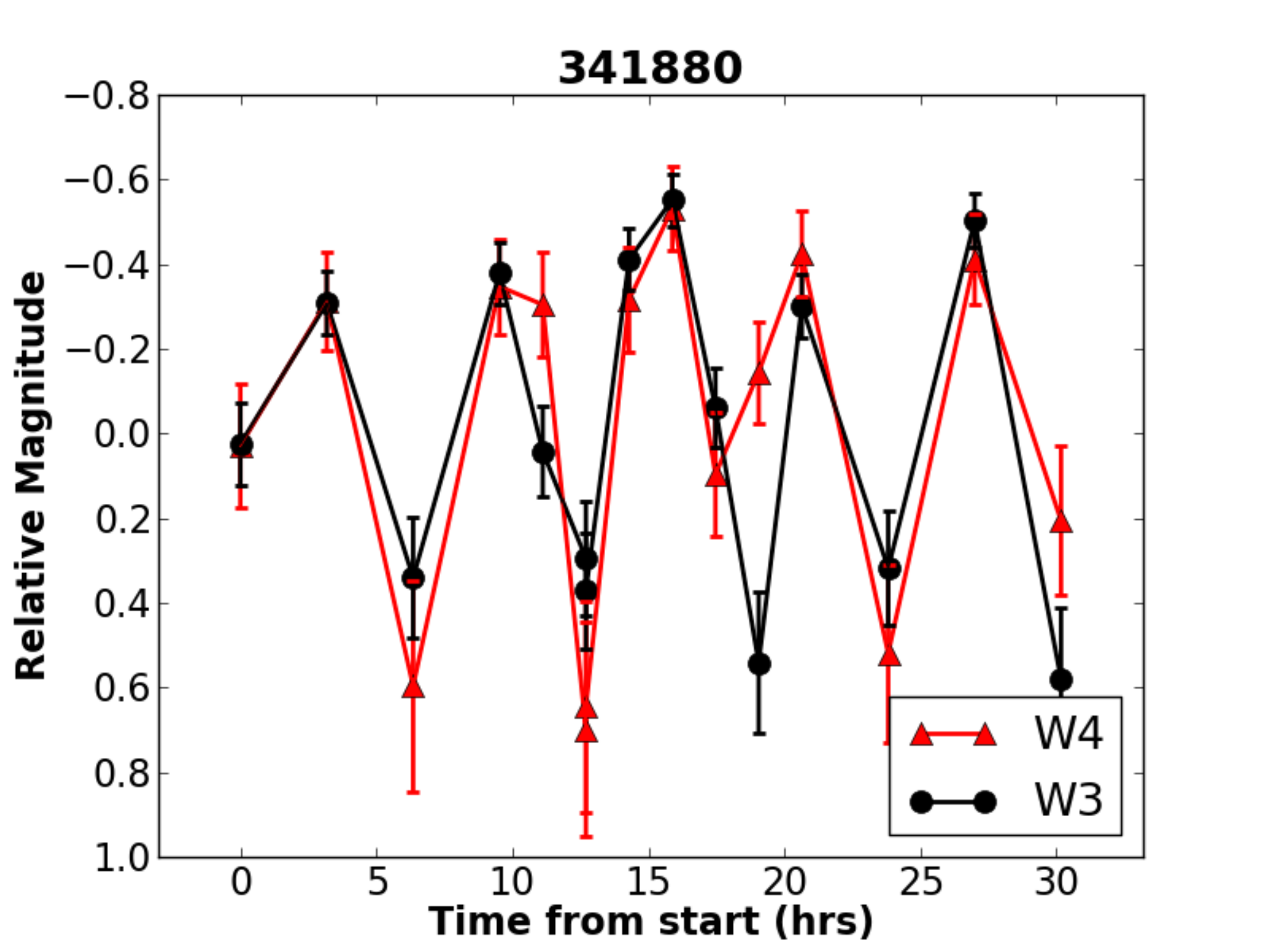}
\includegraphics[width=3.5in,height=2.6in]{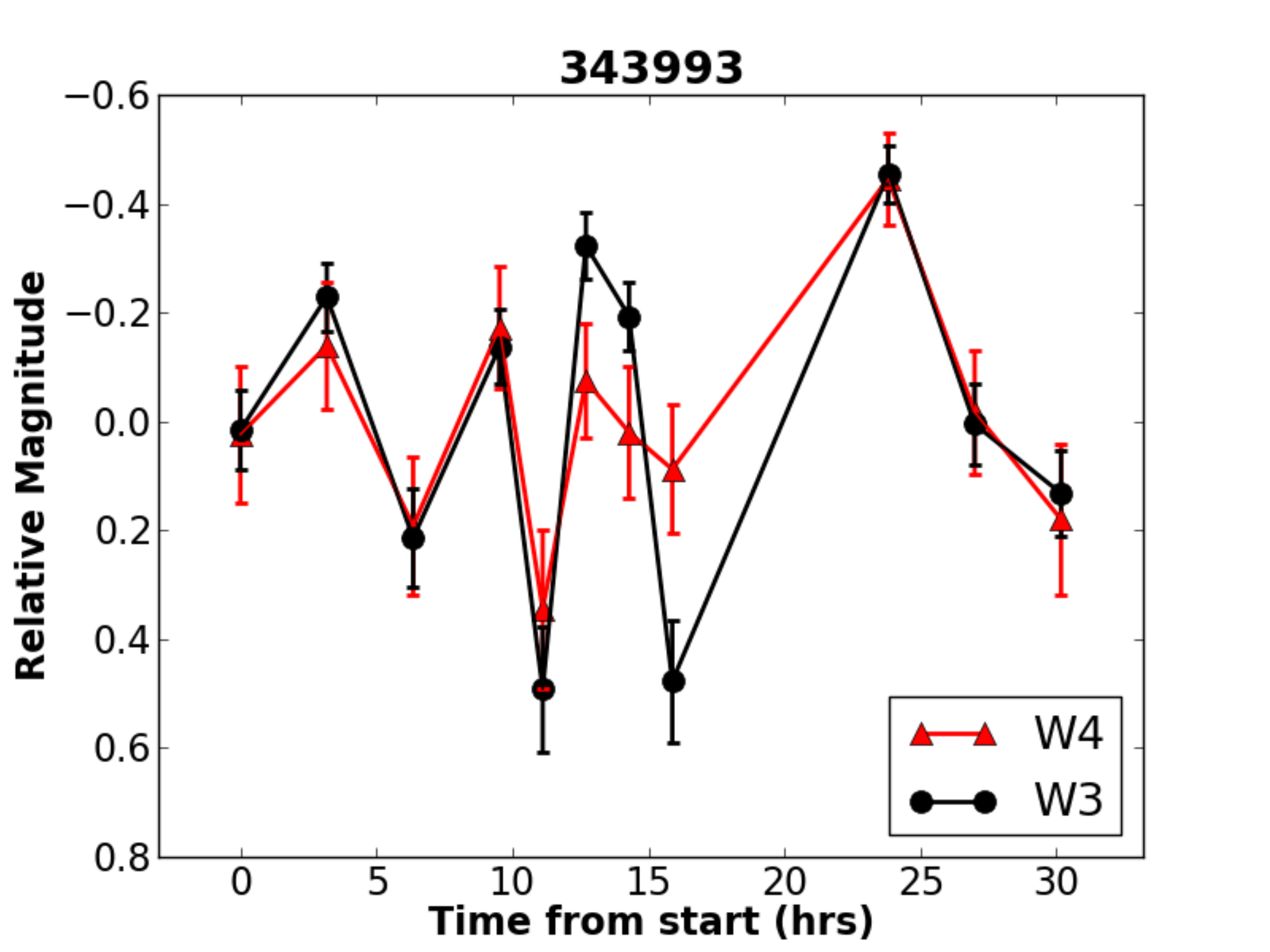}
\caption{\label{Fig:TrojanCandidates} Candidate binary Trojans from our survey identified by their anomalously high lightcurve photometric ranges, including known binaries 624 Hektor, 17365 (1978 VF$_{11}$), and 29314 Eurydamas.}
\end{figure} 

\begin{figure}
\figurenum{3}
\includegraphics[width=3.5in,height=2.6in]{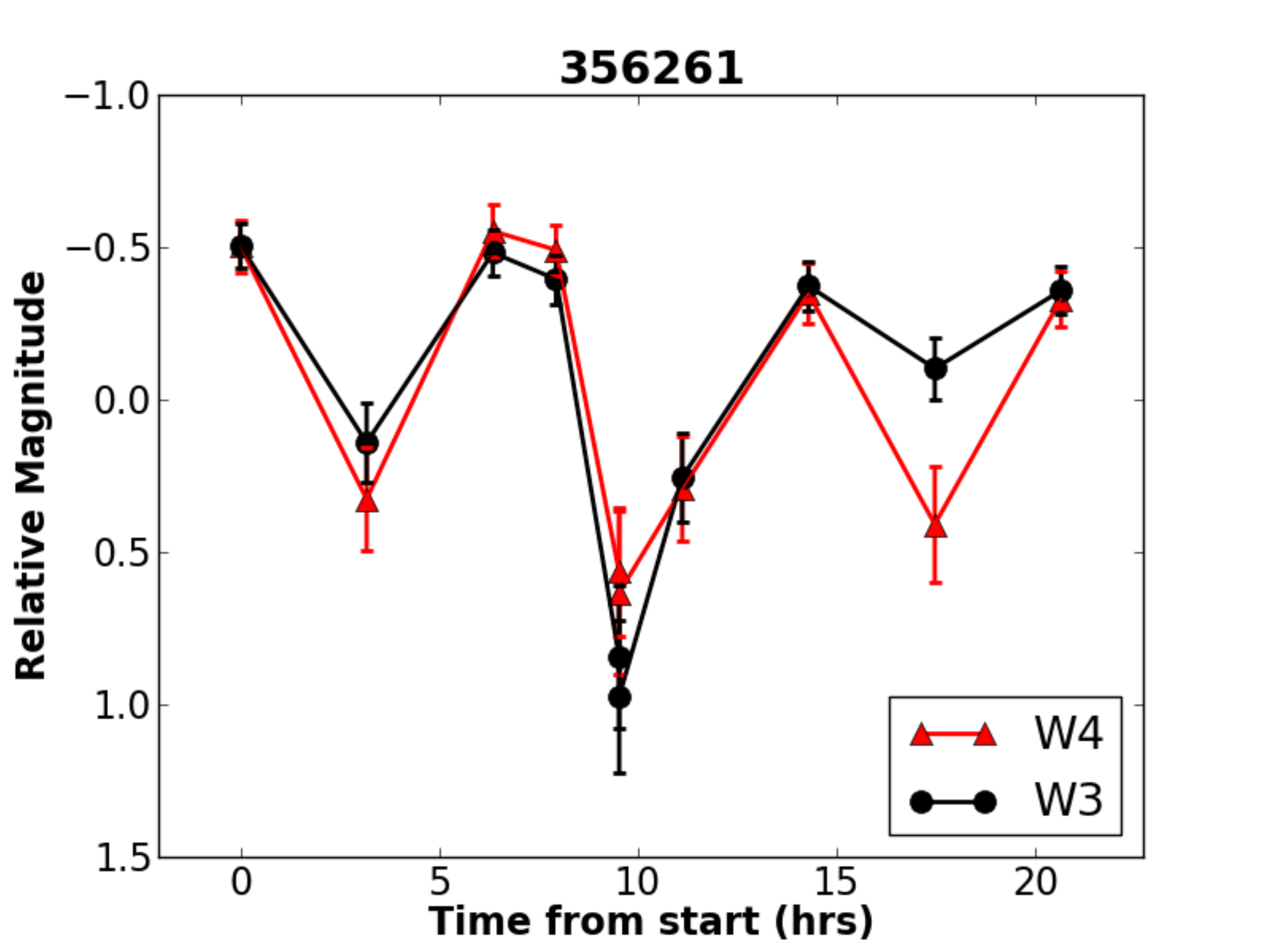}
\caption{\label{Fig:TrojanCandidates} Candidate binary Trojans from our survey identified by their anomalously high lightcurve photometric ranges, including known binaries 624 Hektor, 17365 (1978 VF$_{11}$), and 29314 Eurydamas.}
\end{figure} 

\begin{figure}
\figurenum{4}
\includegraphics[width=3.5in,height=2.6in]{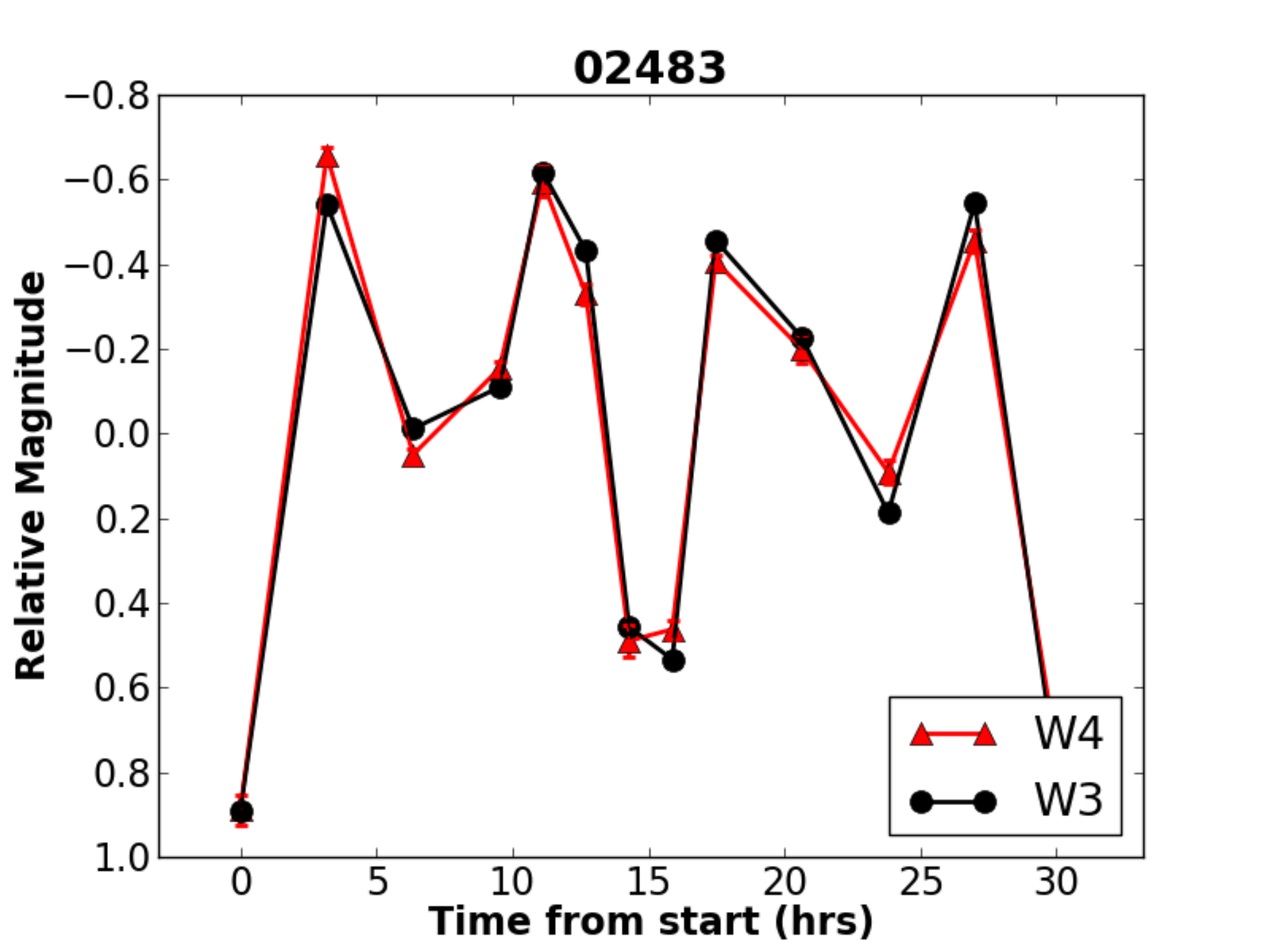}
\includegraphics[width=3.5in,height=2.6in]{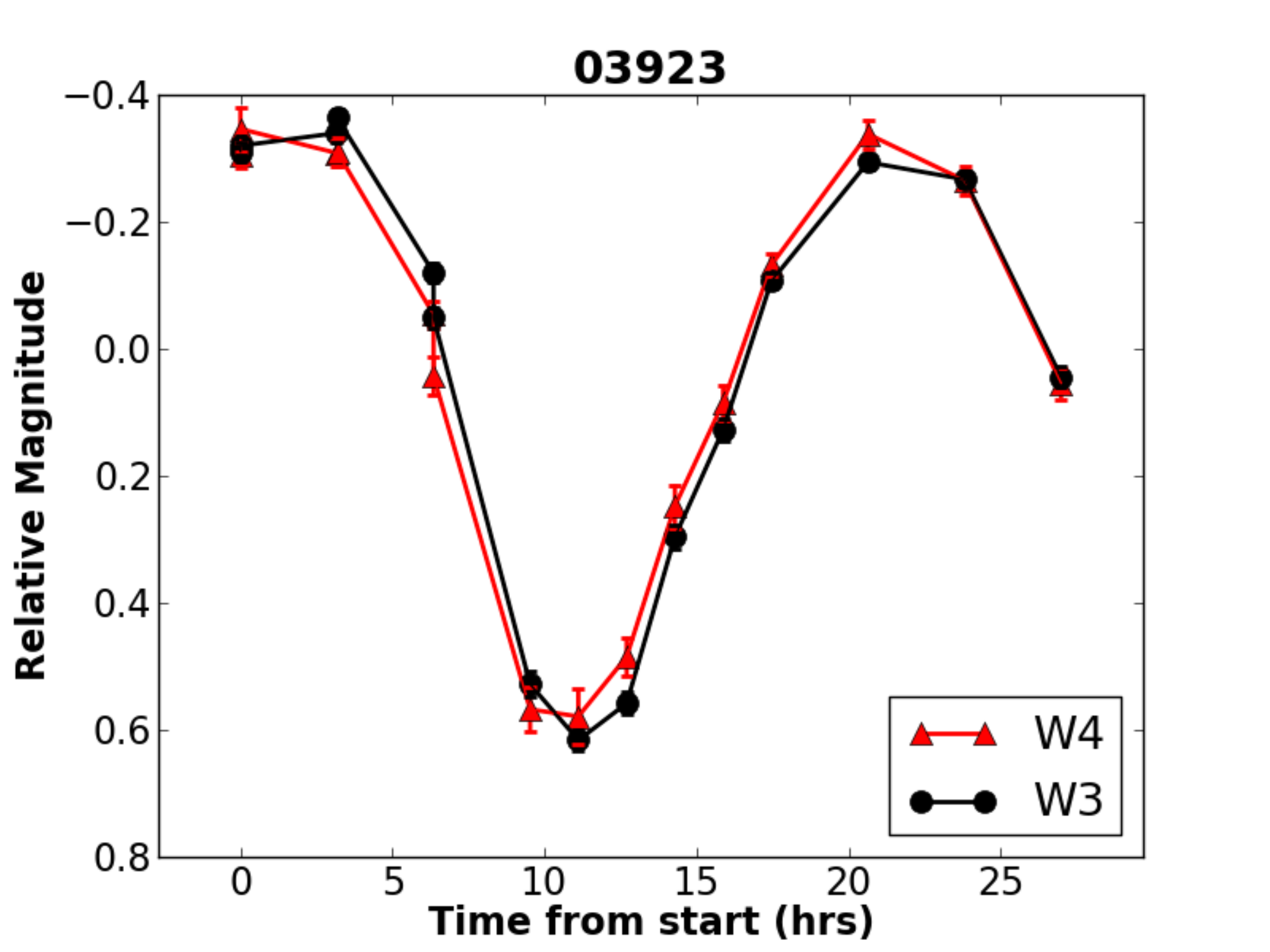}
\includegraphics[width=3.5in,height=2.6in]{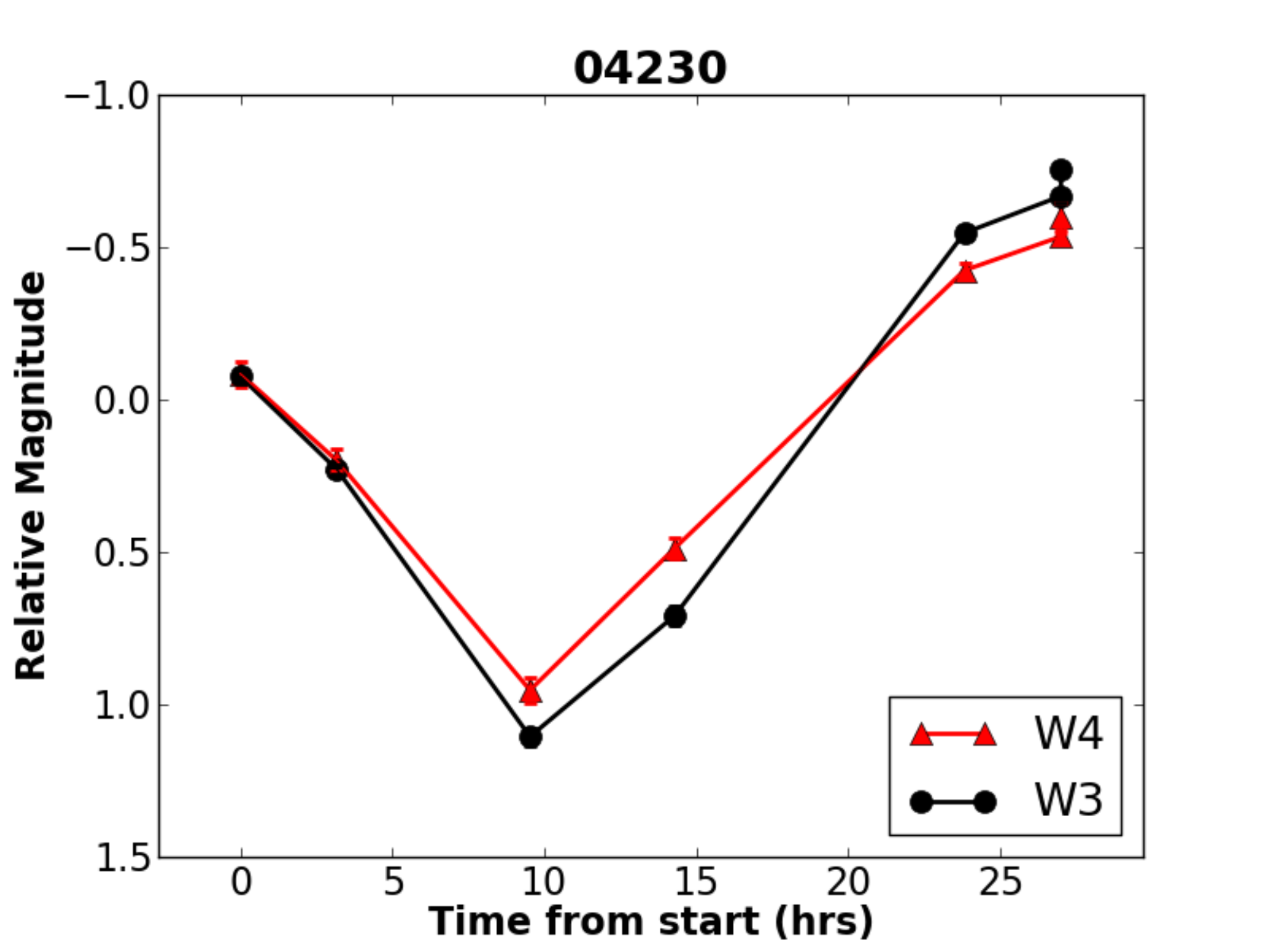}
\includegraphics[width=3.5in,height=2.6in]{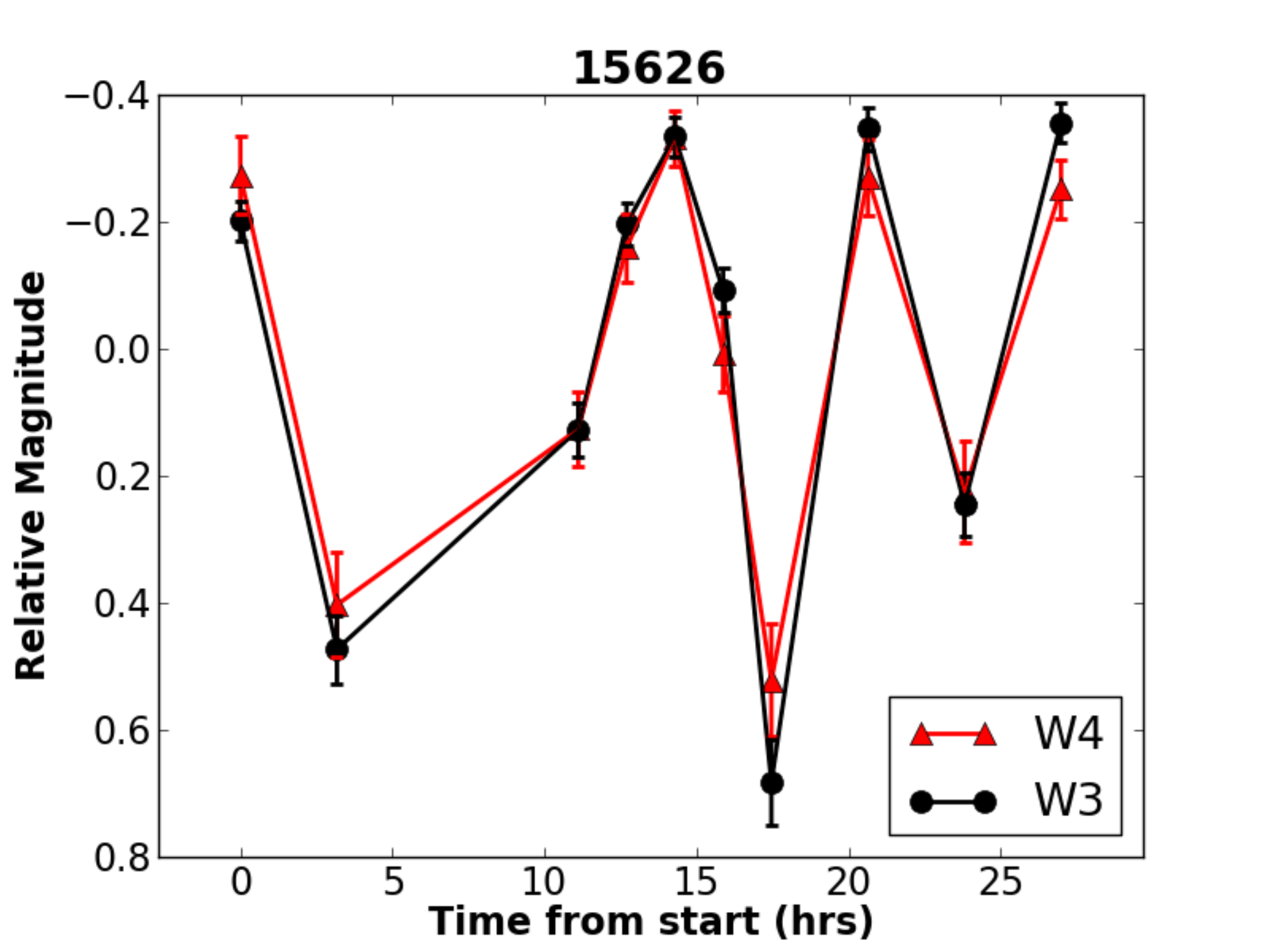}
\includegraphics[width=3.5in,height=2.6in]{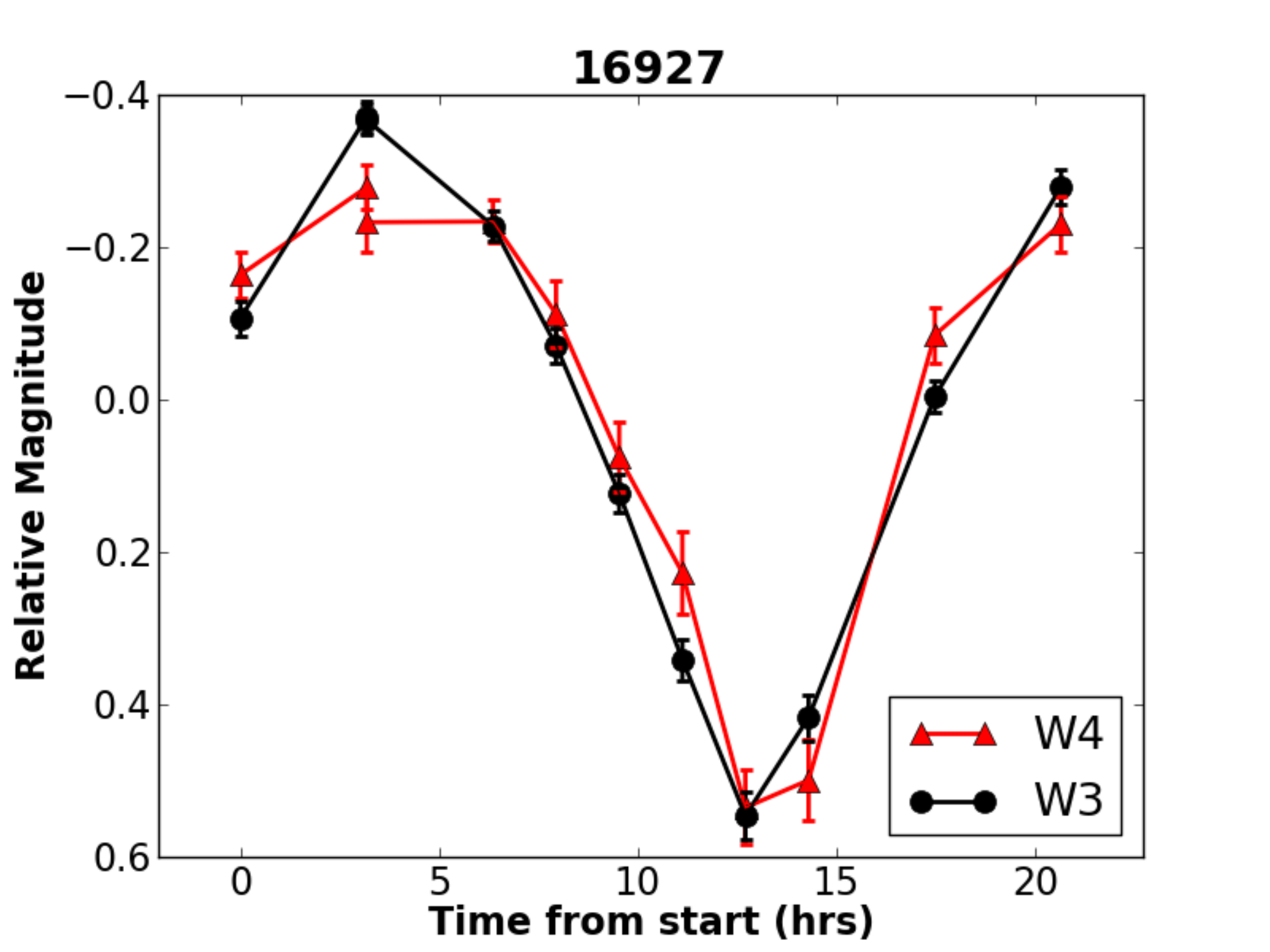}
\includegraphics[width=3.5in,height=2.6in]{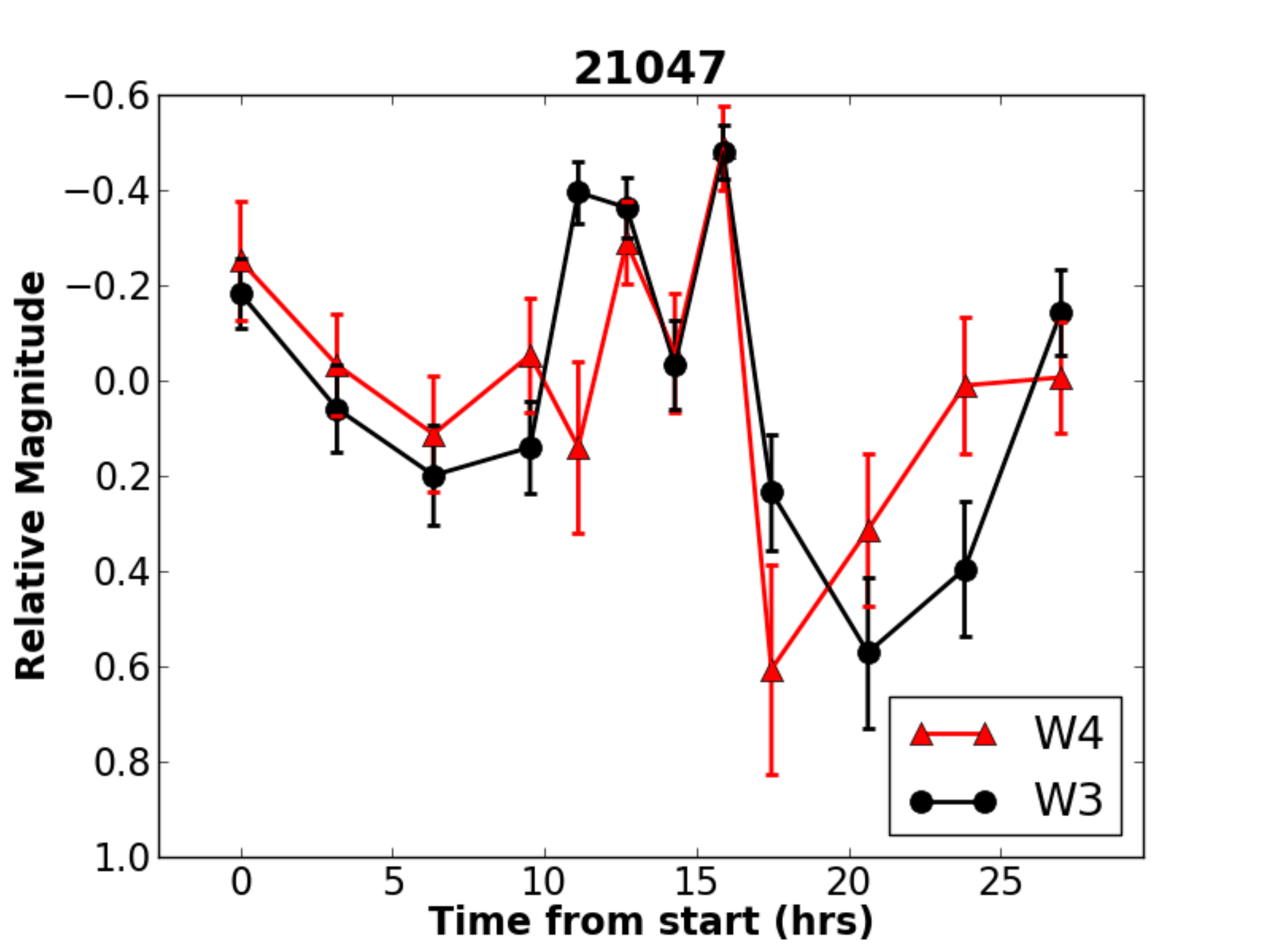}
\caption{\label{Fig:HildaCandidates} Candidate binary Hildas from our survey identified by their anomalously high lightcurve photometric ranges.}
\end{figure} 

\begin{figure}
\figurenum{4}
\includegraphics[width=3.5in,height=2.6in]{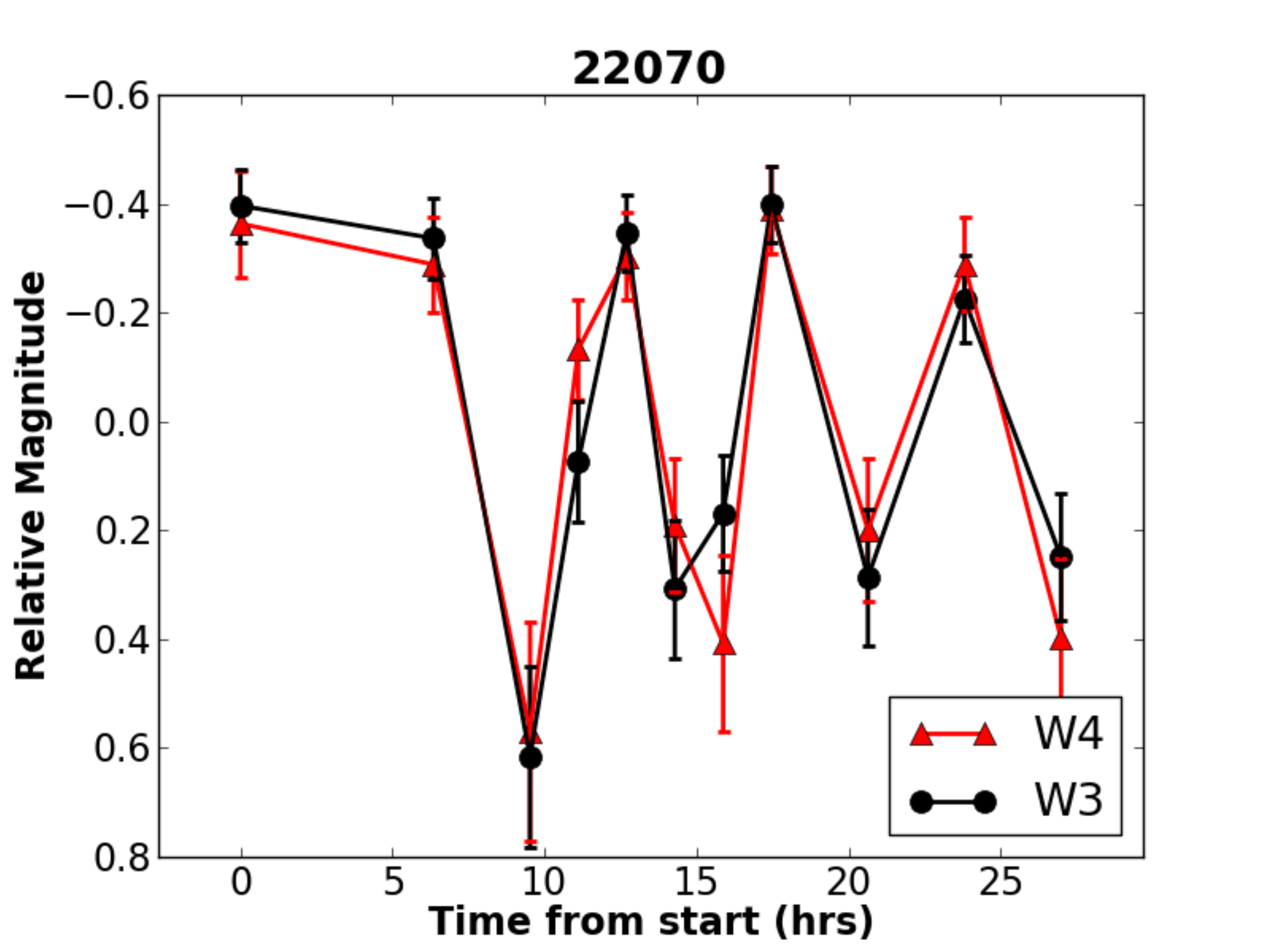}
\includegraphics[width=3.5in,height=2.6in]{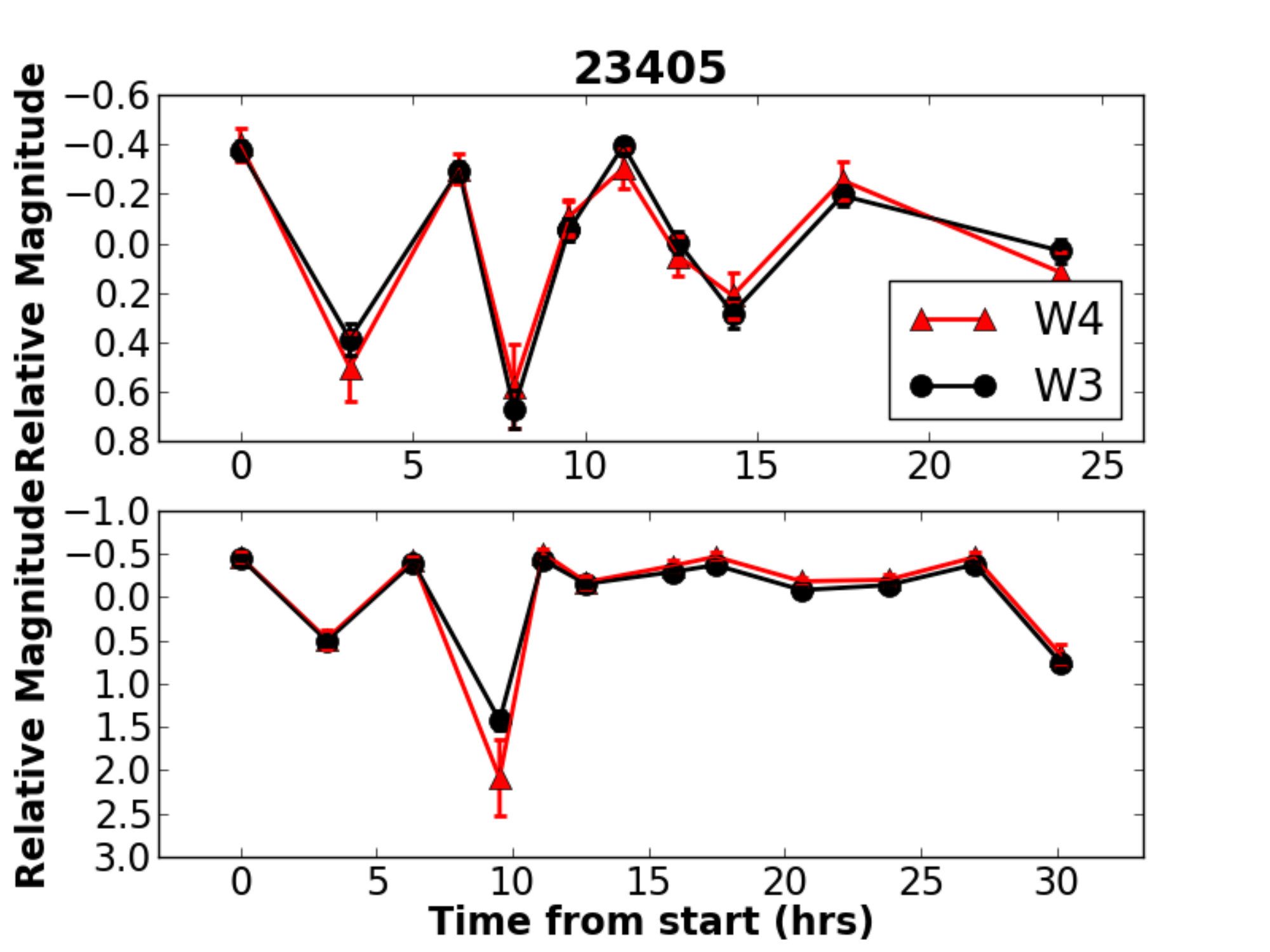}
\includegraphics[width=3.5in,height=2.6in]{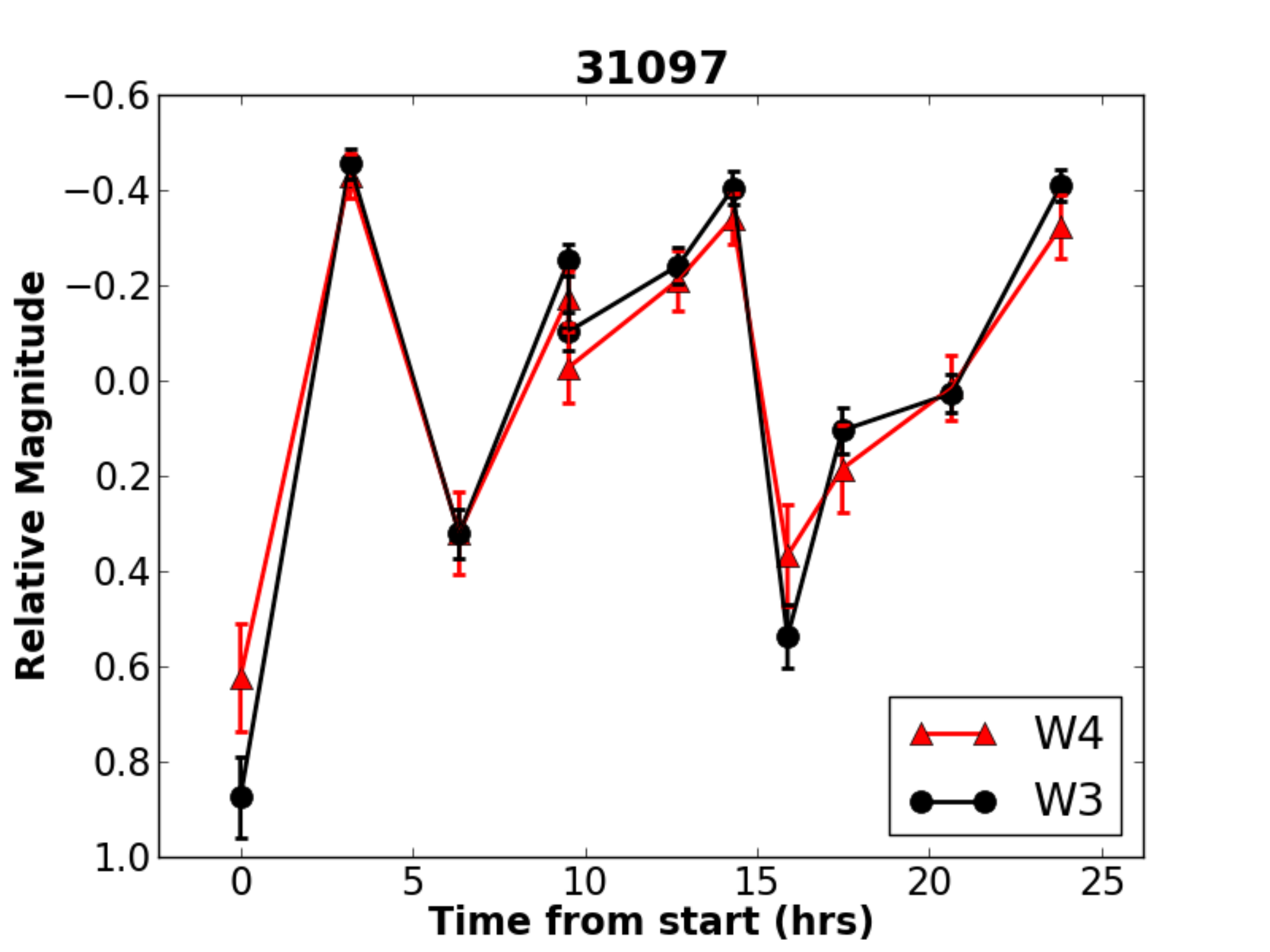}
\includegraphics[width=3.5in,height=2.6in]{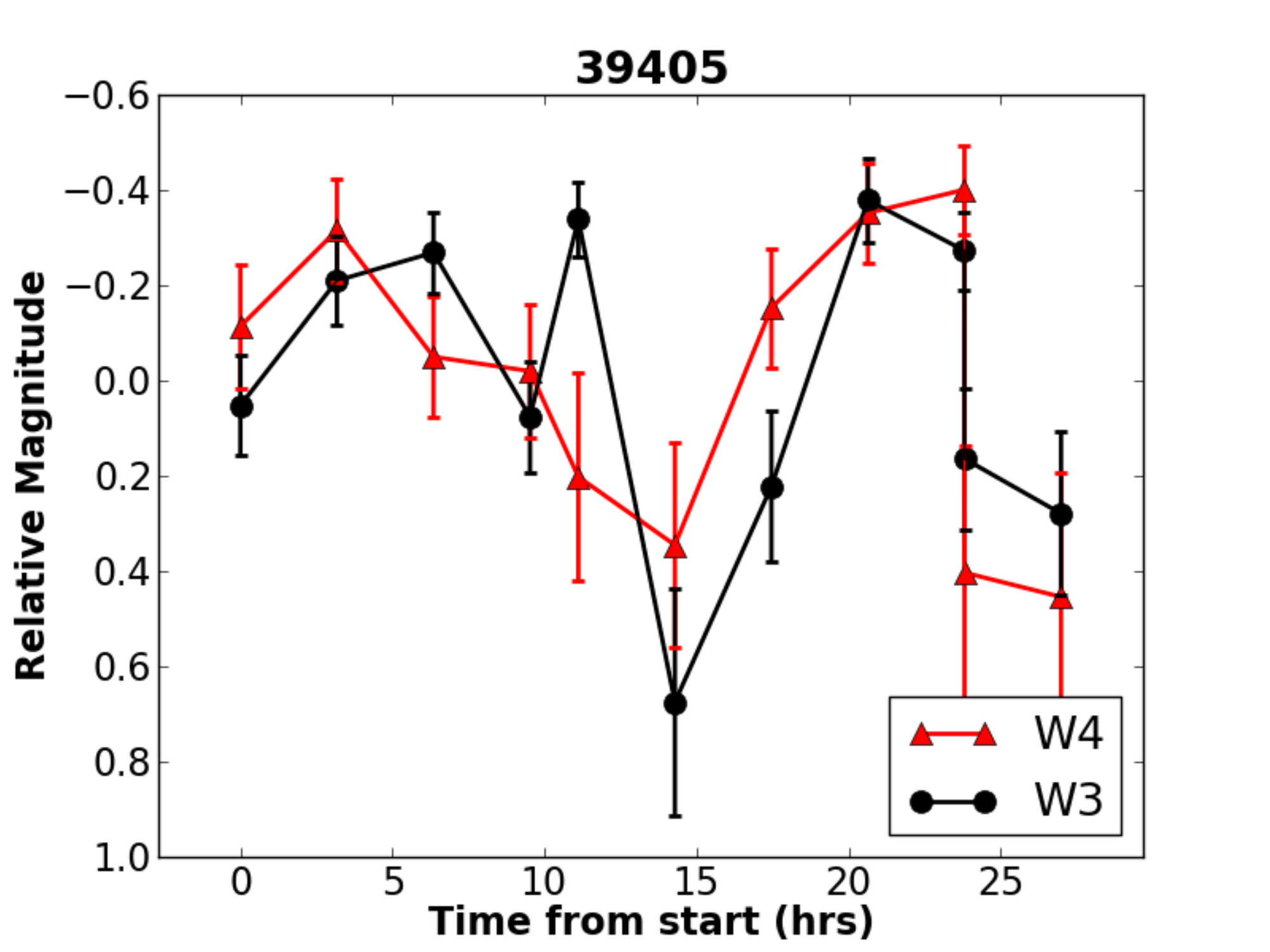}
\includegraphics[width=3.5in,height=2.6in]{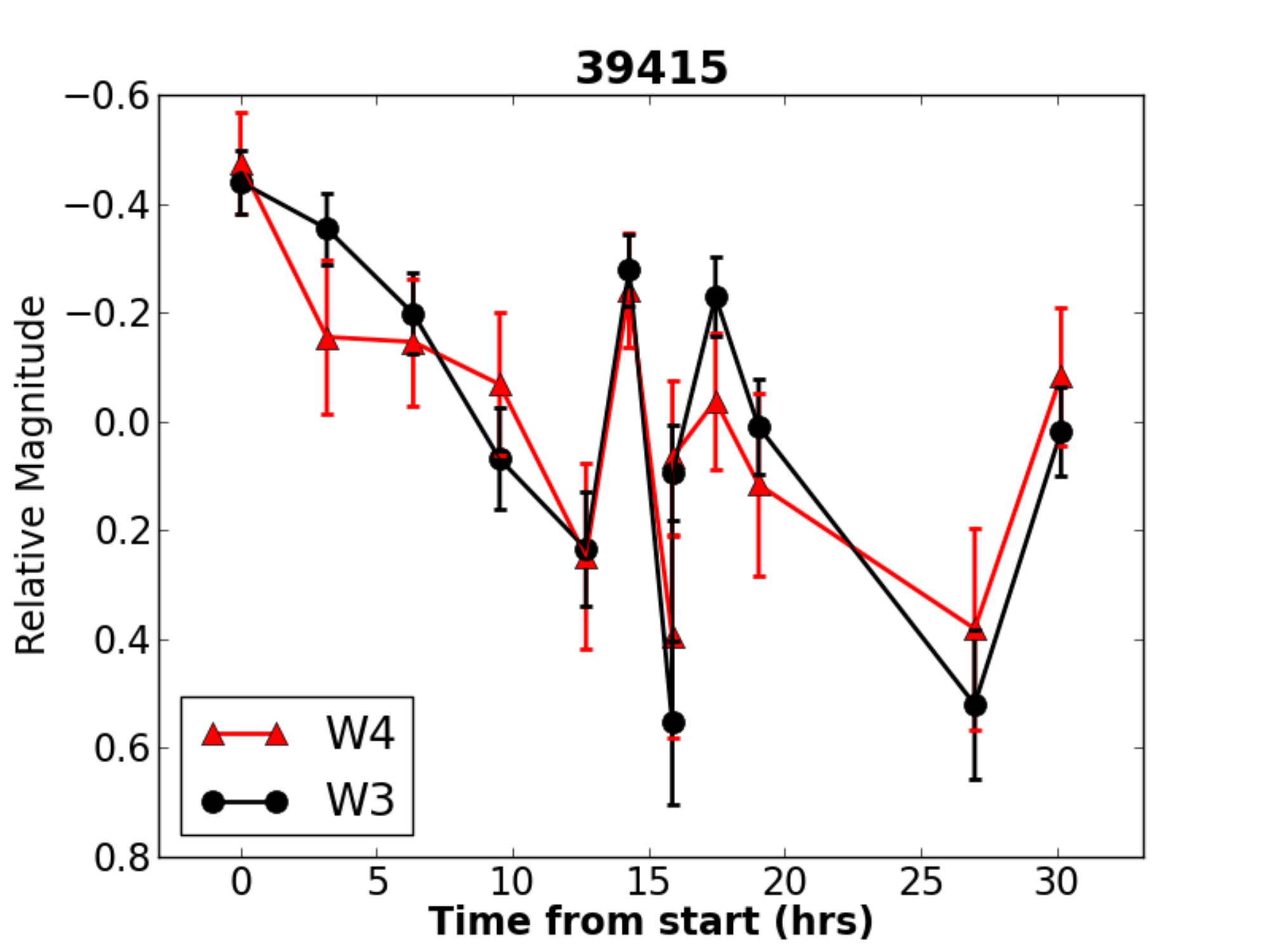}
\includegraphics[width=3.5in,height=2.6in]{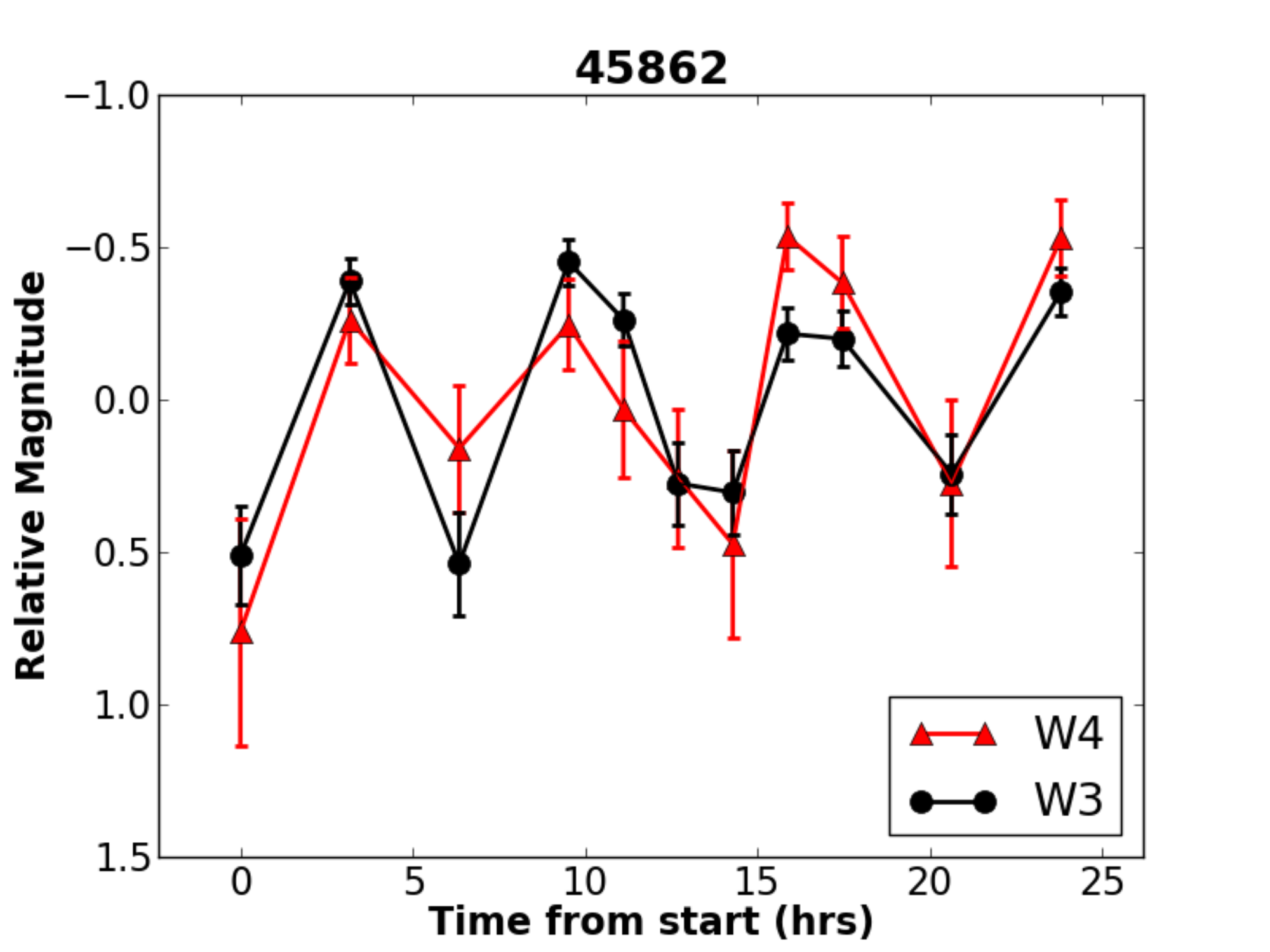}
\caption{\label{Fig:HildaCandidates} Candidate binary Hildas from our survey identified by their anomalously high lightcurve photometric ranges.}
\end{figure} 

\begin{figure}
\figurenum{4}
\includegraphics[width=3.5in,height=2.6in]{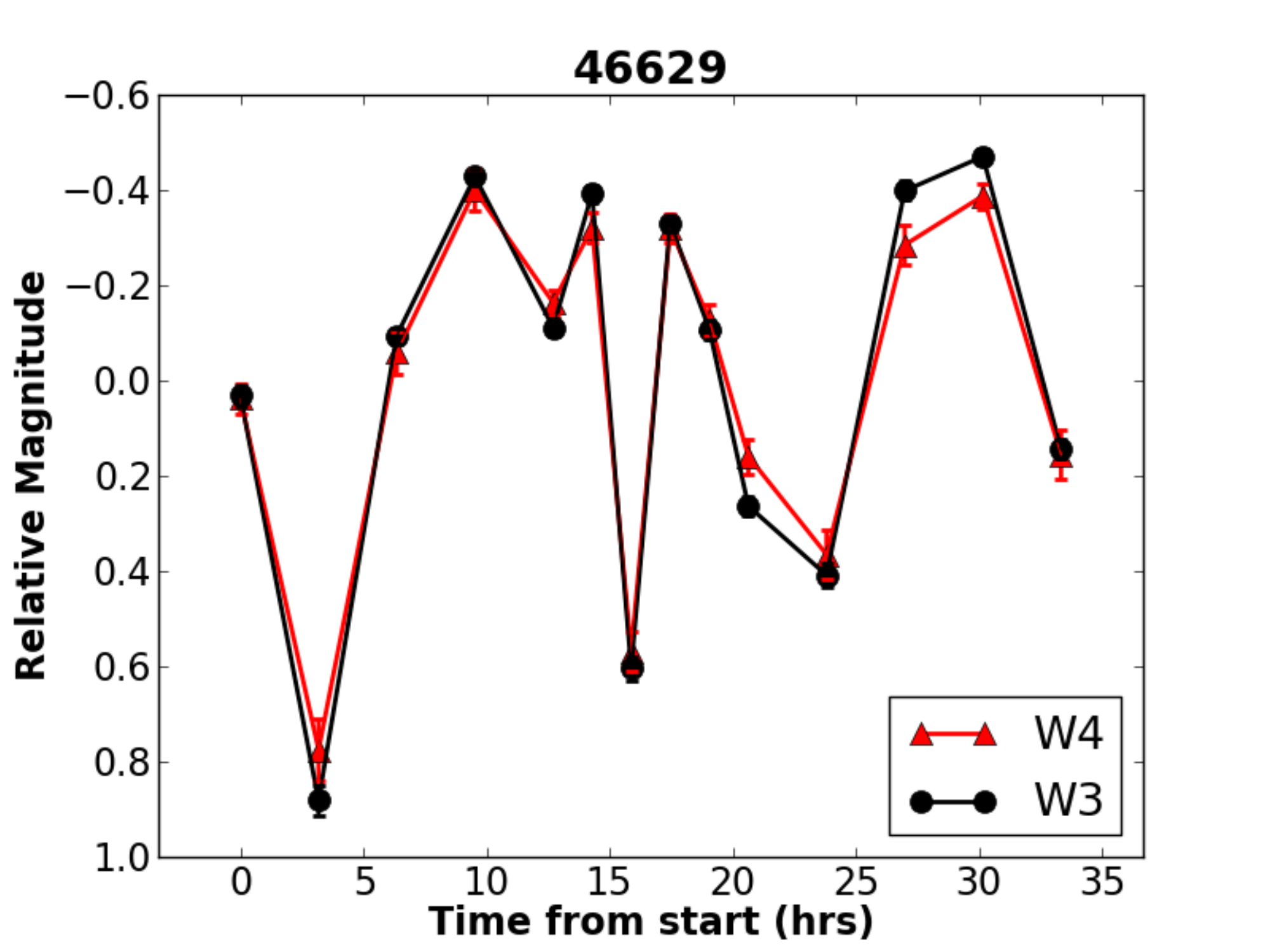}
\includegraphics[width=3.5in,height=2.6in]{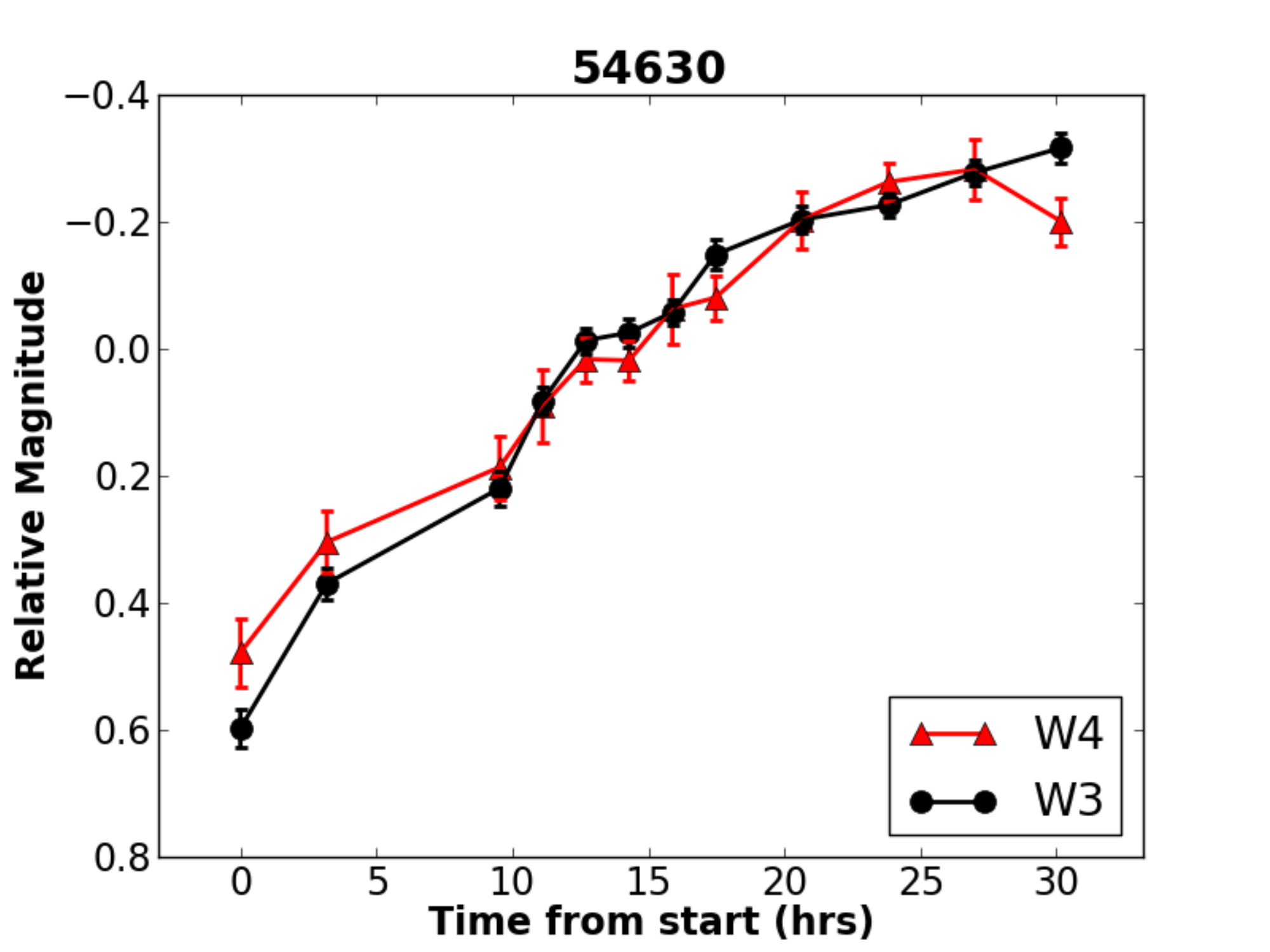}
\includegraphics[width=3.5in,height=2.6in]{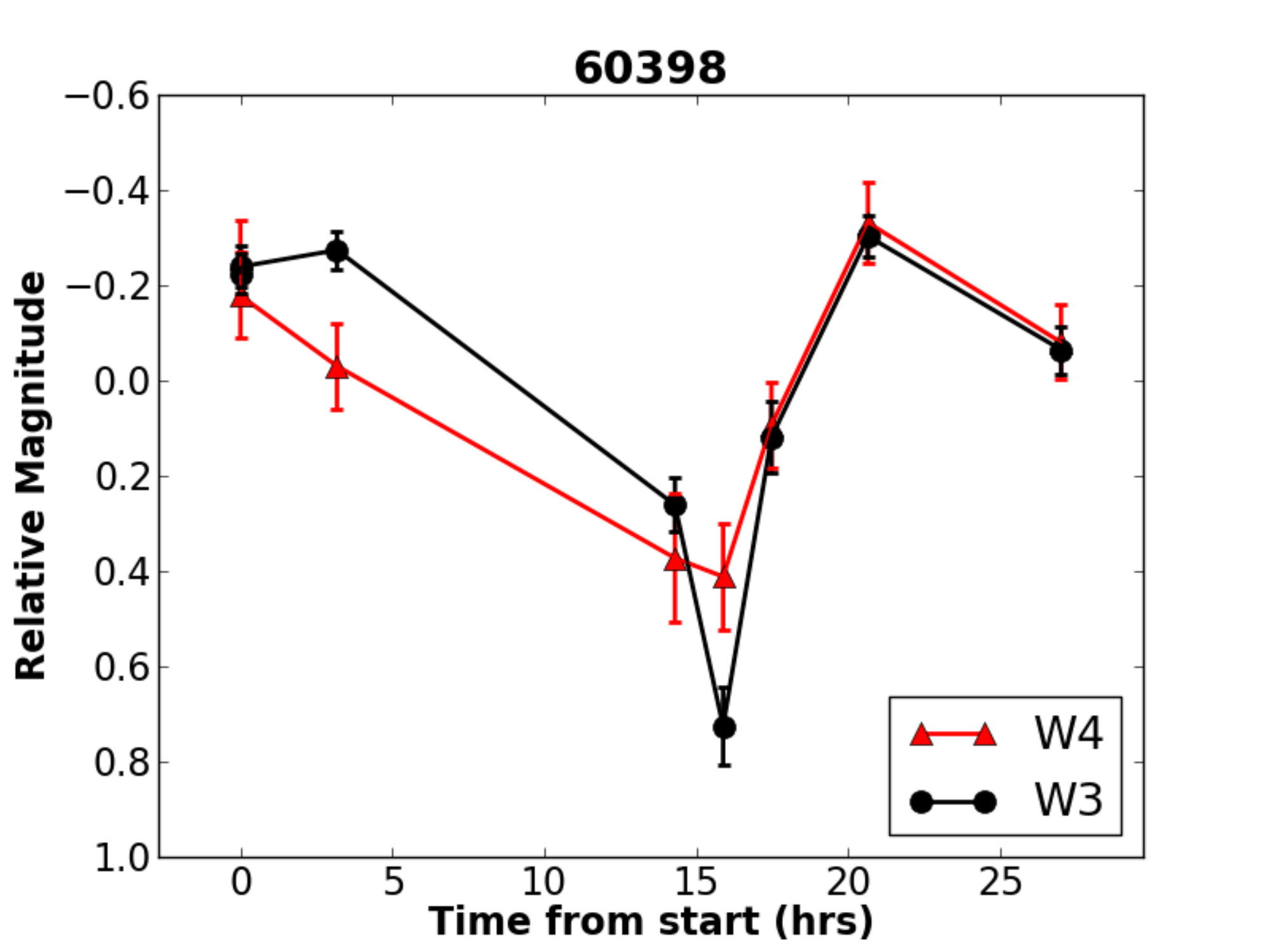}
\includegraphics[width=3.5in,height=2.6in]{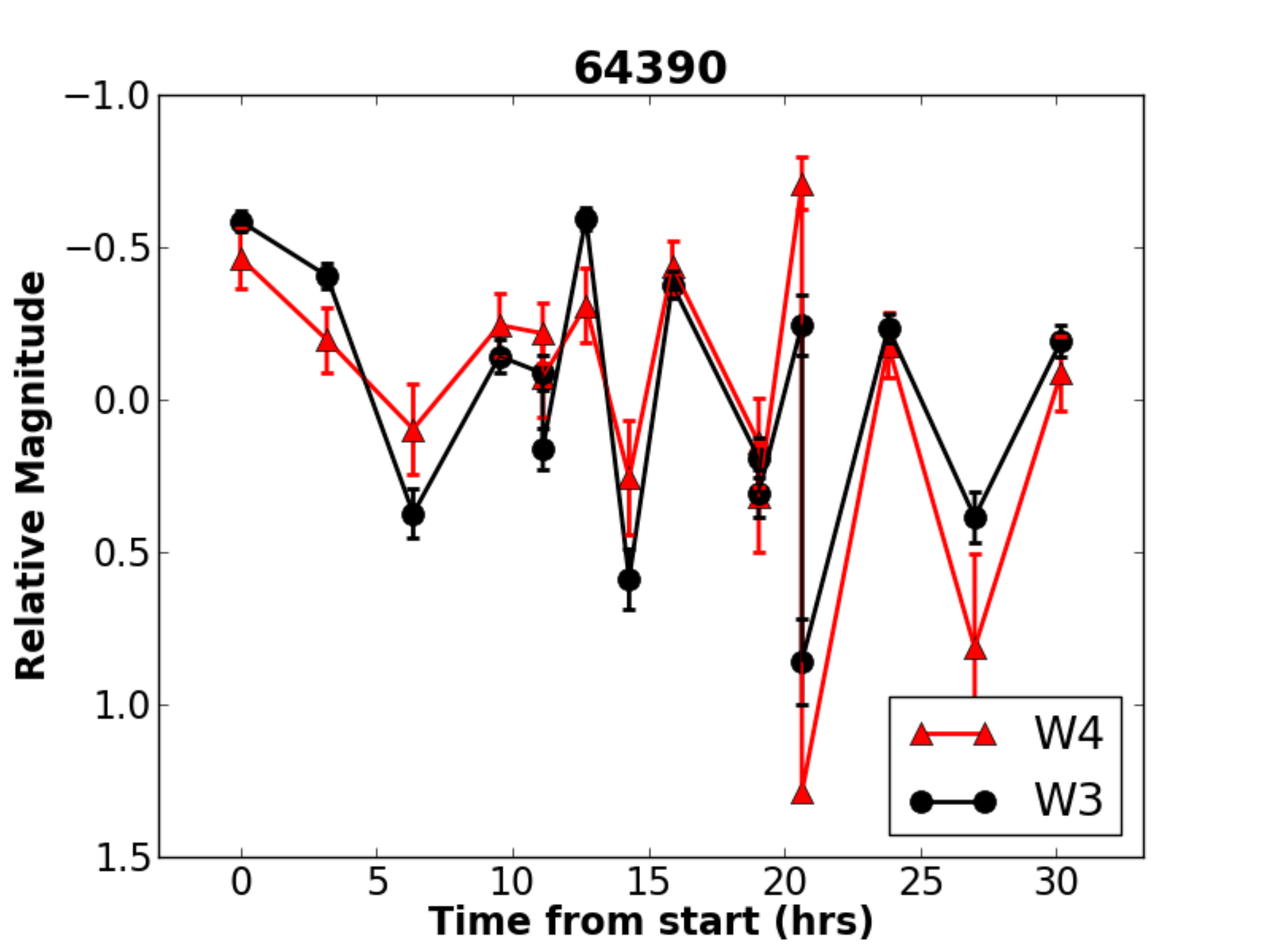}
\includegraphics[width=3.5in,height=2.6in]{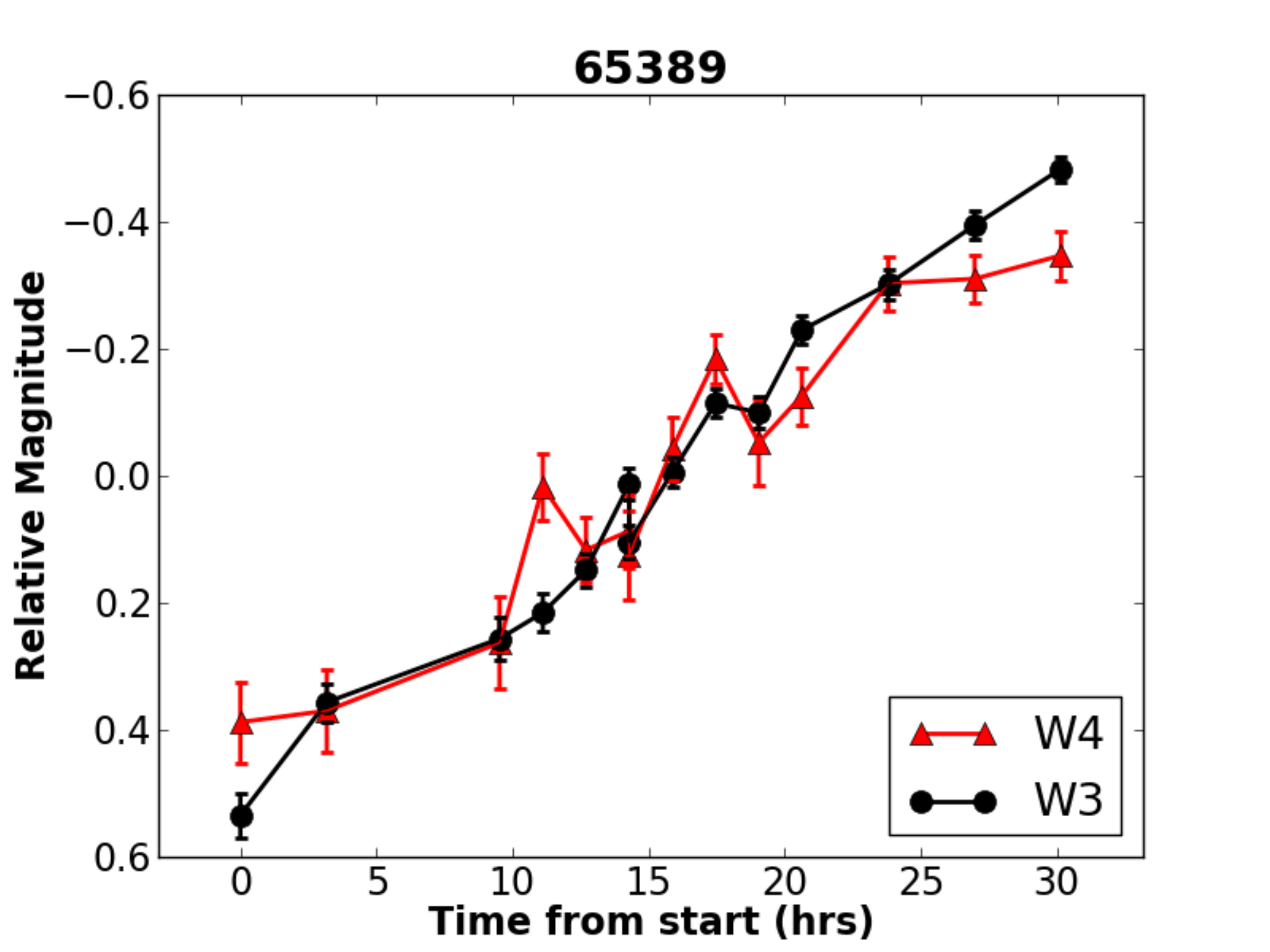}
\includegraphics[width=3.5in,height=2.6in]{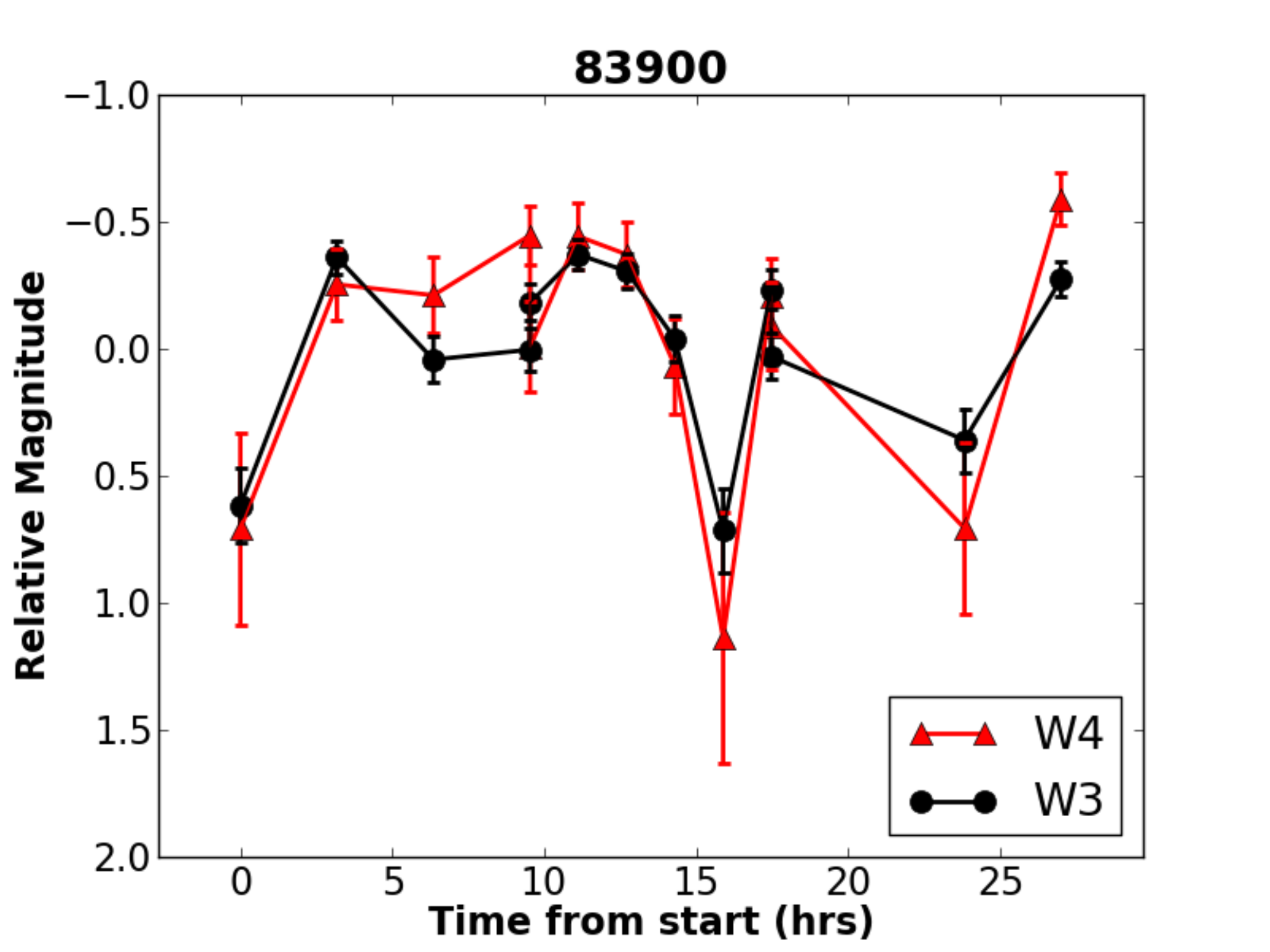}
\caption{\label{Fig:HildaCandidates} Candidate binary Hildas from our survey identified by their anomalously high lightcurve photometric ranges.}
\end{figure} 
\begin{figure}
\figurenum{4}
\includegraphics[width=3.5in,height=2.6in]{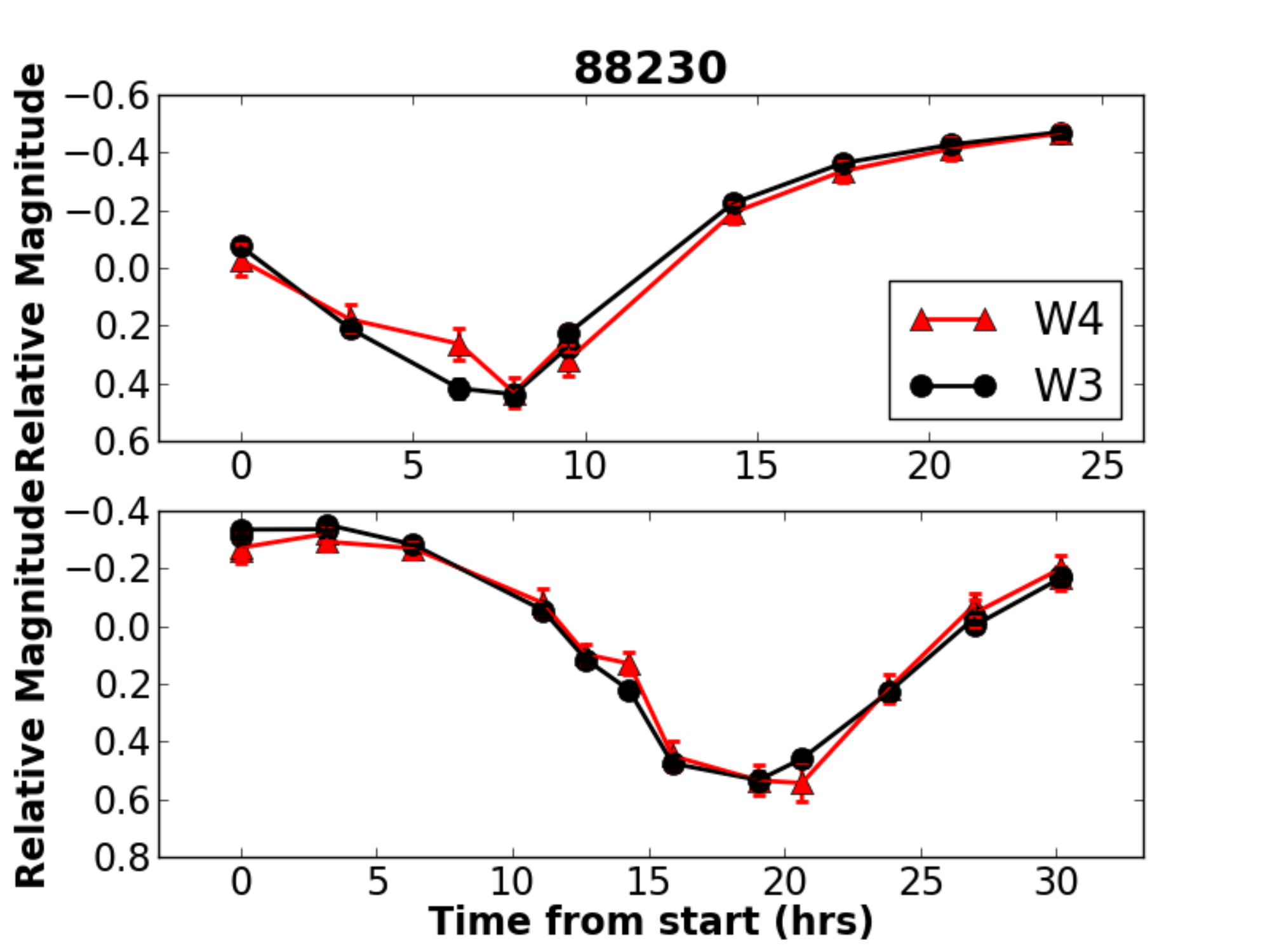}
\includegraphics[width=3.5in,height=2.6in]{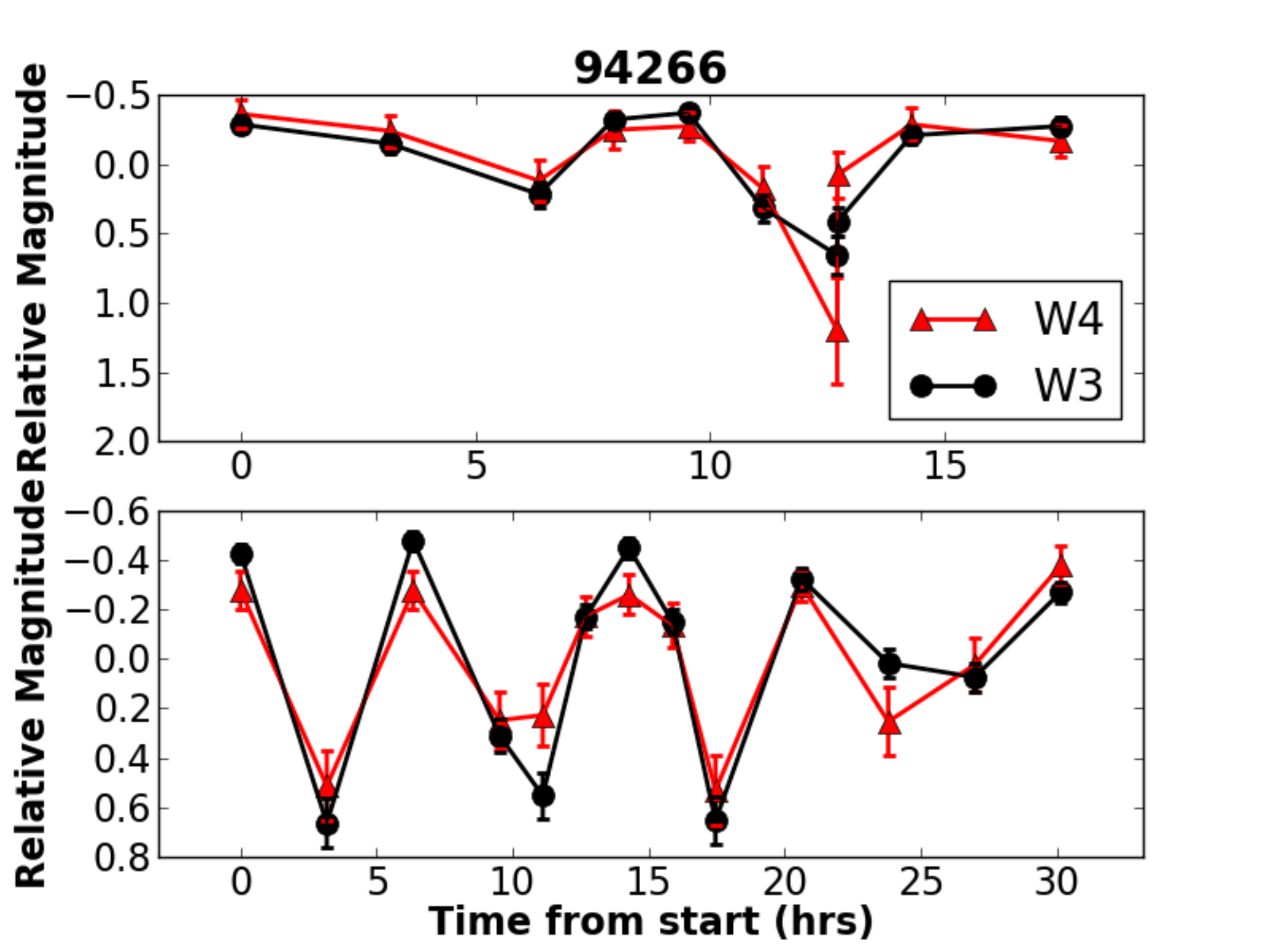}
\includegraphics[width=3.5in,height=2.6in]{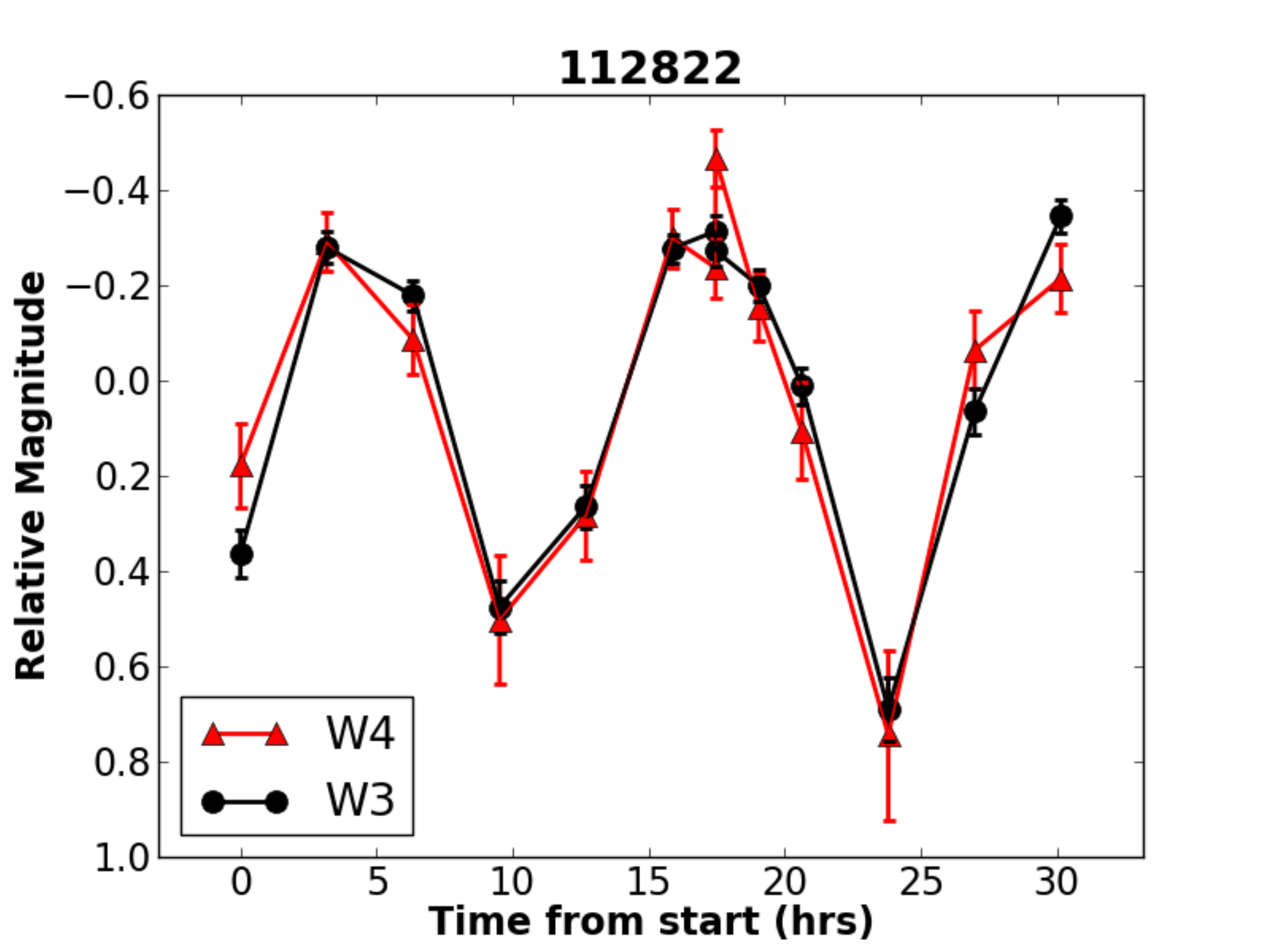}
\includegraphics[width=3.5in,height=2.6in]{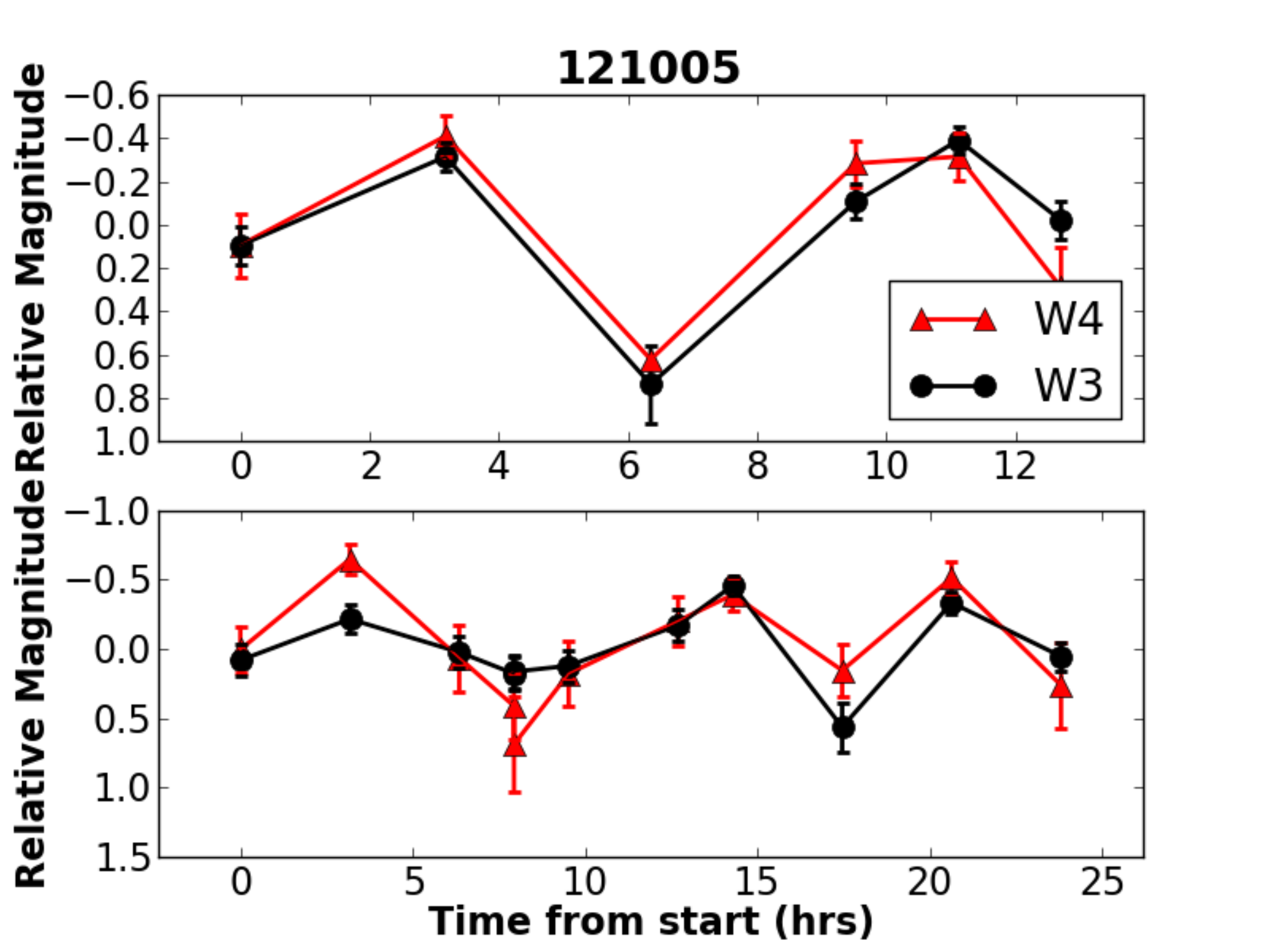}
\includegraphics[width=3.5in,height=2.6in]{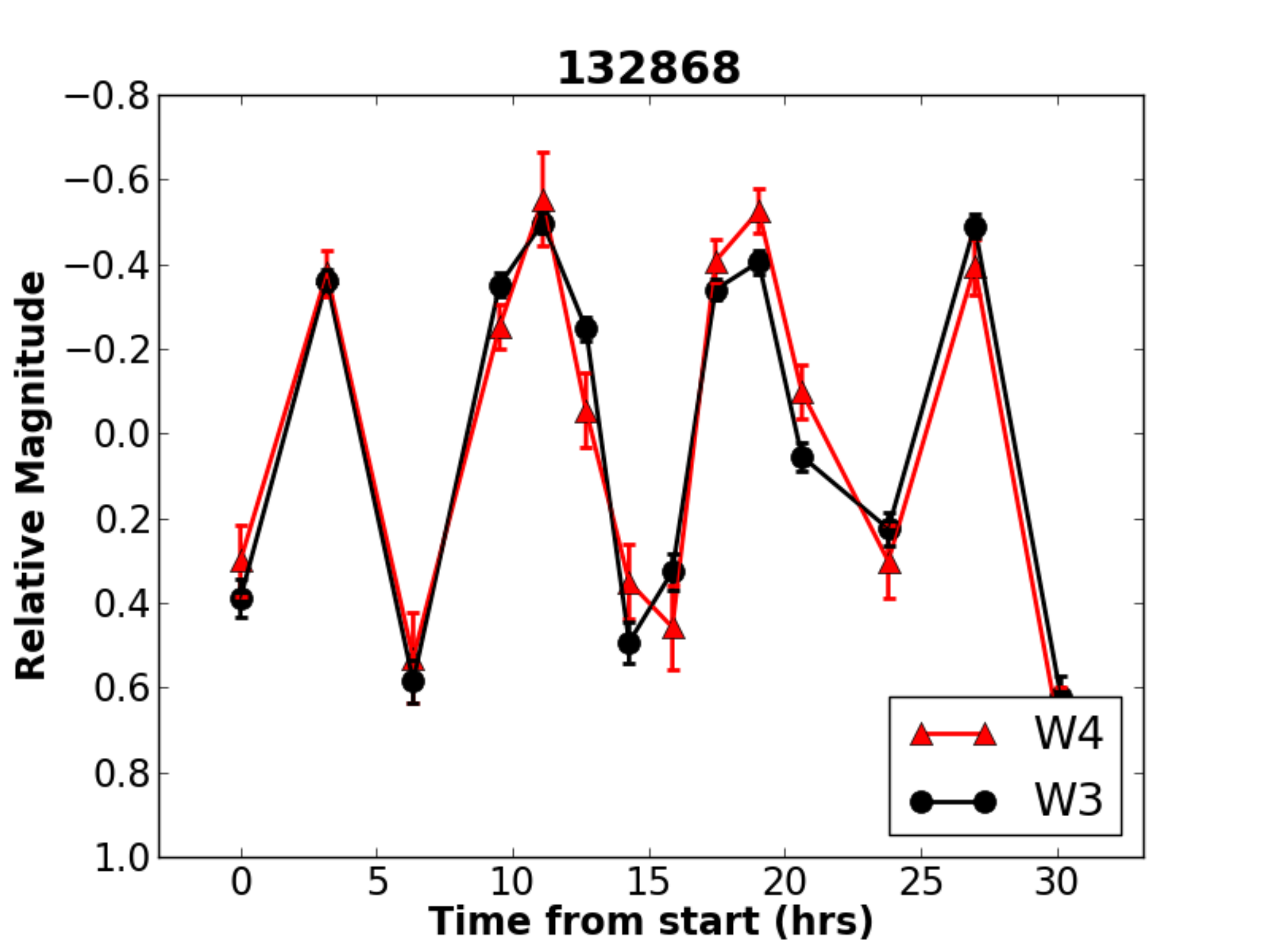}
\includegraphics[width=3.5in,height=2.6in]{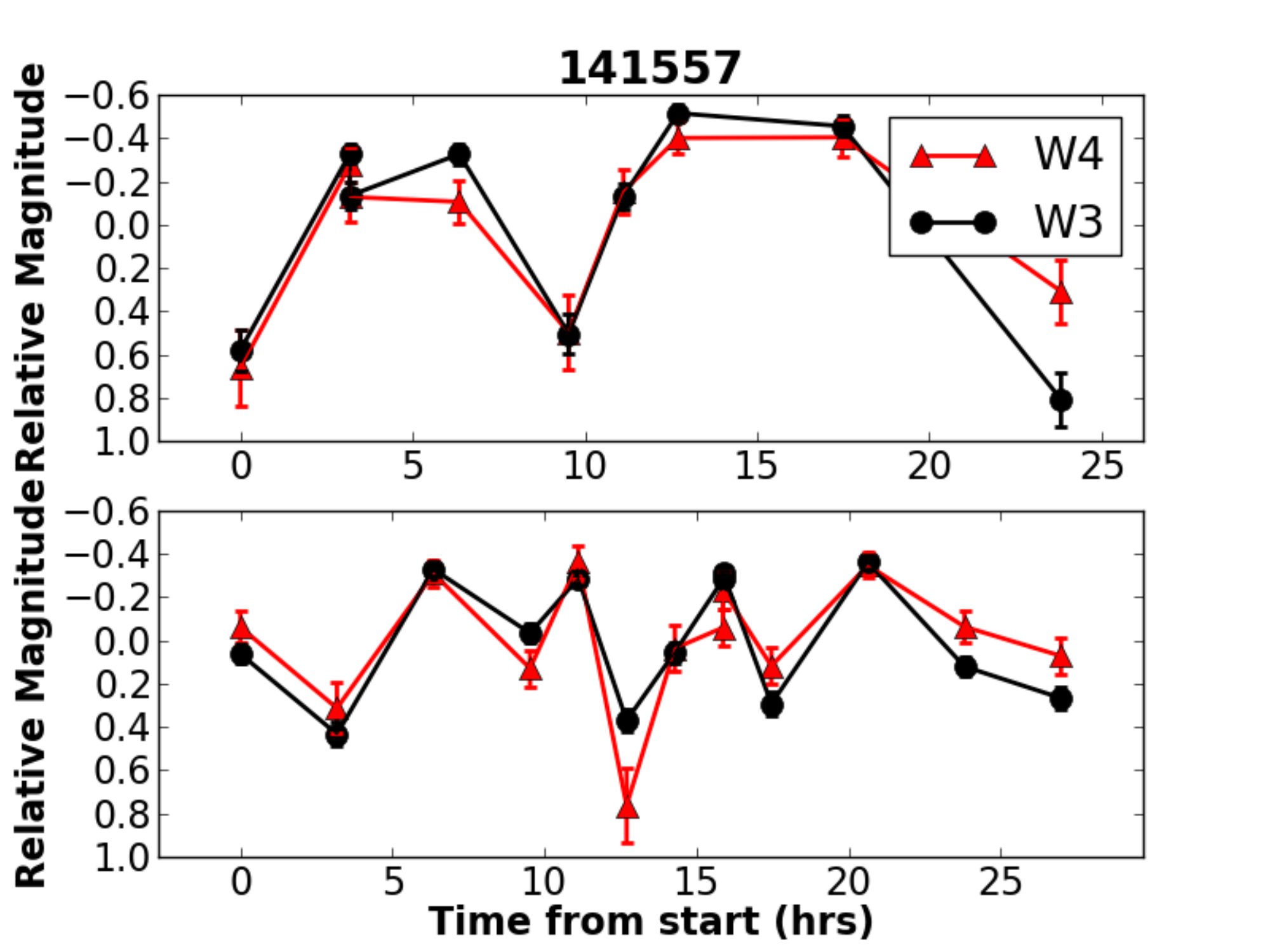}
\caption{\label{Fig:HildaCandidates} Candidate binary Hildas from our survey identified by their anomalously high lightcurve photometric ranges.}
\end{figure} 

\begin{figure}
\figurenum{4}
\includegraphics[width=3.5in,height=2.6in]{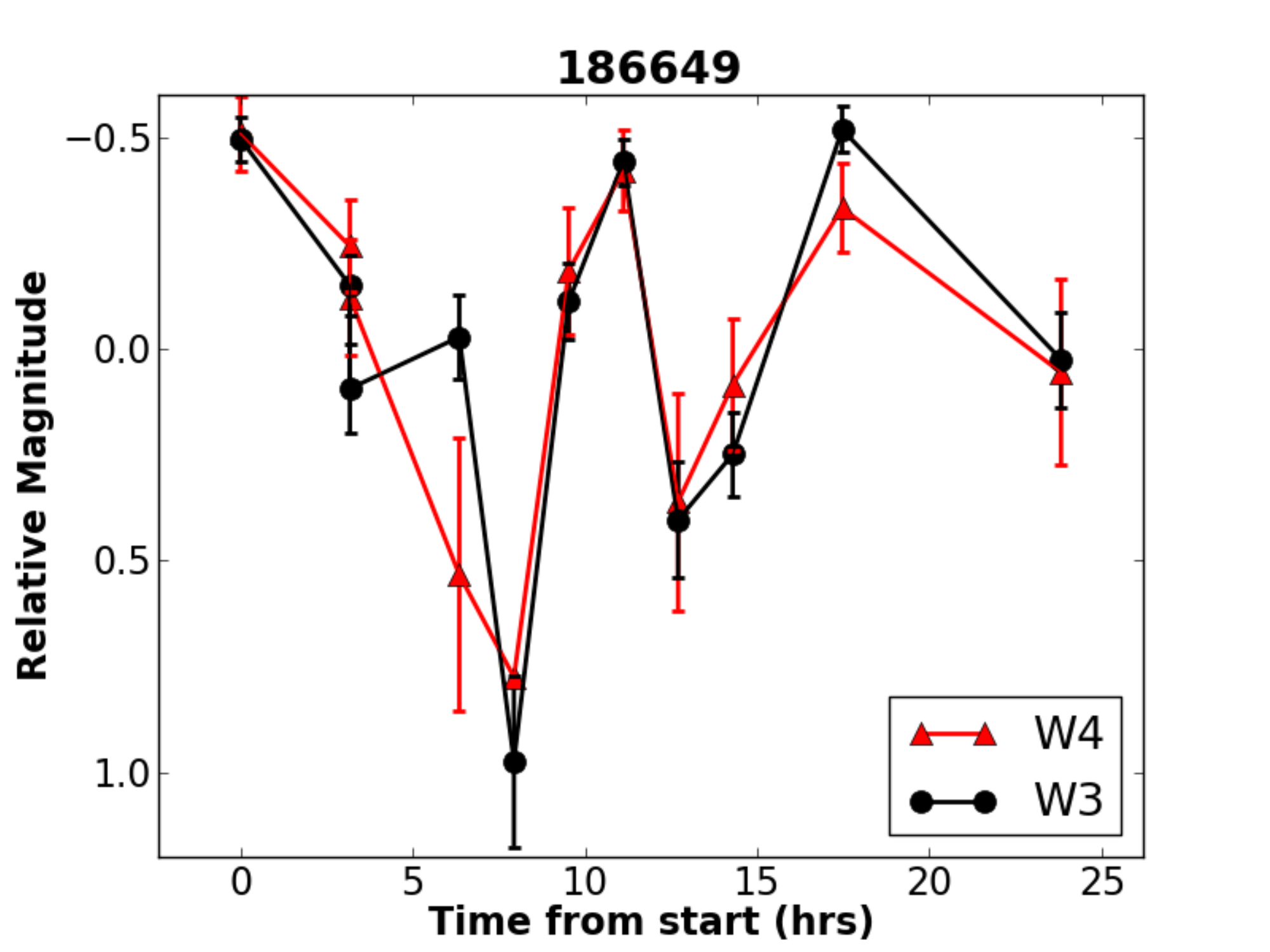}
\includegraphics[width=3.5in,height=2.6in]{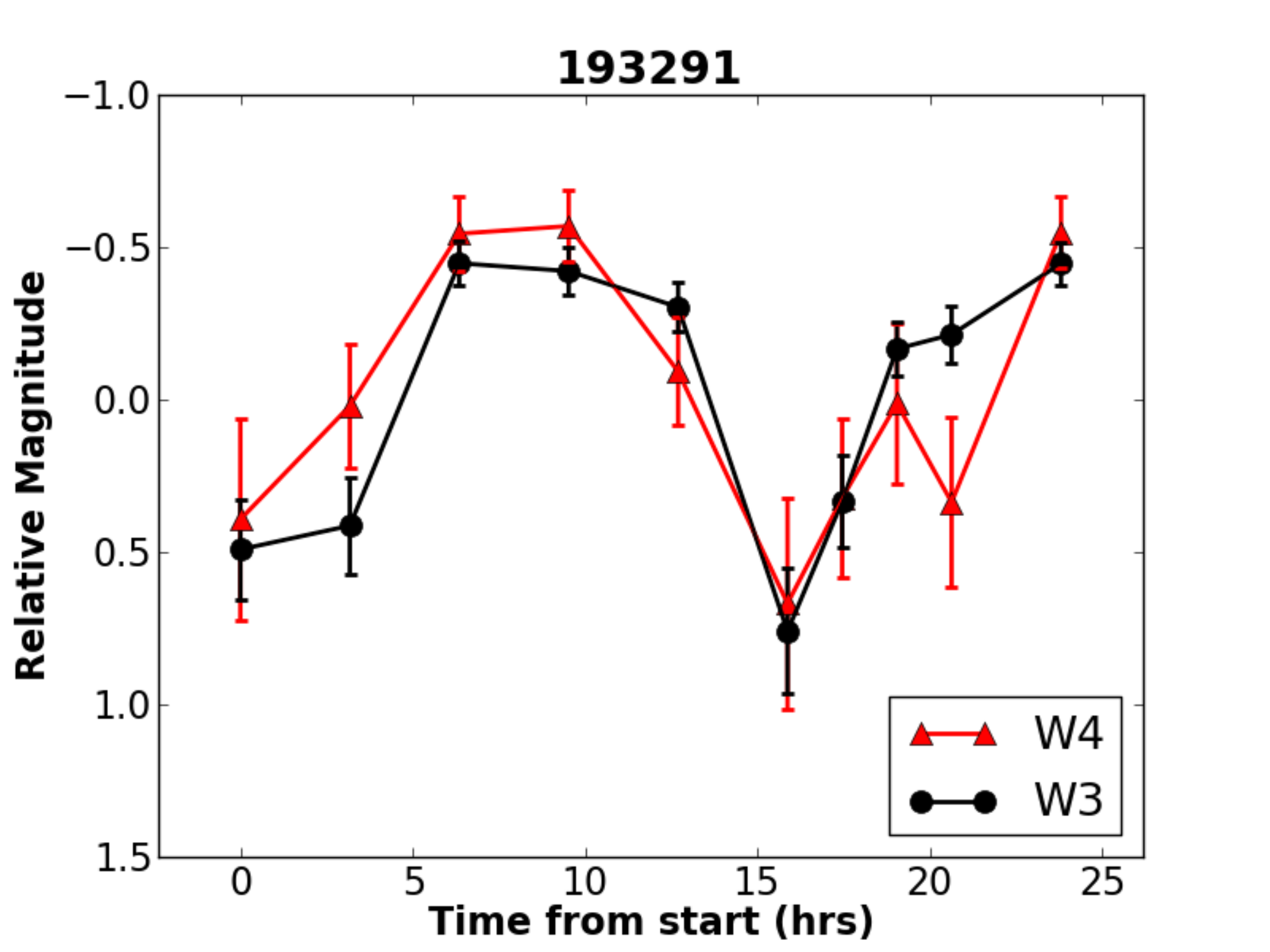}
\includegraphics[width=3.5in,height=2.6in]{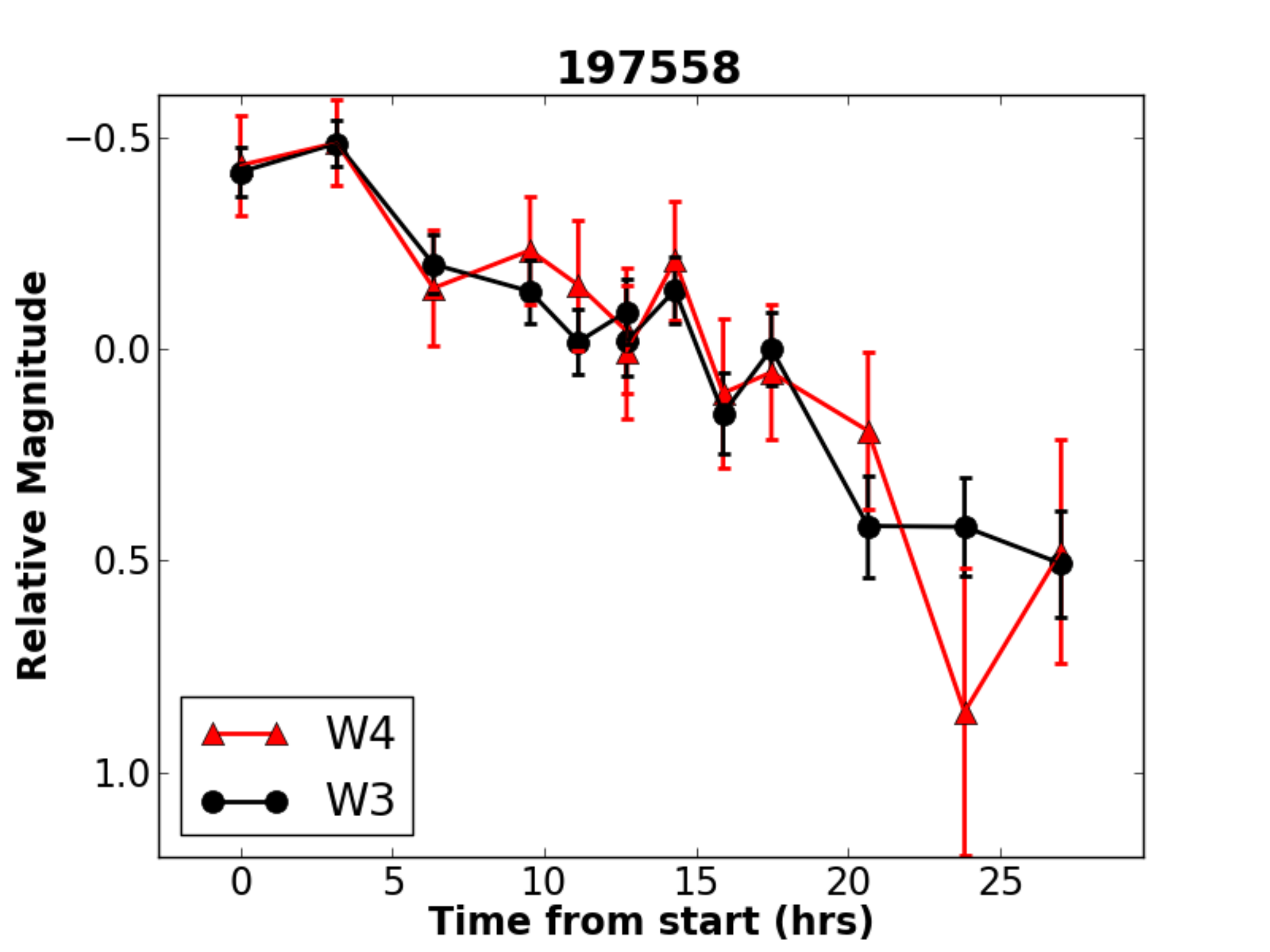}
\includegraphics[width=3.5in,height=2.6in]{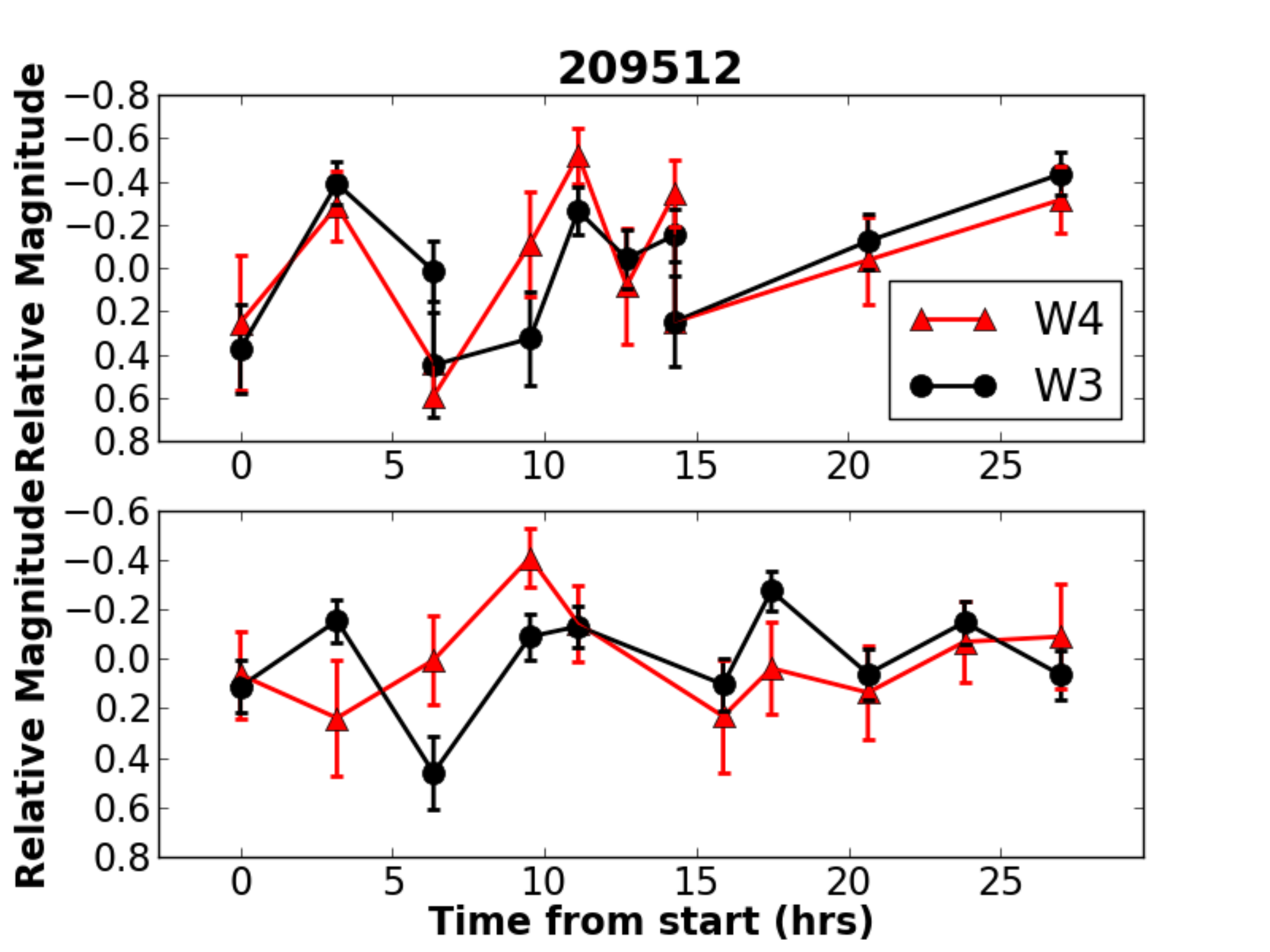}
\includegraphics[width=3.5in,height=2.6in]{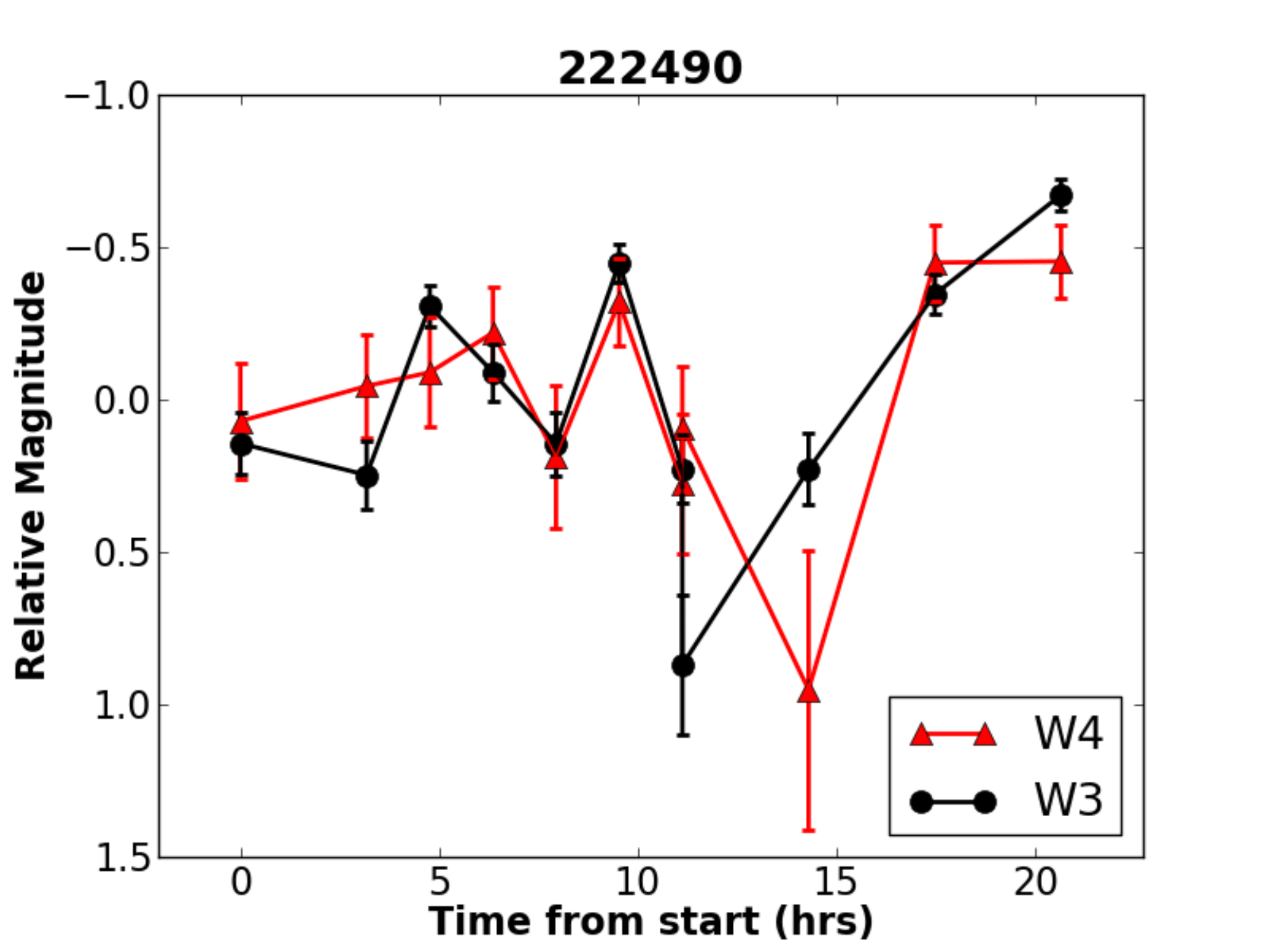}
\includegraphics[width=3.5in,height=2.6in]{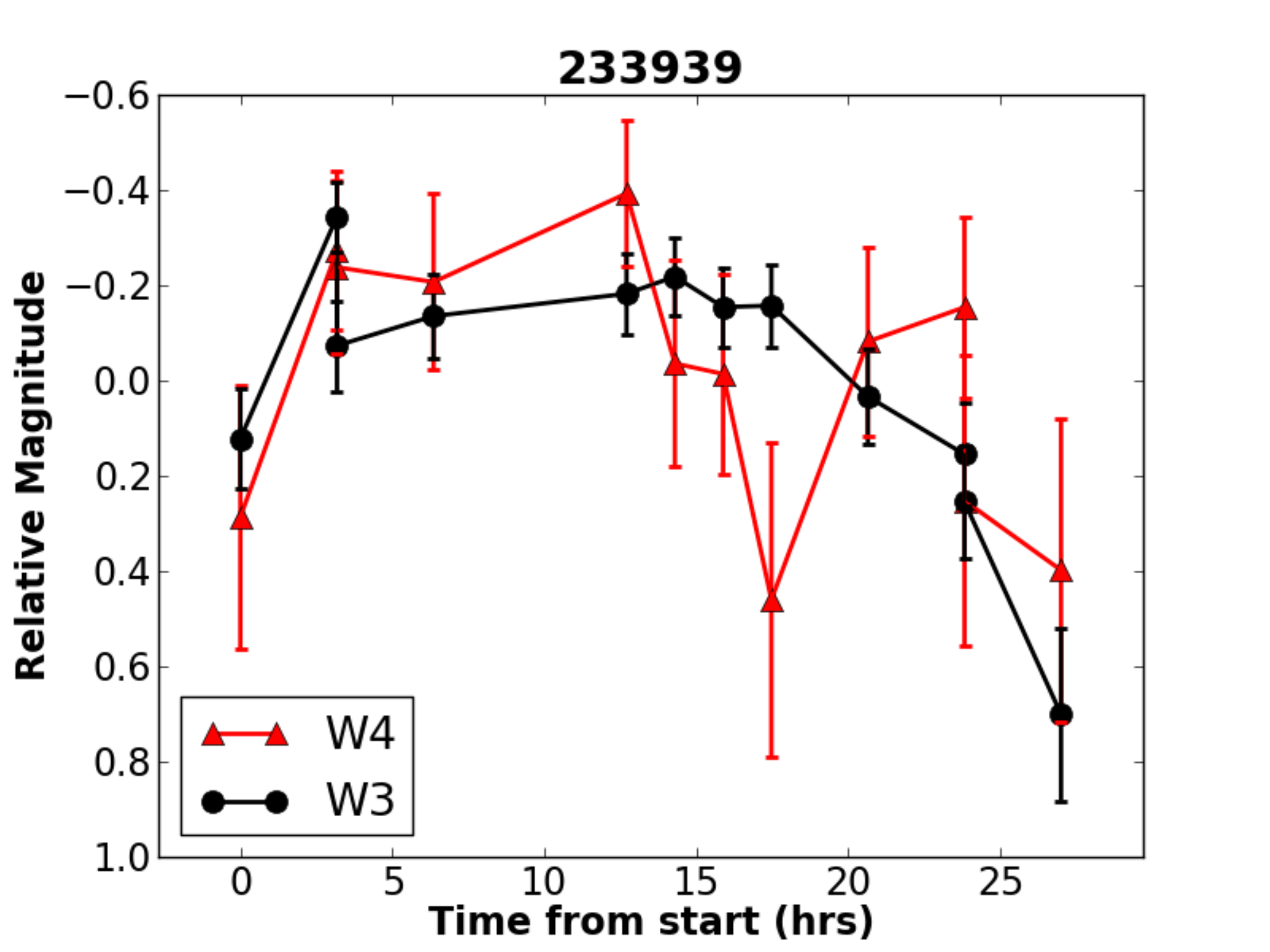}
\caption{\label{Fig:HildaCandidates} Candidate binary Hildas from our survey identified by their anomalously high lightcurve photometric ranges.}
\end{figure} 

\begin{figure}
\figurenum{4}
\includegraphics[width=3.5in,height=2.6in]{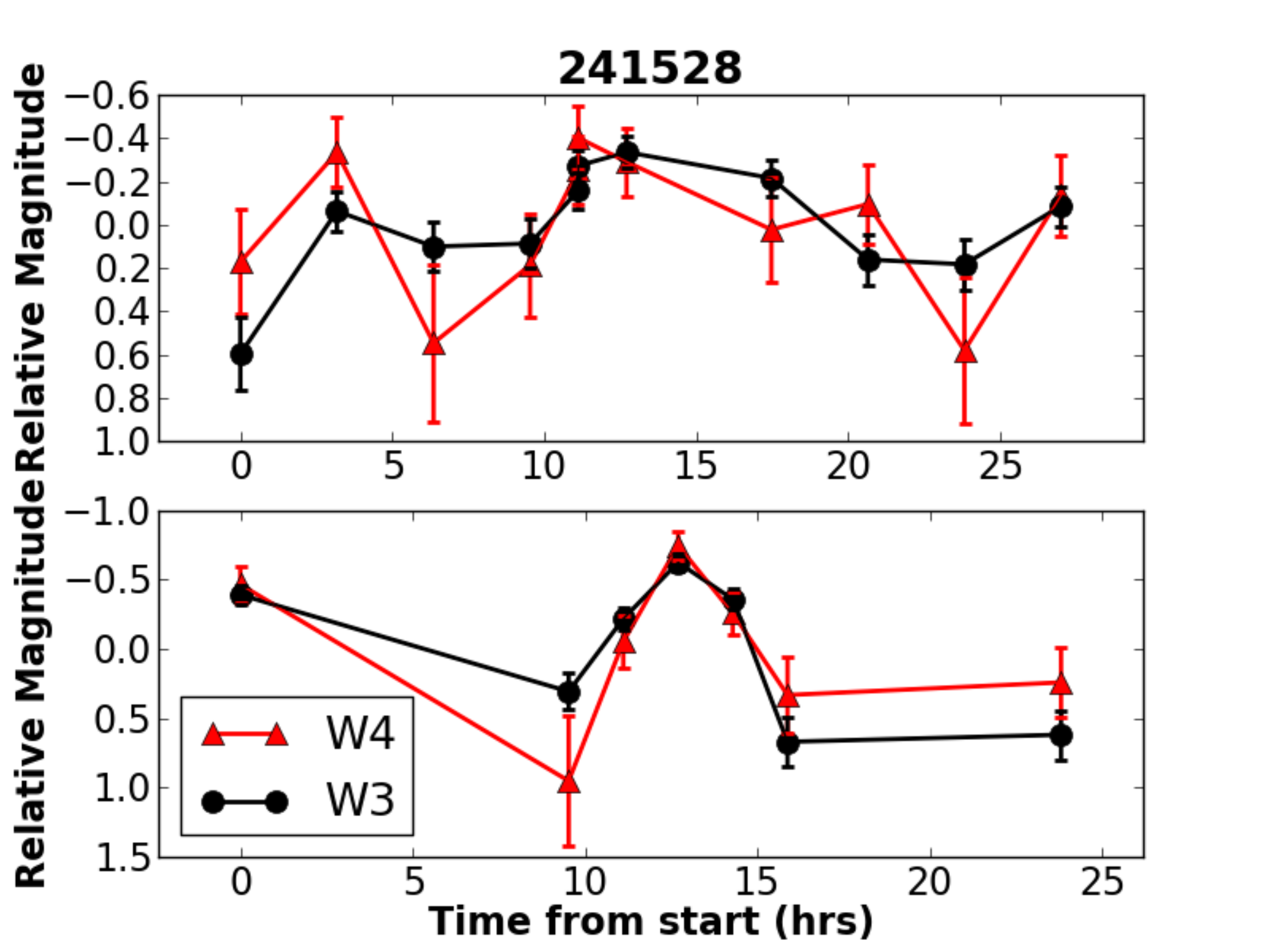}
\includegraphics[width=3.5in,height=2.6in]{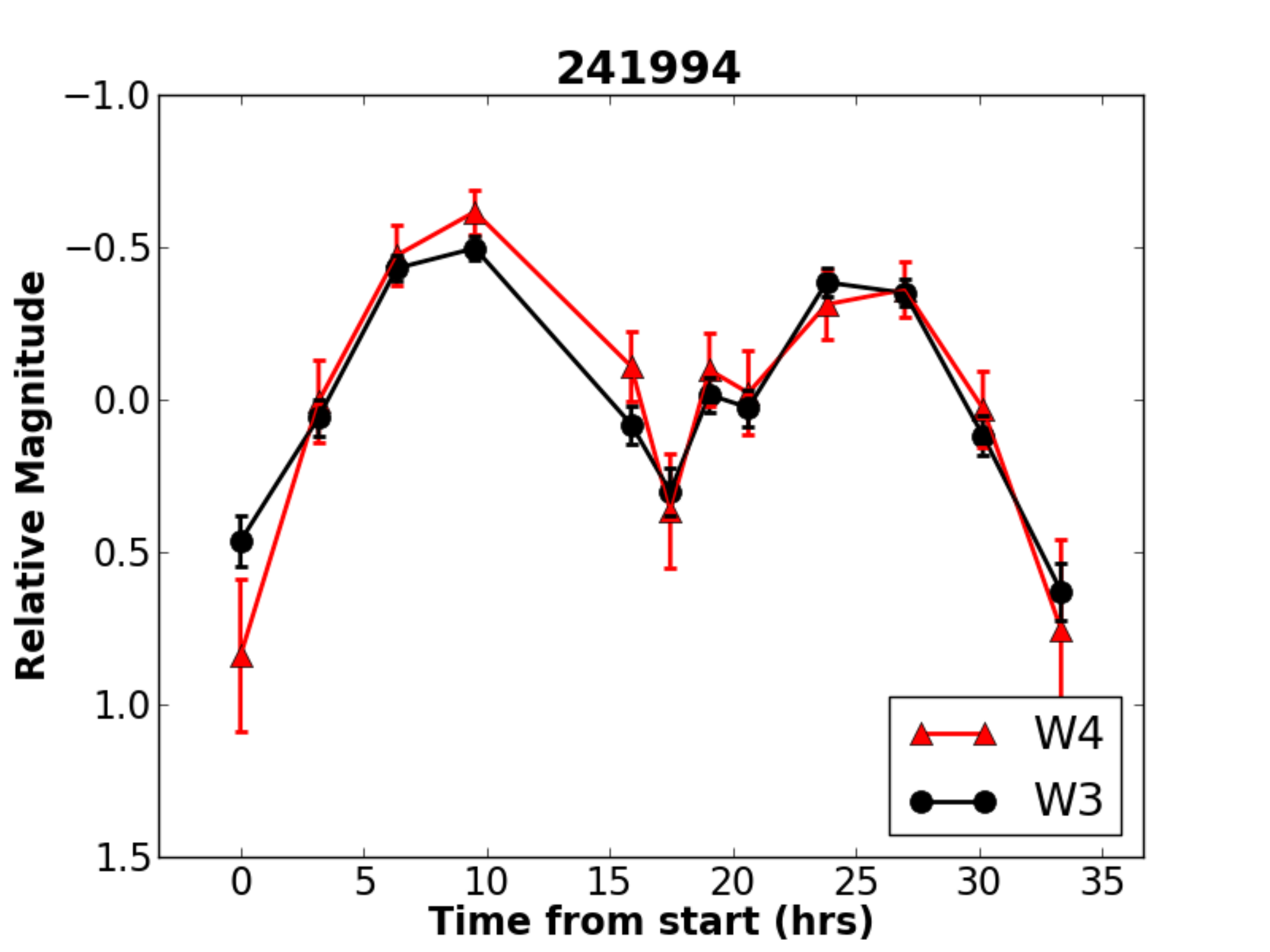}
\includegraphics[width=3.5in,height=2.6in]{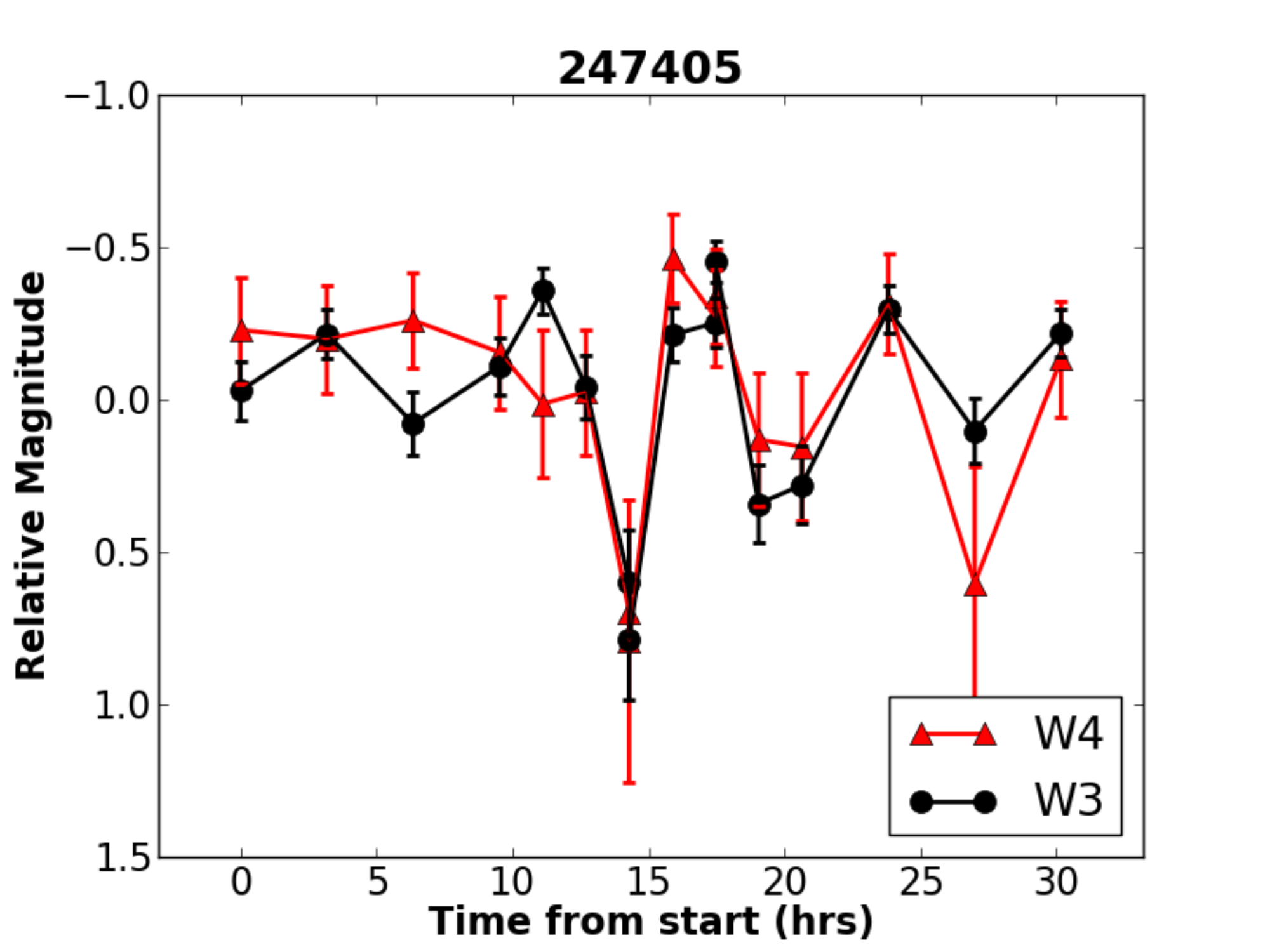}
\includegraphics[width=3.5in,height=2.6in]{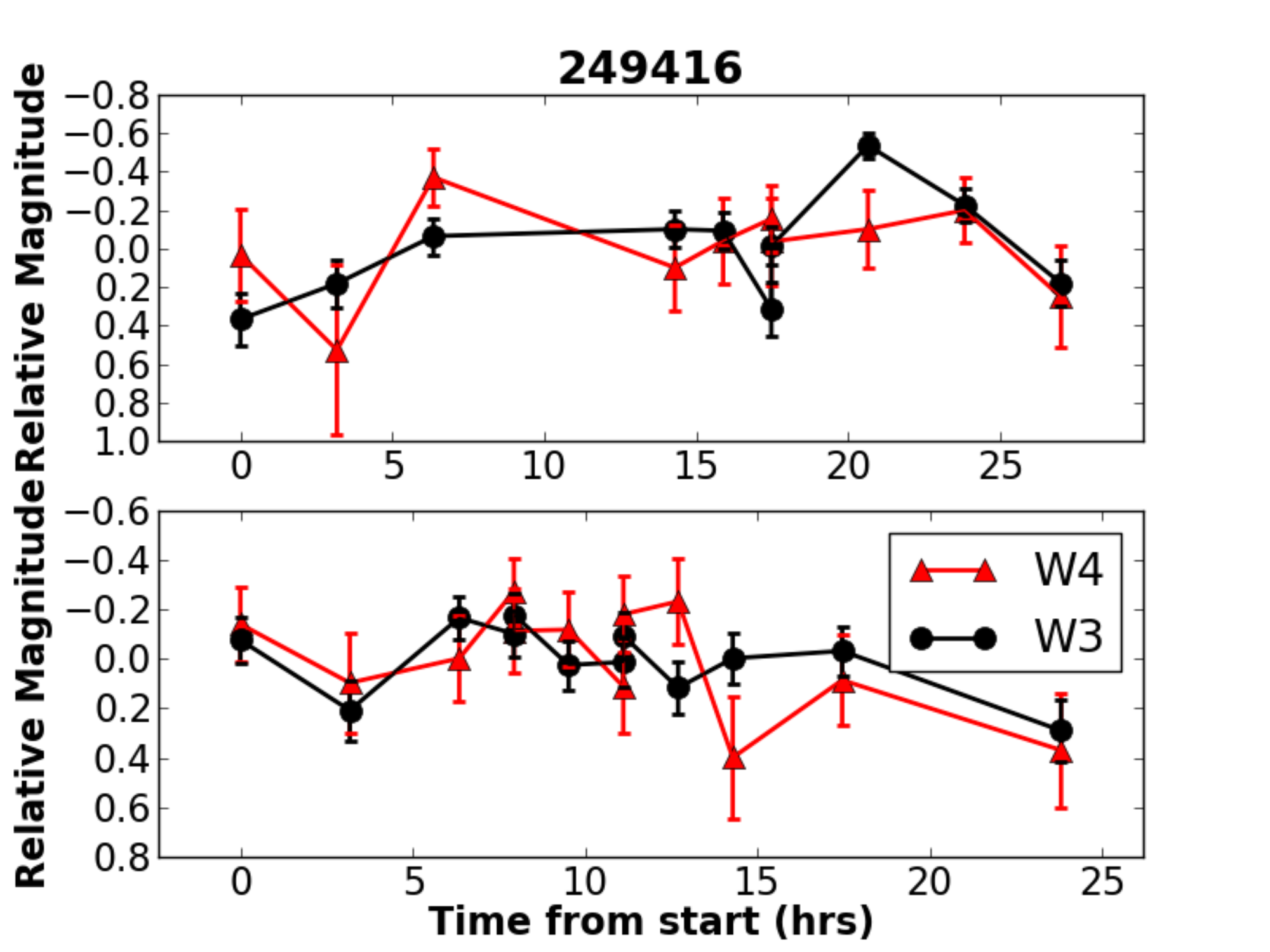}
\includegraphics[width=3.5in,height=2.6in]{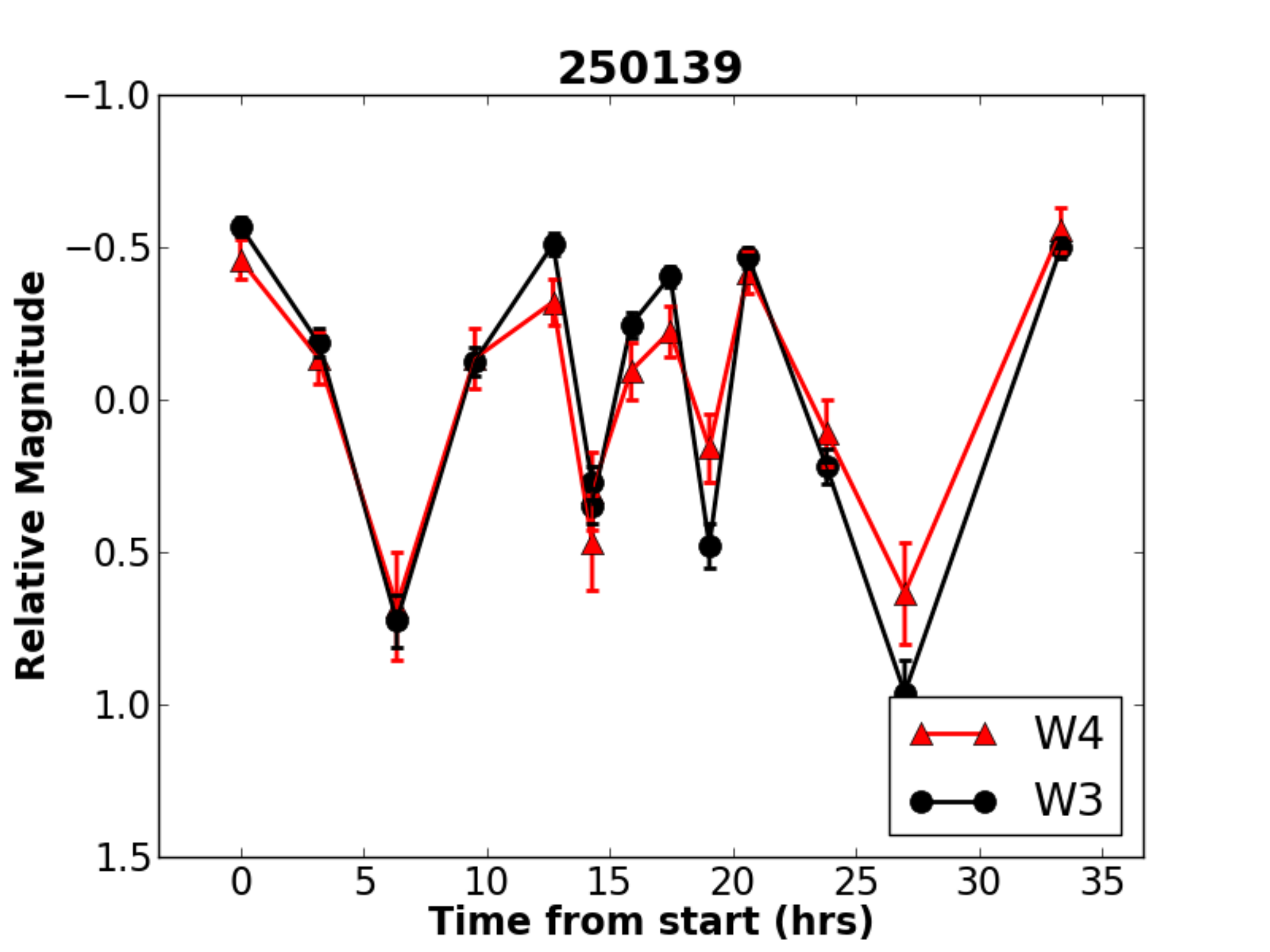}
\includegraphics[width=3.5in,height=2.6in]{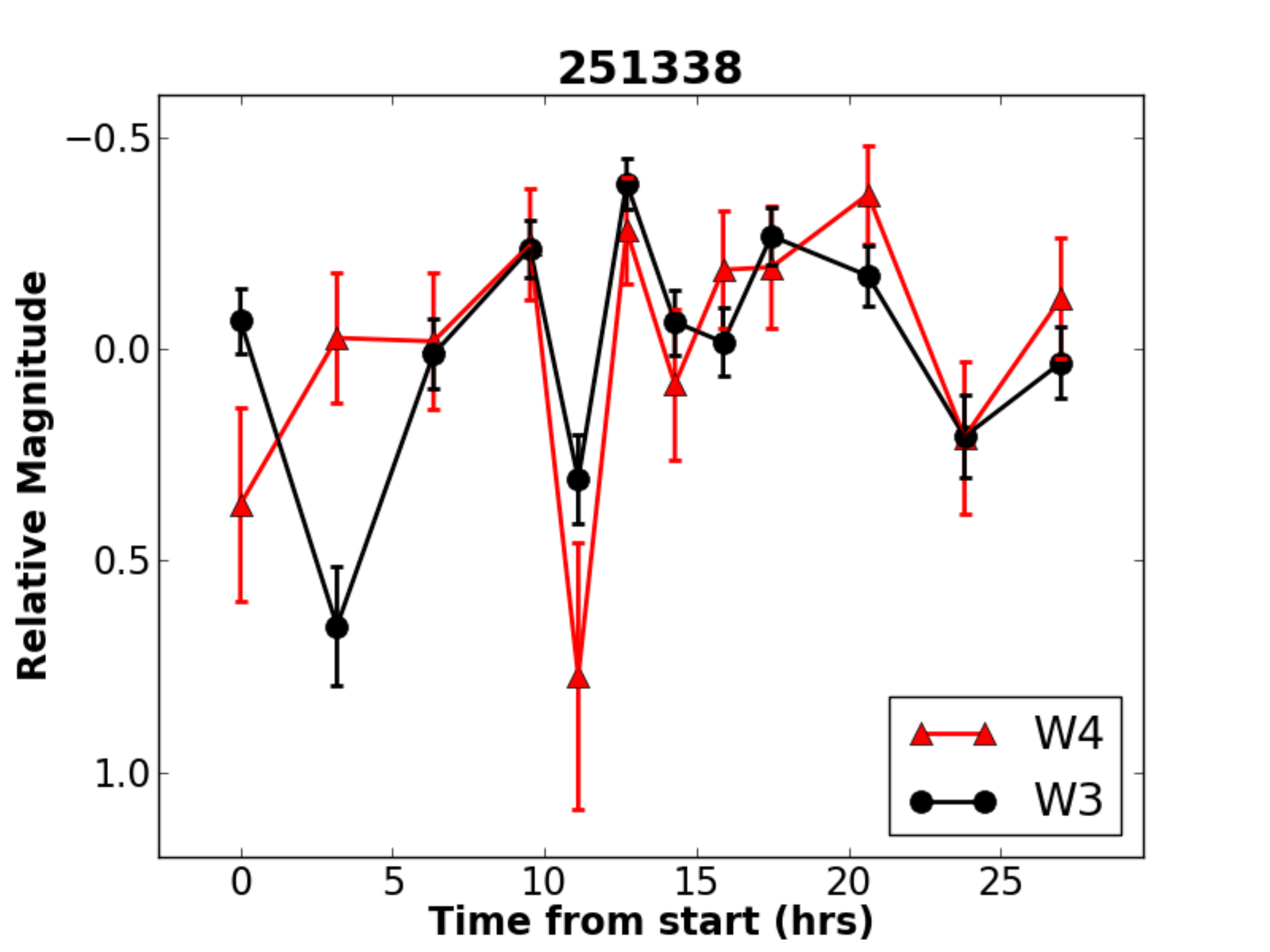}
\caption{\label{Fig:HildaCandidates} Candidate binary Hildas from our survey identified by their anomalously high lightcurve photometric ranges.}
\end{figure} 

\begin{figure}
\figurenum{4}
\includegraphics[width=3.5in,height=2.6in]{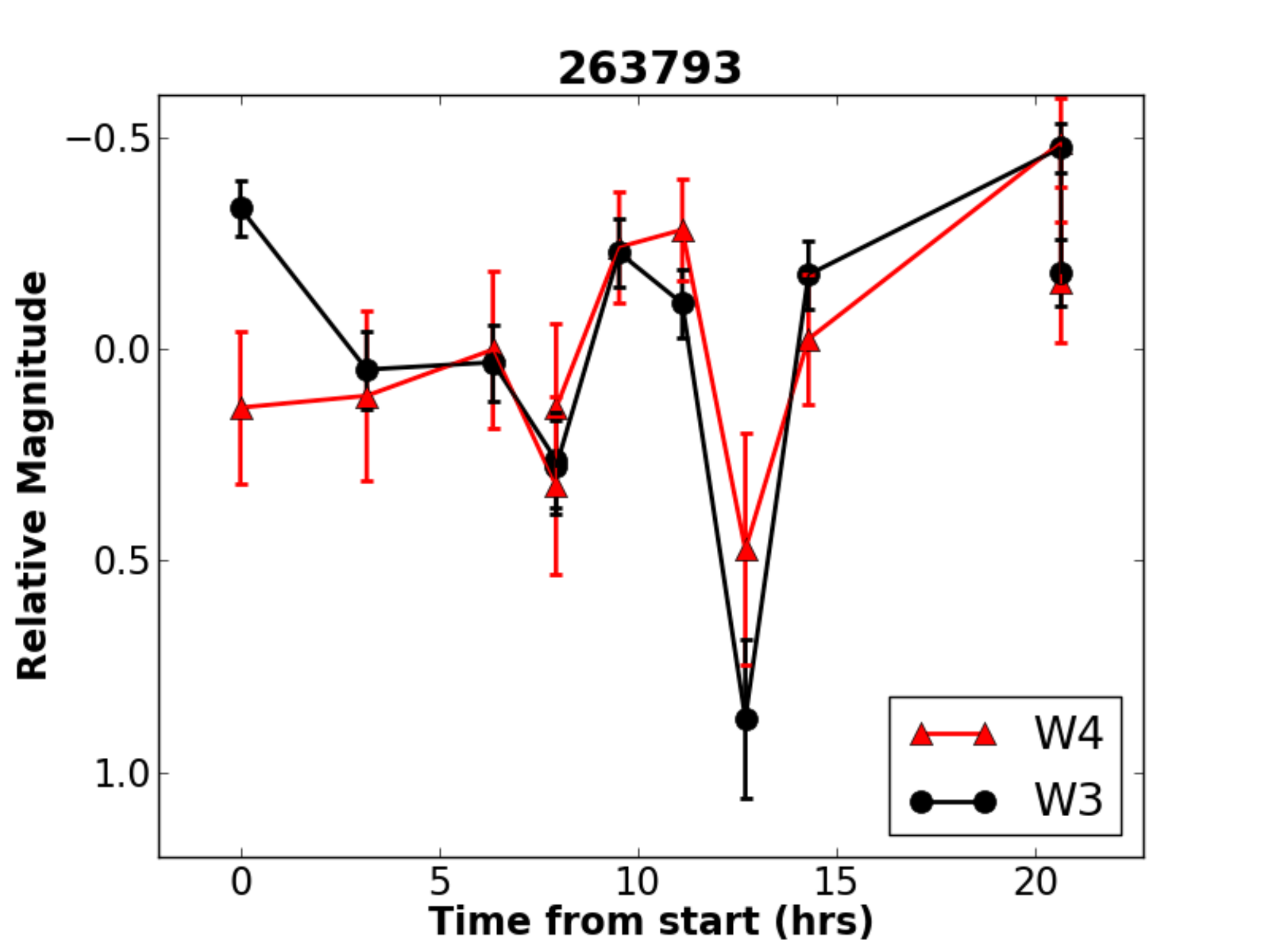}
\includegraphics[width=3.5in,height=2.6in]{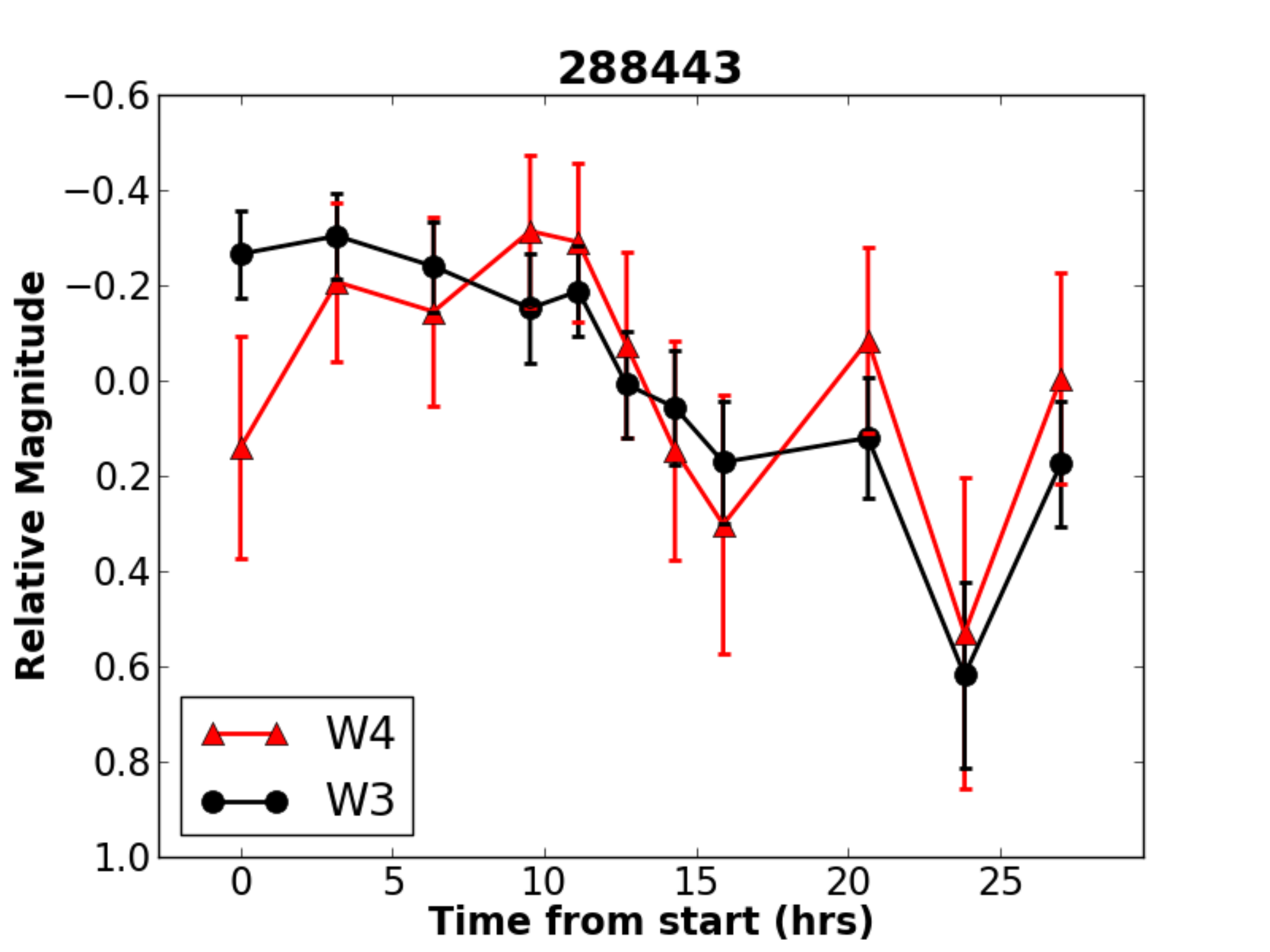}
\includegraphics[width=3.5in,height=2.6in]{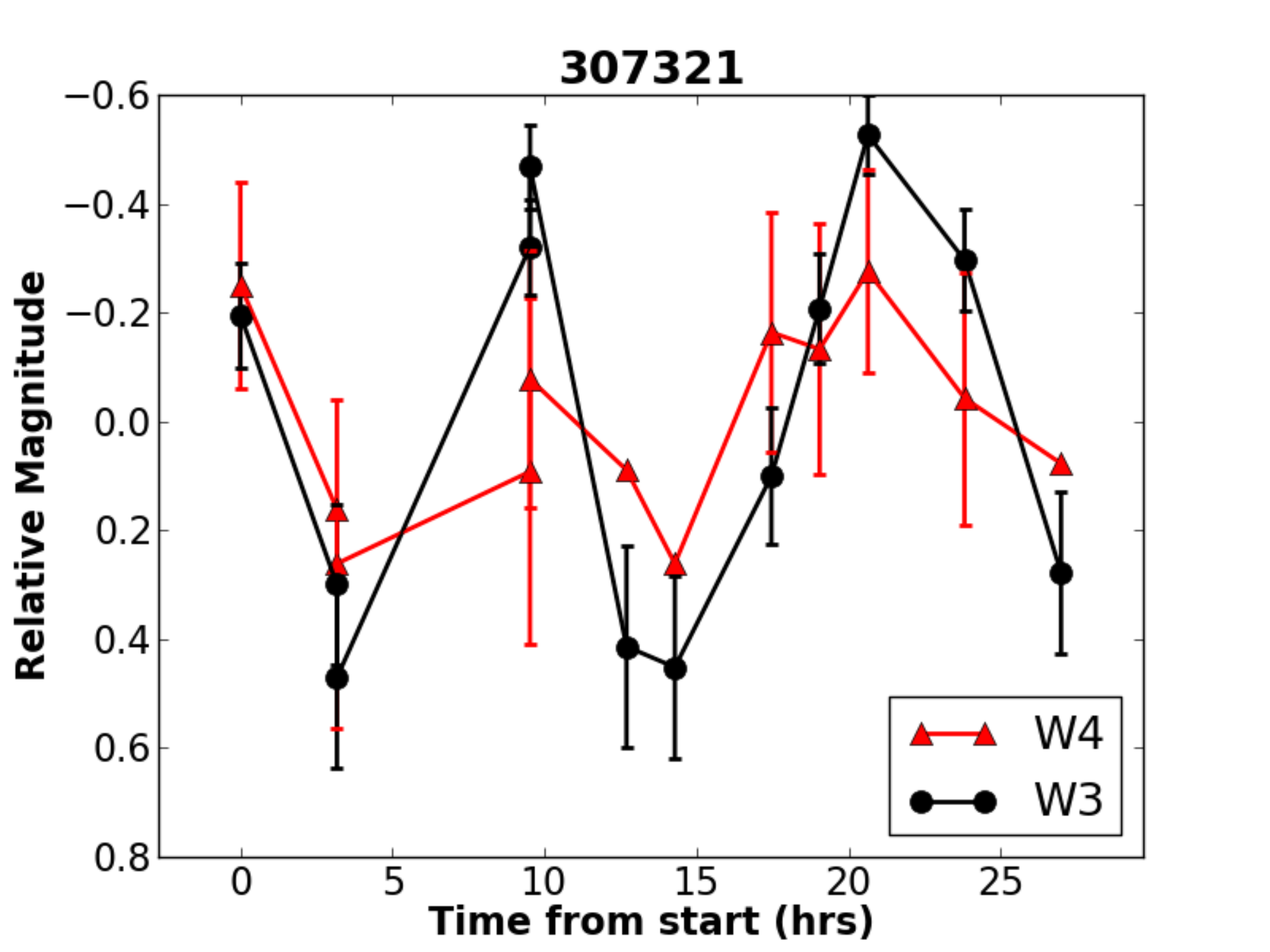}
\includegraphics[width=3.5in,height=2.6in]{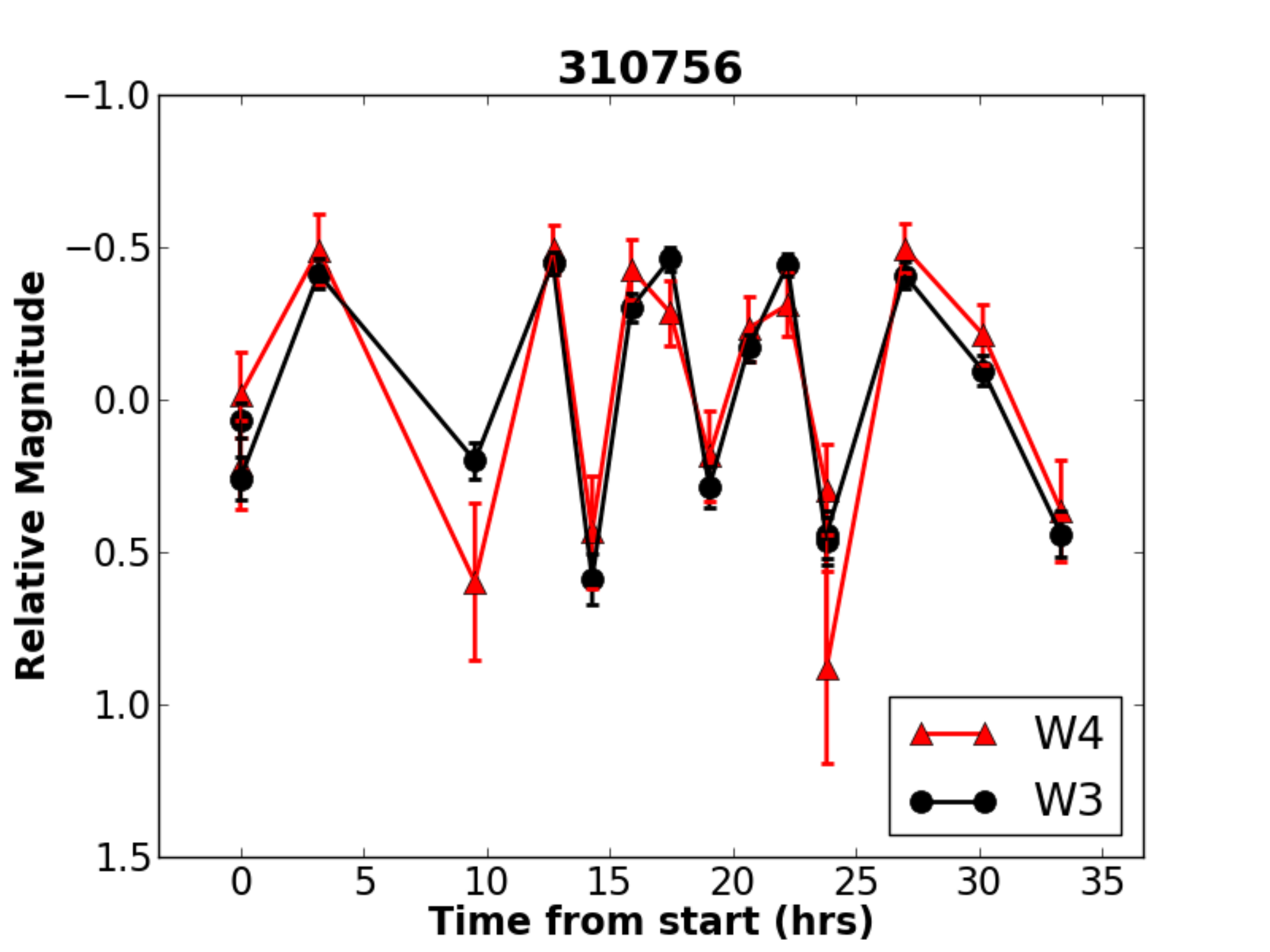}
\includegraphics[width=3.5in,height=2.6in]{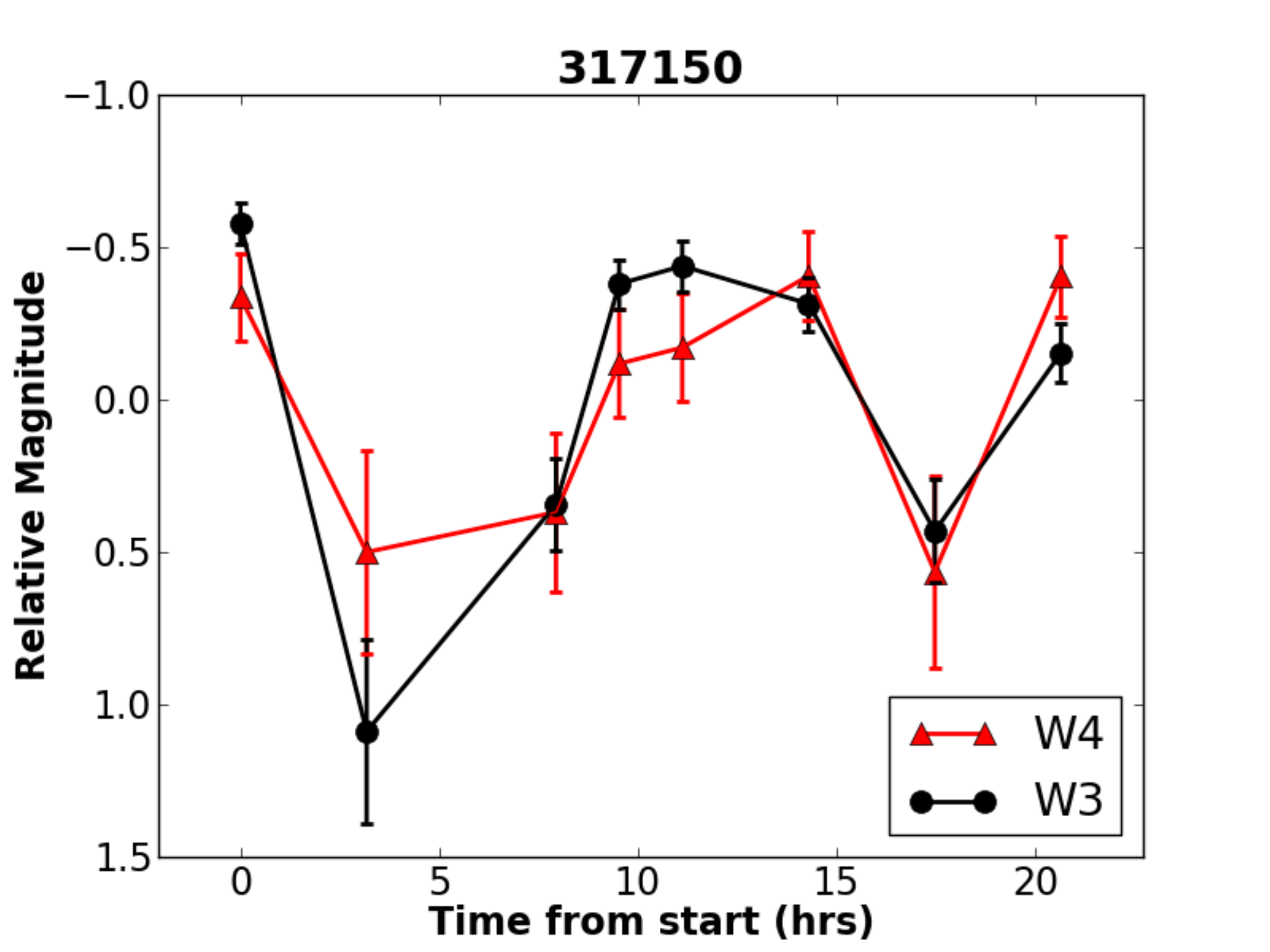}
\includegraphics[width=3.5in,height=2.6in]{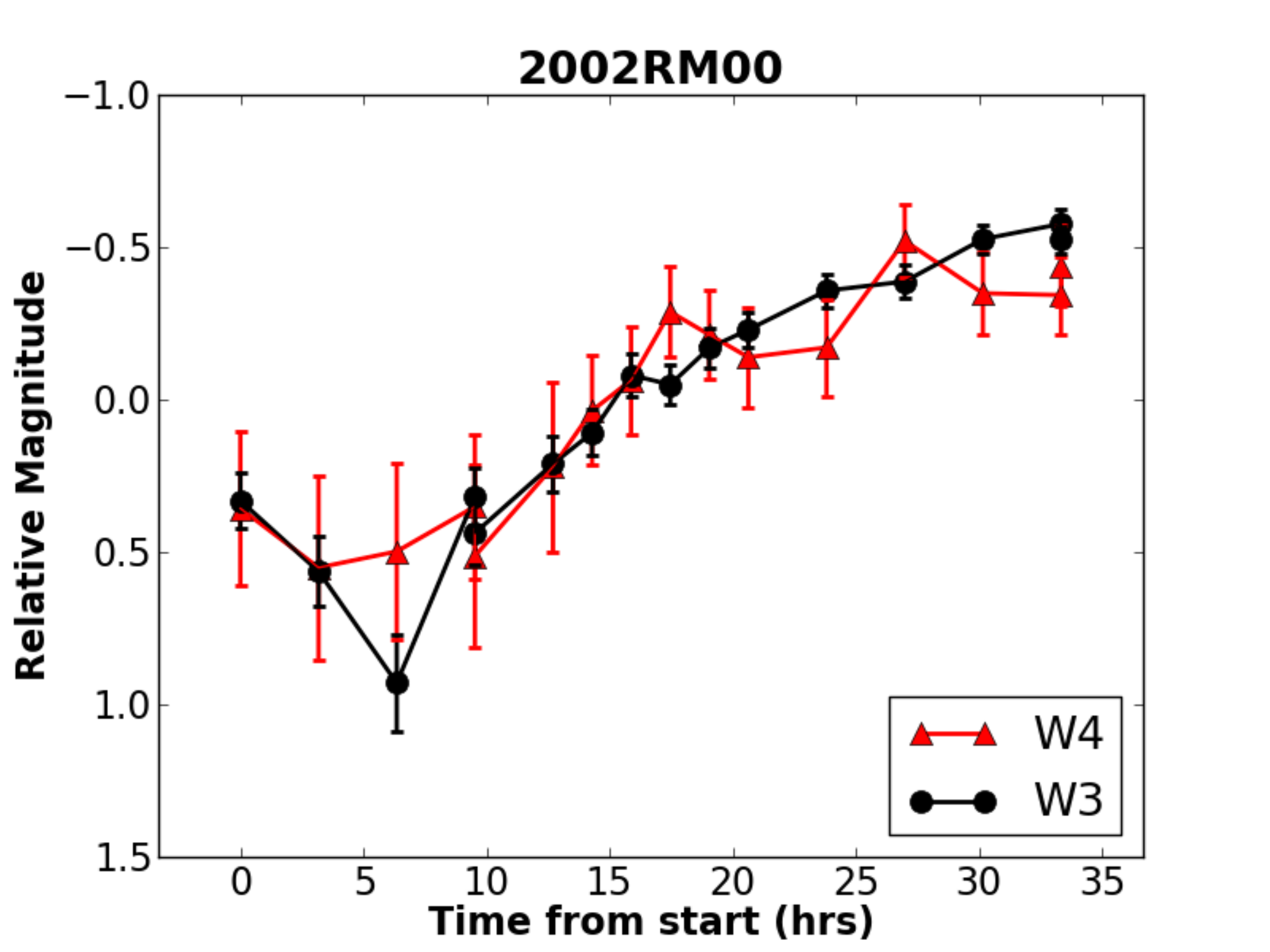}
\caption{\label{Fig:HildaCandidates} Candidate binary Hildas from our survey identified by their anomalously high lightcurve photometric ranges.}
\end{figure} 

\begin{figure}
\figurenum{4}
\includegraphics[width=3.5in,height=2.6in]{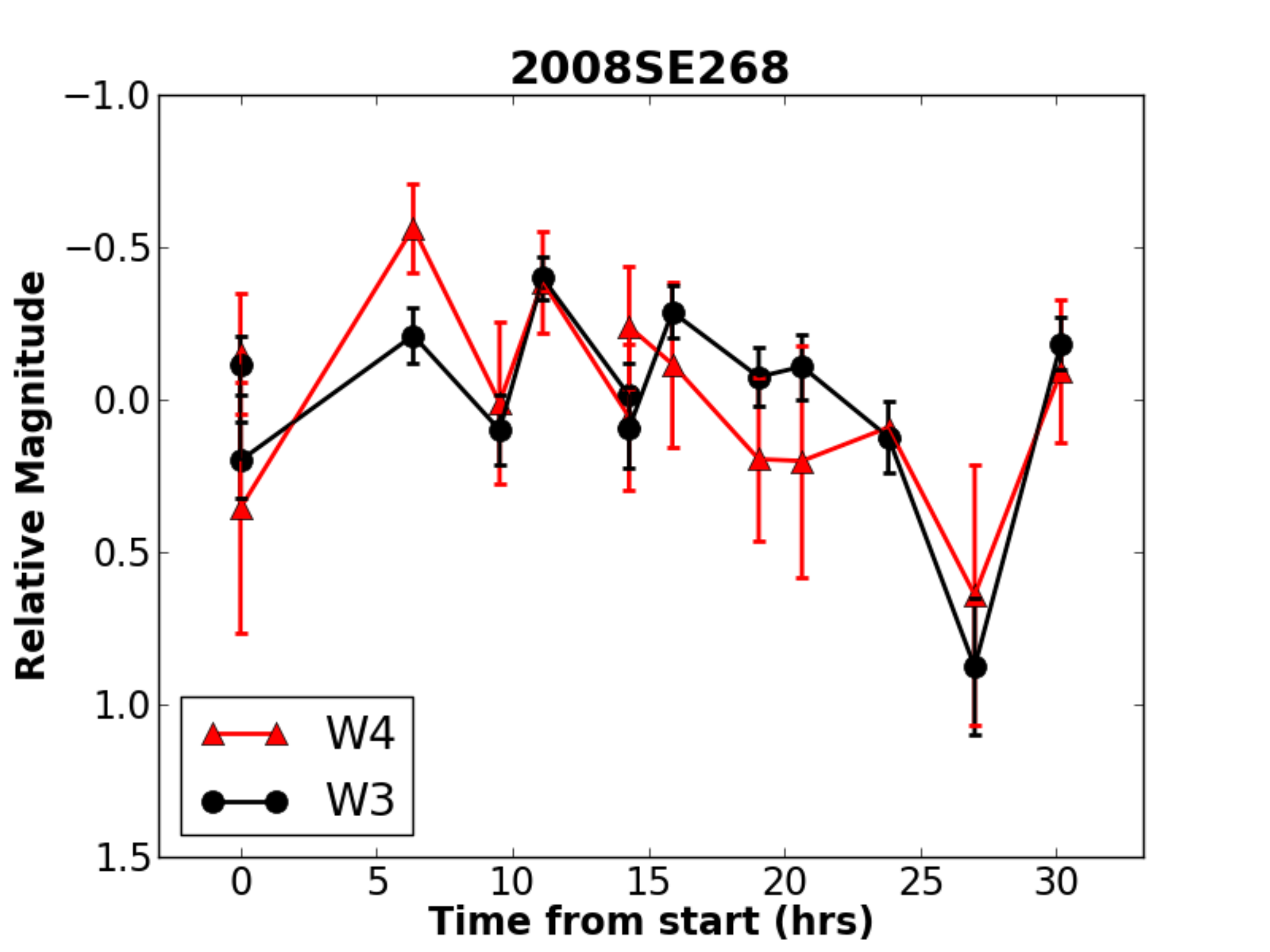}
\includegraphics[width=3.5in,height=2.6in]{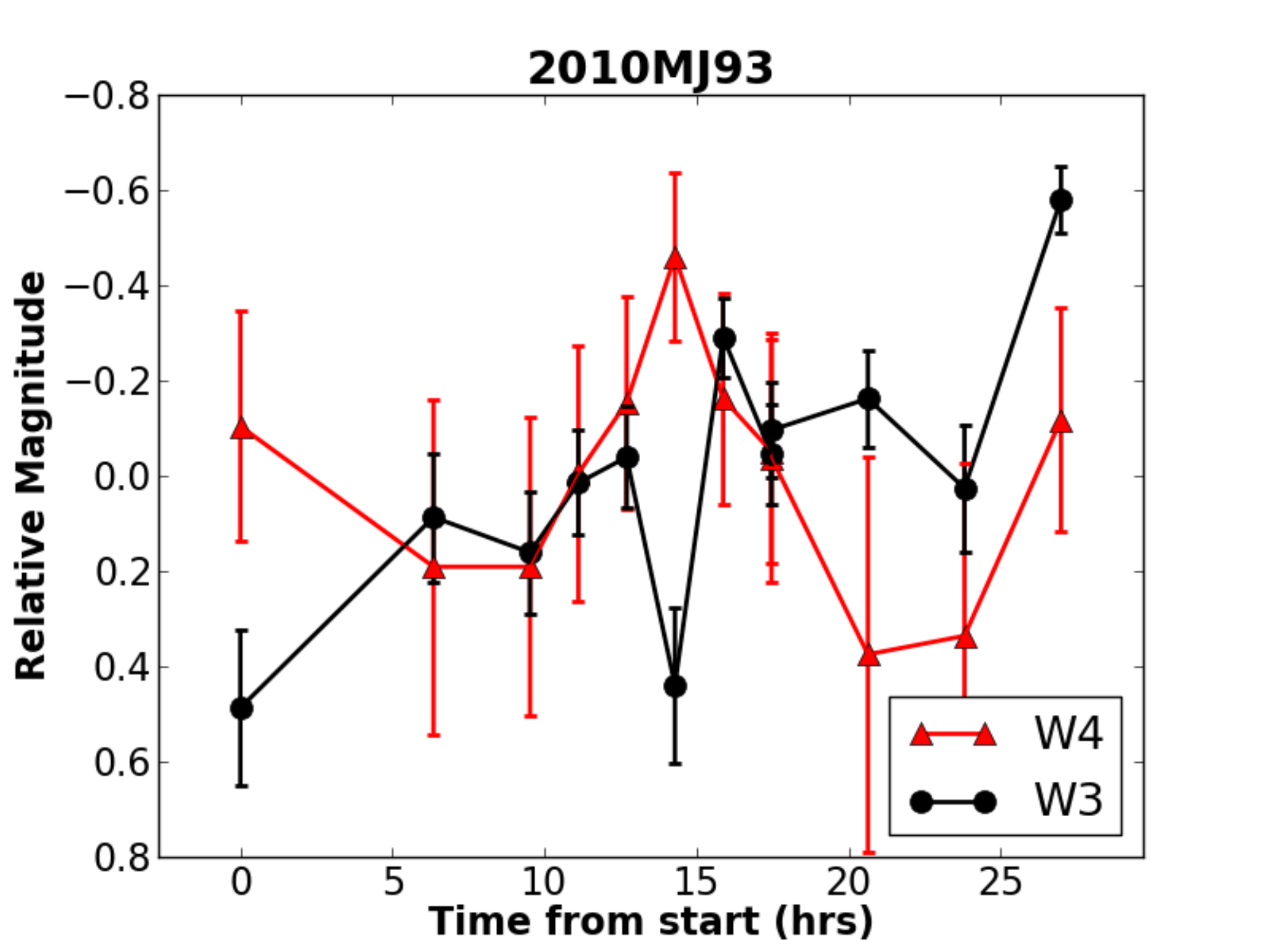}
\includegraphics[width=3.5in,height=2.6in]{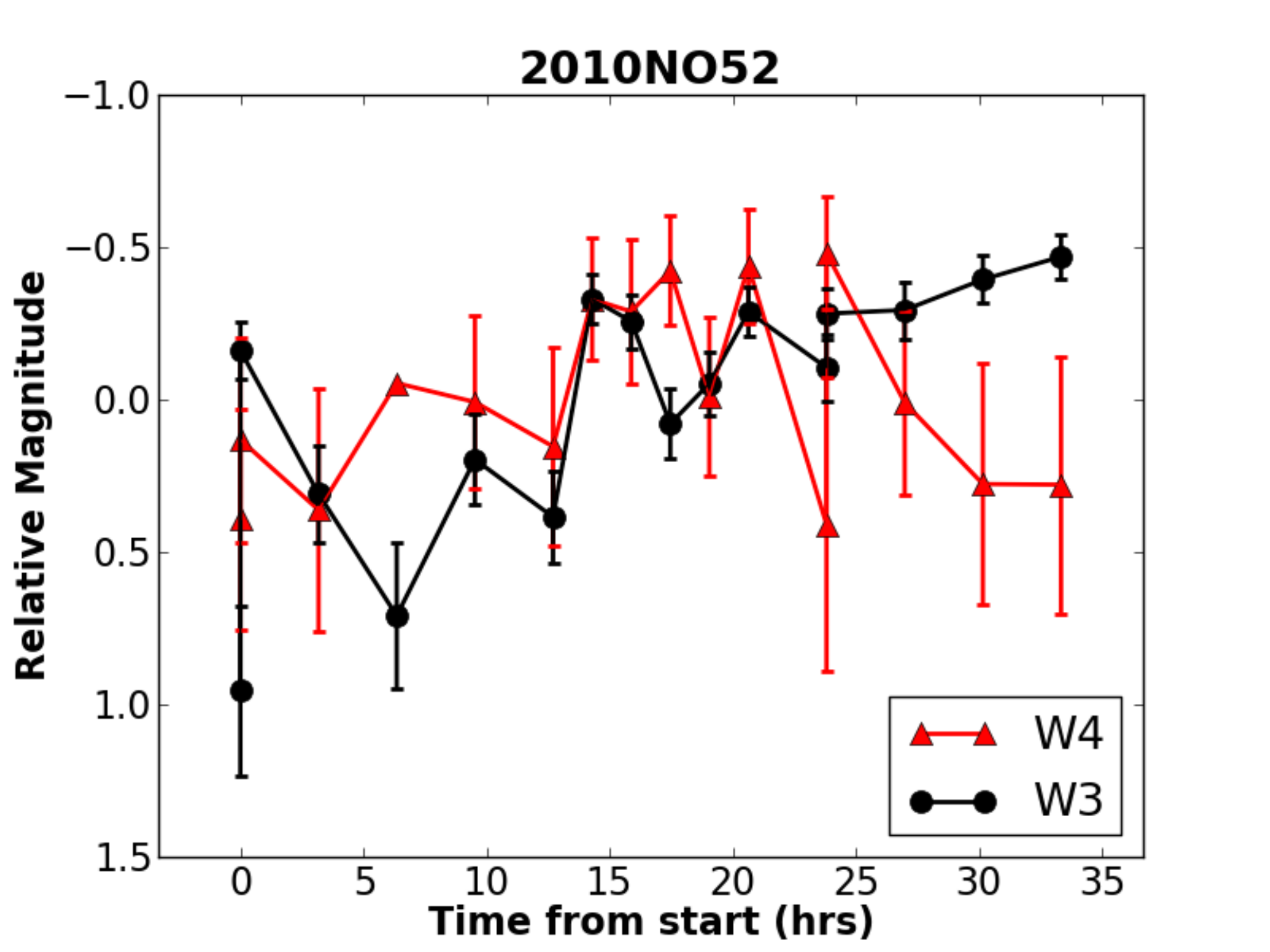}
\includegraphics[width=3.5in,height=2.6in]{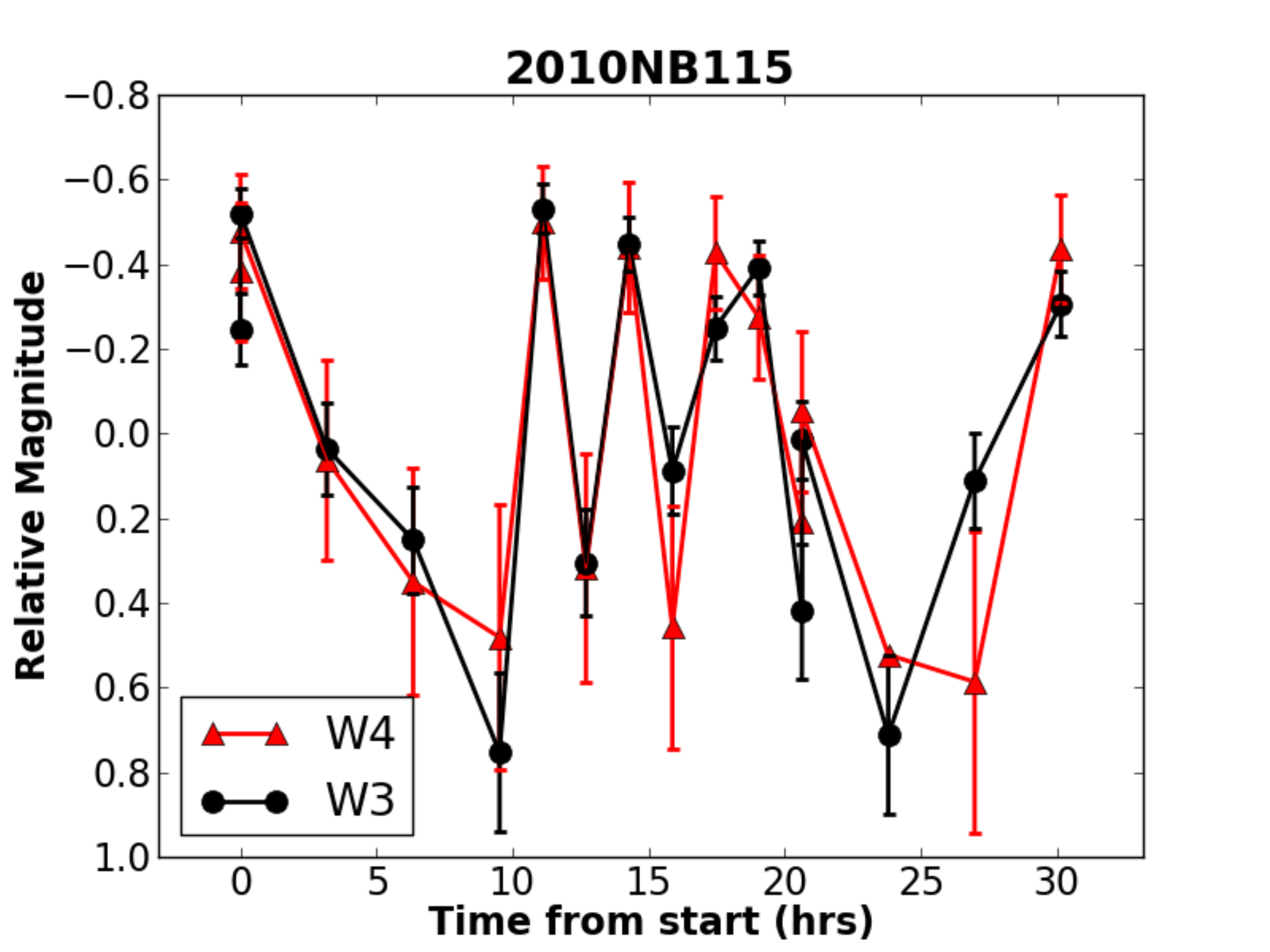}
\includegraphics[width=3.5in,height=2.6in]{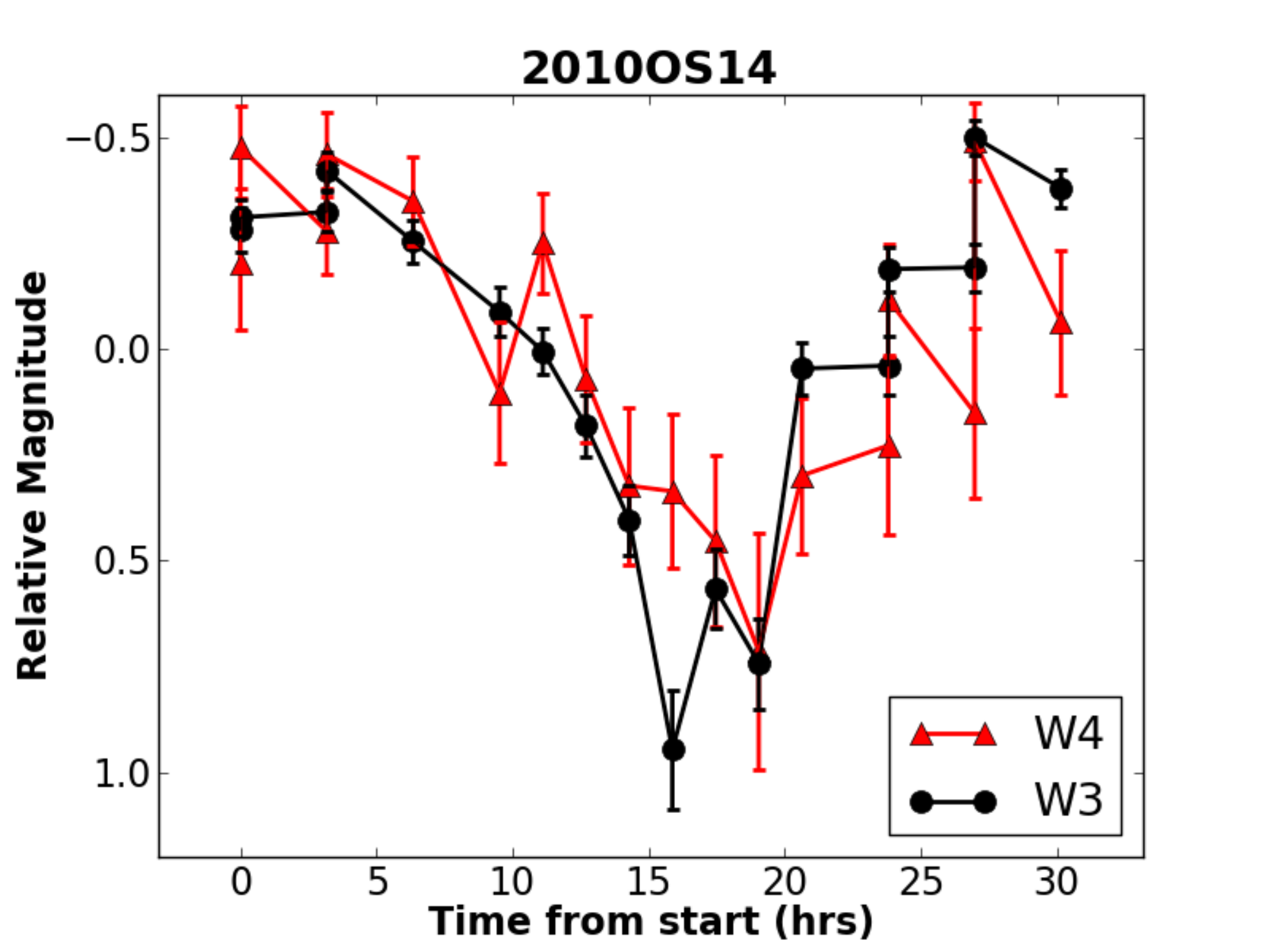}
\caption{\label{Fig:HildaCandidates} Candidate binary Hildas from our survey identified by their anomalously high lightcurve photometric ranges.}
\end{figure}

\end{document}